\documentclass[a4paper,11pt]{article}
\pdfoutput=1 % if your are submitting a pdflatex (i.e. if you have
\usepackage{jheppub}
\usepackage[T1]{fontenc}

\newcommand{\bea}{\begin{eqnarray}}  
\newcommand{\eea}{\end{eqnarray}}  

\newcommand{\ba}{\begin{array}}
\newcommand{\ea}{\end{array}}

\title{
The Full Lepton Flavor of the Littlest Higgs Model with T--parity
}

\author[a]{Francisco del Aguila,}
\author[b]{Lluis Ametller,}
\author[a]{Jose Ignacio Illana,}
\author[a]{Jose Santiago,}
\author[b]{Pere Talavera,}
\author[a]{and Roberto Vega-Morales~}

\affiliation[a]{CAFPE and Departamento de F{\'\i}sica Te{\'o}rica y del Cosmos, Universidad
de Granada, E\textendash{}18071 Granada, Spain}
\affiliation[b]{Departament de F\'{\i}sica, Universitat Polit{\`e}cnica de Catalunya, 
E-08034 Barcelona, Spain}

\emailAdd{faguila@ugr.es}
\emailAdd{lluis.ametller@upc.edu}
\emailAdd{jillana@ugr.es}
\emailAdd{jsantiago@ugr.es}
\emailAdd{pere.talavera@icc.ub.edu}
\emailAdd{rvegamorales@ugr.es}

\abstract{We re-examine lepton flavor violation (LFV) in the Littlest Higgs model with T--parity (LHT) 
	including the full T--odd (non-singlet) lepton and Goldstone sectors. 
	The heavy leptons induce two independent sources of LFV associated with the couplings 
	necessary to give masses to the T--odd mirror fermions and to their partners in right-handed 
	$SO(5)$ multiplets, respectively. 
	The latter, which have been neglected in the past, can be decoupled from gauge mediated 
	processes but not from Higgs mediated ones and must therefore also be included in a 
	general analysis of LFV in the LHT. 
	We also further extend previous analyses by considering on-shell $Z$ and 
	Higgs LFV decays together with the LFV processes at low momentum transfer. 
	We show that current experimental limits can probe the LHT parameter space up to 
	global symmetry breaking scales $f \sim 10$ TeV. 
	For lower $f$ values $\gtrsim 1$ TeV, $\mu-e$ transitions require the misalignment 
	between the heavy and the Standard Model charged leptons to be $\lesssim 1 \, \%$. 
	Future LFV experiments using intense muon beams should be sensitive to misalignments 
	below the per mille level. 
	For $\tau$ LFV transitions, which could potentially be observed at Belle II and the LHC as well as 
	future lepton colliders, we find that generically they can not discriminate between the LHT and 
	supersymmetric models though in some regions of parameter space this may be possible.}

\begin{document}
\maketitle
\flushbottom

\section{Introduction}
\label{Introduction}

The discovery of a Higgs boson at the Large Hadron Collider (LHC)~\cite{Aad:2012tfa,Chatrchyan:2012xdj} with Standard Model 
(SM) like properties~\cite{Falkowski:2013dza,Khachatryan:2016vau} appears to have settled the
nature of electroweak symmetry breaking and the focus now turns to
uncovering the mechanism responsible for stabilizing the electroweak
scale.~Generically, models which solve the `hierarchy problem'
predict, in accordance with the naturalness principle, new particles
at $\sim$ TeV.\footnote{There are exceptions such as the proposed
	`relaxion' mechanism~\cite{Graham:2015cka}.}
As the precision and energy of LHC measurements increases without uncovering signs of physics beyond the SM 
\cite{Falkowski:2015fla,Englert:2015hrx,Sirunyan:2017nrt,Aaboud:2018bun}, combined with constraints from 
electroweak precision data and other low energy experiments~\cite{deBlas:2013qqa,Falkowski:2014tna,deBlas:2015aea,Berthier:2015gja,Buckley:2015lku,deBlas:2016ojx,Falkowski:2017pss,Ellis:2018gqa,Grojean:2018dqj,Almeida:2018cld}, 
the apparent fine tuning of the electroweak scale is brought into
sharper focus.~In the coming years this will motivate a rigorous
search program at the LHC and low energy experiments in order to
ensure no stone is left unturned in the search for a solution to the
hierarchy problem and reconciliation of the naturalness principle.  

The two most strongly motivated contenders for a natural theory of electroweak
symmetry breaking are Supersymmetry and composite Higgs models where the Higgs boson arises as a 
pseudo Goldstone boson.~Among the latter, Little Higgs Models 
\cite{ArkaniHamed:2001nc,ArkaniHamed:2001ca,ArkaniHamed:2002qy} with T--parity 
\cite{Cheng:2003ju,Cheng:2004yc,Cheng:2005as} provide an elegant and calculable framework at the lowest orders 
and up to scales $\sim 10$ TeV thanks to the collective symmetry breaking mechanism which protects the Higgs 
mass from quadratic divergences at one loop.~The discrete T--parity also ensures that new particles cannot be 
singly produced or contribute at tree level to observables involving only SM particles, thus significantly relaxing 
direct and indirect constraints~\cite{Hubisz:2005tx}.~We focus on the particular case of the Littlest Higgs Model 
with T-Parity (LHT)~\cite{Low:2004xc} which has received considerable attention in many phenomenological studies 
\cite{Hubisz:2004ft,Hubisz:2005bd,Chen:2006cs,Blanke:2006sb,Buras:2006wk,Belyaev:2006jh,Blanke:2006eb,Hill:2007nz,Hill:2007zv,Han:2008wb,Goto:2008fj,delAguila:2008zu,Blanke:2009am,delAguila:2010nv,Goto:2010sn,Zhou:2012cja,Reuter:2013iya,Yang:2016hrh,Dercks:2018hgz}. (See for a review \cite{Schmaltz:2005ky,Perelstein:2005ka,Panico:2015jxa}.)

The stringent experimental limits on Flavor Changing Neutral Currents (FCNC)
also pose an additional `flavor hierarchy problem' to any SM extension at the TeV scale.~In the SM it is the 
Glashow-Iliopoulos-Maiani (GIM) mechanism which implies loop-suppressed and very small FCNC~\cite{Glashow:1970gm}. 
Thus the GIM mechanism naturally explains the observed suppression of rare flavor violating processes.\footnote{The SM 
	does not explain the large ratios between fermion masses of different families nor the observed charged current mixings, 
	but FCNC are naturally suppressed.}
By the same token, such stringent bounds strongly constrain any TeV extension of the SM.~Indeed, in the absence of any 
complementary mechanism aligning the new physics with the SM Yukawa couplings, present bounds on FCNC restrict the 
scale of flavor violation to be greater than $\sim 10^{3-5}$ TeV for couplings ${\cal{O}}(1)$ depending on the quark flavor 
transition~\cite{Isidori:2010kg}, or even higher for some lepton transitions~\cite{Raidal:2008jk} (see~\cite{Pich:2018njk} for 
a recent review).~Thus any extension of the SM must allow for small mixings and ideally provide an explanation for the alignment 
of the new flavor violating sources with the SM Yukawa couplings \cite{Georgi:1986ku} as for example in Planck scale supergravity 
theories~\cite{AlvarezGaume:1983gj} where this alignment results from the universal origin of the soft breaking sfermion masses. 
In general this is not the case for composite Higgs models although they can accommodate the required (small) misalignment 
(see, however, \cite{Kaplan:1991dc,Csaki:2008qq,Santiago:2008vq,Csaki:2008eh,delAguila:2010vg}). 
~Nonetheless, models with T--parity allow for a lower scale of new physics and a lower scale of flavor violation, mitigating the flavor 
hierarchy problem.\footnote{In this respect T--parity plays a similar role to R-parity in Supersymmetry.}
As the bounds are particularly precise in the leptonic case and the lepton mixing is unrelated to the mixing in the quark sector, 
one can study the implications of lepton flavor constraints independently, 
as we do in the following.

A number of phenomenological studies have examined the possibility
of lepton flavor violation (LFV) in low-energy experiments in the LHT 
\cite{delAguila:2008zu,Blanke:2009am,delAguila:2010nv,Goto:2010sn,Yang:2016hrh}.~These have focused on LFV induced by the 
misalignment between the mass eigenstates of the light SM--like leptons and the heavy T--odd `mirror' leptons.~However, as we 
emphasized recently~\cite{delAguila:2017ugt}, there is an additional independent source of LFV in the LHT.~This LFV is induced by 
the misalignment between the T--odd mirror leptons and the T--odd right-handed `partner' leptons required to maintain the $SO(5)$ 
global symmetry protecting the Higgs mass from dangerous quartic divergences at higher orders~\cite{Cheng:2004yc}.~This new 
source of LFV has been overlooked in the past because the mass of the partner leptons is independent of the $SU(5)$ breaking 
and they can be decoupled in LFV processes mediated by gauge bosons.~However, they do not decouple in Higgs-mediated LFV 
processes~\cite{delAguila:2017ugt} which require that all T--odd non-singlet leptons (and full Goldstone sector) be included to 
obtain a finite amplitude at one loop and avoid reintroducing a fine-tuning in the Higgs mass.~The need to consider the full set 
of new T--odd non-singlet particles opens the possibility that the contributions from the partner leptons can significantly 
alter the allowed parameter space in the LHT as compared to previous studies in which they were neglected.~This motivates a 
re-examination of LFV in the LHT in order to determine the full parameter space which is allowed by the stringent experimental constraints. 

In this paper we revise previous studies in the LHT on lepton transitions with small momentum transfer.~We extend previous 
analyses in two ways; first we include the full set of T--odd non-singlet leptons which introduces the new source of LFV; 
second we include high-energy LFV observables from LEP and the LHC as they can become competitive in certain regions of 
parameter space and in particular once the LHC enters a high luminosity 
phase.\footnote{High-$Q^2$ Drell-Yan LFV production at the LHC will be considered elsewhere.}
We also discuss the implications of these new contributions to the (flavor conserving) muon magnetic moment, 
$a_\mu$.\footnote{The typical size of these contributions is more than two orders of magnitude smaller than the present experimental 
	error.~Moreover, the apparent discrepancy between the measured value and the SM prediction of $a_\mu$ is coming under further 
	scrutiny \cite{Borsanyi:2017zdw,Blum:2018mom}.~On the other hand, the electric dipole moment of the electron $d_e$ cancels 
	at the order we consider. 
        However, the stringent experimental limit on this CP violating observable, 
		$|d_e| < 1.1 \times 10^{-29}\ e \cdot$cm 
		at 90 \% C.L.~\cite{Andreev:2018ayy}, imposes a non-trivial bound on the corresponding combination of CP violating phases in 
		the model at higher orders \cite{Panico:2018hal} (see also \cite{Chupp:2017rkp,Panico:2017vlk} and references there in). This 
		bound can be always fulfilled, although eventually 
		at the price of fine tuning. A more detailed discussion is gathered in the Appendix.\label{EDM}} 
(For a discussion on the interplay between the muon anomalous magnetic moment and LFV see 
\cite{Lindner:2016bgg}.) 
Analogous considerations hold in the quark sector, but we leave a study of this to ongoing work. 

The right-handed multiplet containing heavy leptons transforming under the $SO(5)$ global symmetry also includes an SM singlet 
whose T--parity can be chosen to be odd, as originally assumed \cite{Cheng:2003ju,Cheng:2004yc}, or even 
\cite{Cheng:2005as,delAguila:2008zu}. 
Although the quantum effects of these singlets depend on this choice~\cite{inpreparation}, they can be safely neglected in the processes 
considered here since, as emphasized in~\cite{delAguila:2017ugt} and demonstrated below, the gauge and Higgs mediated 
amplitudes involving two SM fermions are one-loop finite in their absence.~This allows for a consistent and quantitative phenomenological 
analysis, at the order to which we work, which leaves out the heavy SM singlet leptons and hence, it is independent 
of their T--parity assignment.\footnote{
		As thoroughly discussed in \cite{Pappadopulo:2010jx} T--parity must be properly defined in the fermion sector not to further break the $SU(5)$ global symmetry when giving large vector-like masses to the (T--odd) mirror fermions. Here we concentrate on the extra T--odd (non-singlet) contributions introduced when assigning the mirror fermion right-handed  counterparts to $SO(5)$ multiplets. These are completed with an extra SM doublet and an extra singlet. The mechanism advocated to give large masses to the former, which are also T--odd and are in particular needed to make the mirror fermion contributions to Higgs decay finite at one loop at order $v^2/f^2$ \cite{delAguila:2017ugt}, are implicitly assumed to give eventually only higher order corrections to the LFV observables considered. In this paper we only discuss the eventually large contributions of the T--odd mirror and partner mirror leptons. While the extra singlets are assumed to be heavier and their contributions smaller \cite{Cheng:2005as}. In Ref. \cite{Pappadopulo:2010jx} it is also proposed a new composite Higgs model with T--parity and with an enlarged global symmetry and scalar sector. These allow to give a mass to (T--odd) mirror fermions without introducing partner mirror fermions, and without breaking the global symmetry and without coupling them to the scalar multiplet containing the Higgs doublet. The role of the corresponding Yukawa couplings  together with the enlarged global symmetry and scalar sector make the phenomenology of this model quite different from the LHT one, being its study beyond the scope of this paper.
}

The phenomenological implications of the LHT are largely dictated by the global symmetry breaking scale $f$ with the new physics 
effects becoming negligible in the large $f$ limit.~Thus, electroweak precision data and LHC limits primarily constrain this parameter 
in the LHT as well as the magnitude of the Yukawa couplings $\kappa$ which generate masses for the T--odd mirror fermions 
(e.g.~\cite{Reuter:2013iya,Dercks:2018hgz} and \cite{Tonini:2014dza,Shim:2018ksu} for recent reviews).~The global symmetry 
breaking scale is constrained to be $f \gtrsim 1$~TeV while for the mirror quark masses we have 
$m_{q_H} = {\sqrt 2} \kappa_q f \gtrsim$~few TeV with less stringent limits on the mirror lepton masses $m_{l_H} = {\sqrt 2} \kappa_l f$. 
Our focus here is on rare LFV processes which provide complementary constraints and in particular imply quite stringent limits on the 
amount of `misalignment' between the heavy lepton flavor sector and the SM one. 
In our quantitative analysis we will take as benchmark values $f = 1.5$ TeV and $m_{q_{Hi}} = 2$ TeV 
\cite{Reuter:2018xfr}, allowing to vary the heavy lepton masses and mixings. 
While the LHT does not incorporate any specific mechanism to explain the number of families or contain any flavor symmetries, 
these could in principle be incorporated in a UV complete model.\footnote{
	An exception is the 
	top quark sector, which must be extended to implement the collective breaking protecting the Higgs mass 
	from the one-loop quadratic divergence due to the top quark Yukawa. Large effects induced by the quark sector 
	will be discussed elsewhere \cite{Pappadopulo:2010jx,Vecchi:2013bja}.} 
With this in mind, the LHT can easily accommodate the SM spectrum at low energies and the small mixings necessary to satisfy 
constraints over a range of parameters.~Below we show that in order to satisfy present limits on rare LFV processes, at least 
some alignment is needed if $f$ is below $\sim 10$ TeV while strong alignment is needed for $f \gtrsim 1$~TeV.~In particular, 
$\mu-e$ transitions require the alignment between the heavy LFV sources and the SM Yukawa couplings of the first two families 
to be $\lesssim 1 \, \%$ while branching ratios for LFV $Z$ and Higgs decays into $\tau \mu,\,\tau e$ can be as large as 
$\sim 10^{-7}$ for $\sim$~TeV masses and sizable mixing values of the T--odd particles.~These limits also serve as guidance 
for constructing UV models which contain the flavor symmetries needed to naturally explain the alignment. 

	The lepton sector with three families of T--odd (non-singlet) 
	mirror and partner mirror leptons involves 4 CP violating phases after fermion 
	field redefinitions. Although the LFV observables considered in the following 
	do depend on these phases, they do not appear to expand significantly 
	the range of variation of the LHT predictions for these observables. 
	(For related parity and time-reversal asymmetries see, for instance, \cite{Goto:2010sn}.) 
	However, this is not the case for the electric dipole moment of the electron, 
	which is a CP--odd observable preserving flavor and then vanishes 
	in the absence of CP violating phases. 
	(See footnote \ref{EDM} and the Appendix.)

Dedicated LFV experiments plan to exploit intense muon beams to increase their sensitivity by several orders of magnitude 
\cite{Baldini:2018uhj}.~Should no signal be observed, this will improve the current limit on the misalignment between the heavy 
leptons and two lightest SM lepton families by an order of magnitude to $\sim 1$\textperthousand.~Although LFV processes involving the third 
family do not significantly constrain the LHT at present, 
Belle II \cite{Liventsev:2018gin}, the LHC~\cite{Bediaga:2018lhg}, and/or future lepton colliders 
\cite{Benedikt:2018qee,Baer:2013cma} may eventually be sensitive. 
However, as we discuss in more detail below, when the effects of both the mirror and partner leptons are included, 
three-body $\tau$ decays will not be able to unambiguously discriminate between the LHT and supersymmetric 
models, in contrast to the case in which the partner leptons are decoupled 
\cite{Blanke:2009am,Blanke:2007db}.

In the next section we summarize the most restrictive LFV processes and the 
LHT parameter space to be confronted with the corresponding experimental 
limits.~In Section \ref{sec:LFVprocessesLHT} various amplitudes for LFV processes 
are computed including two and three body $\tau$ and $\mu$ decays as well as 
$\mu \rightarrow e$ transitions and $Z\to\bar{\ell}\ell'$ decays.~The relevant details 
of the LHT are reviewed in Ref.~\cite{delAguila:2017ugt} where one can find the 
Feynman rules necessary for calculating the amplitude for LFV Higgs decays while 
those needed for computing amplitudes of gauge mediated processes are completed 
here.~The predictions of the LHT are then confronted with experimental data in Section 
\ref{sec:ConfrontingLFVandLHT}, 
	emphasizing the changes of the LHT predictions when the 
	partner mirror lepton contributions are also included. Contrary to the case 
	without partner mirror leptons, they can not be distinguished from the 
	supersymmetric ones in general. 
Prospects at future facilities using intense muon beams as well as Belle II, 
LHC and future lepton colliders are reviewed and examined in Section \ref{Futureprospects}. 
Finally, Section \ref{Conclusions} is devoted to discussion and summary. 
A study of the muon magnetic moment 
and of the electric dipole moment of the electron are 
also included in the Appendix. 

%-------------------------------- EXPERIMENTAL CONSTRAINTS 
%                                  AND LHT PARAMETRIZATION ---------------------------------%

\section{Lepton flavor violating limits and LHT parametrization} 
\label{sec:LFVandLHT} 

LFV processes provide a powerful probe of the flavor structure of
new physics models.~We gather in Table~\ref{Limits} the present limits on lepton flavor changing transitions mediated by 
electroweak gauge and Higgs bosons.
%
%%%%%%%%LFV constraints Table%%%%%%%%%%%%%%%%%s
\begin{table}
	\hspace*{-.1cm}
	\begin{center}
		\begin{tabular}{|c|c|c|c|c|c|}\hline
			%%%%%%%%%%%%%%%%%%%%%%%
			Branching Ratio& $90\%$~C.L.~Bound&Ref.& 
			Branching Ratio& $90\%$~C.L.~Bound&Ref.\\\hline
			%%%%%%%%%%%%%%%%%%%%%%%
			$\mu \to e\ \gamma$ &~$4.2 \times 10^{-13}$~&\cite{Adam:2013mnn}&
			$\mu \to e\ \overline{e}\ e$ &~$1.0\times 10^{-12}$~&\cite{Bellgardt:1987du}\\\hline
			%%%%%%%%%%%%%%%%%%%%%%%
			Conversion Rate&  & &&&\\\hline
			$\mu \to e\ {\rm{(Au)}}$ &~$7.0\times 10^{-13}$~&\cite{Bertl:2006up} &&&\\\hline
			$\mu \to e\ {\rm{(Ti)}}$ &~$4.3\times 10^{-12}$~&\cite{Bertl:2006up} &&&\\\hline\hline     
			%%%%%%%%%%%%%%%%%%%%%%%
			Branching Ratio& ~~ &~~&~~~& ~~ &~~\\\hline
			$\tau \to e\ \gamma$~ &~$3.3 \times 10^{-8}$
			&\cite{Aubert:2009ag}&
			$\tau \to \mu\ \overline{e}\ \mu$ &~$1.7\times 10^{-8}$~&\cite{Tanabashi:2018oca}\\\hline
			$\tau \to \mu\ \gamma$~&~$4.4 \times 10^{-8}$~
			&\cite{Aubert:2009ag}&
			$\tau \to e\ \overline{\mu}\ e$ &~$1.5\times 10^{-8}$~&\cite{Tanabashi:2018oca}\\\hline
			&&&
			$\tau \to \mu\ \overline{e}\ e$ &~$1.8\times 10^{-8}$~&\cite{Tanabashi:2018oca}\\\hline
			&&&
			$\tau \to e\ \overline{\mu}\ \mu$ &~$2.7\times 10^{-8}$~&\cite{Tanabashi:2018oca}\\\hline
			&&&
			$\tau \to e\ \overline{e}\ e$ &~$2.7\times 10^{-8}$~&\cite{Tanabashi:2018oca}\\\hline
			&&&
			$\tau \to \mu\ \overline{\mu}\ \mu$ &~$2.1\times 10^{-8}$~&\cite{Tanabashi:2018oca}\\\hline\hline
			%%%%%%%%%%%%%%%%%%%%%%%
			~~& $95\%$~C.L.~Bound &~~&~~& $95\%$~C.L.~Bound&~~\\\hline
			$Z \to \mu\ e$ &$7.3\times 10^{-7}$&\cite{Nehrkorn:2017fyt}&
			$h \to \mu\ e$ &$3.5\times 10^{-4}$&\cite{Khachatryan:2016rke}\\\hline
			$Z \to \tau\ e$ &$9.8 \times 10^{-6}$&\cite{Akers:1995gz}&
			$h \to \tau\ e$ &$6.2 \times 10^{-3}$&\cite{Sirunyan:2017xzt}\\\hline
			$Z \to \tau\ \mu$ &$1.2 \times 10^{-5}$&\cite{Abreu:1996mj}&
			$h \to \tau\ \mu$ &$2.5 \times 10^{-3}$&\cite{Sirunyan:2017xzt}\\\hline
			%      $\mu^- \to e^- \bar{\nu}_\mu \nu_e$ &~$1.2\times 10^{-2}$~&~~\cite{Agashe:2014kda}~~\\\hline
		\end{tabular}
		\caption{Limits from lepton flavor violating processes mediated by electroweak gauge and Higgs bosons. 
			Although is flavor conserving, we also compute the contribution to the muon magnetic moment 
			(see Appendix), whose current experimental value $a_\mu = (116592091\pm63)\times 10^{-11}$
			~\cite{Tanabashi:2018oca}.}
		\label{Limits}
	\end{center}
\end{table}
%%%%%%%%%%%%%%%%%%%%%%%%%s
%
As we discuss in detail below, the corresponding LHT amplitudes  
are one-loop suppressed, as required by T--parity, and result from effective 
operators of dimension 6.~They are therefore universally suppressed~\cite{delAguila:2008zu,delAguila:2010nv,delAguila:2017ugt} 
by a factor of $(1 /4\, \pi)^2 \times (v / f)^2$.~Thus, although they are quite stringent, this suppression factor alone largely accounts 
for many of the bounds in Table~\ref{Limits} to be easily satisfied for $f\gtrsim 1$~TeV.~The strongest constraint results from the 
most recent measurements of $\mu \rightarrow e \gamma$~\cite{Adam:2013mnn} with bounds involving $\tau$ leptons in general 
being less restrictive and not very sensitive to the T--odd masses under consideration.~Only in the case of LFV Higgs decays, 
which are proportional to the light SM lepton masses, are channels involving $\tau$ leptons  the more sensitive ones.

The LHT has recently been reviewed in~\cite{delAguila:2017ugt} to which we refer the reader for details.~Here we give the expressions 
for the LFV amplitudes mediated by gauge bosons including the full T--odd sector while those for $h \to \overline{\ell} \ell'$ decays are 
given in~\cite{delAguila:2017ugt} and used here for our phenomenological analysis.~These can all be written in terms of form factors 
which, apart from the common suppression factor mentioned above, have a generic dependence on the different mixing matrix elements 
similar to those mediated by Higgs bosons~\cite{delAguila:2017ugt}.~As we will discuss more explicitly below, we can write the two-body decays 
$\ell \to \ell' \gamma$, $Z\to \overline{\ell} \ell'$, and $h \to \overline{\ell} \ell'$ in terms of three-point form factors with the generic form,
\begin{align}
	\label{ffs}
	{\cal{F}}_3 =\ & \sum_i V^\dagger_{\ell' i} V_{i \ell}\ {F}_3(m_{\ell_{H i}}, ...)
	\nonumber \\ 
	+\ & \sum_{i,j,k} V^\dagger_{\ell' i} \frac{m_{\ell_{H i}}}{M_{W_H}} W^\dagger_{ij} W_{jk} 
	\frac{m_{\ell_{H k}}}{M_{W_H}} V_{k \ell}
	\ {G}_3(m_{\tilde{\nu}^c_j}, ...)\ , 
\end{align}
where $V_{i \ell}$ are the matrix elements of the $3\times 3$ unitary mixing matrix parametrizing the misalignment between the SM 
left-handed charged leptons $\ell$ with the heavy mirror ones $l_H$.~The $W_{j k}$ are the matrix elements of the $3\times 3$ 
unitary mixing matrix parametrizing the misalignment between the mirror leptons and their partners $\tilde{l}^c$ in the $SO(5)$ 
(right-handed) multiplets~\cite{delAguila:2017ugt}.~Note that the mass eigenstates $l_H$ and $\tilde{l}^c$ are both $SU(2)_L$ 
doublets with vector-like masses.~The dots in the loop functions stand for the masses of the heavy boson fields running in the 
loop, which can be the T--odd heavy gauge bosons $W_H,  A_H, Z_H$ or the electroweak triplet scalar $\Phi$.~All heavy fields 
are T--odd and have masses 
$\sim f$.~The first term in Eq.\,(\ref{ffs}) corresponds to the contribution from the mirror leptons that has been considered in 
previous studies (in the $Q^2=0$ limit) whereas the second term contains the new source of LFV induced by the partner leptons 
usually assumed to be decoupled.~The explicit dependence of ${G}_3$ on the mirror lepton masses $m_{\ell_{H}}$ only occurs 
for the $h\to \overline{\ell} \ell'$ process (and is related to the 
non--decoupling behavior of these amplitudes~\cite{delAguila:2017ugt}).

At the order to which we work, the three-body decays $\ell \to \ell'\overline{\ell'}\ell',\, \ell'\overline{\ell''}\ell''$ also receive contributions 
from some of the three-point form factors in Eq.\,(\ref{ffs}).~However, as we examine below, the `double flavor' violating three-body 
decay $\ell \to \ell'\overline{\ell''}\ell'$ does not receive contributions from these three-point form factors.\footnote{As we discuss below, 
	unlike the single flavor violating decays $\ell \to \ell'\overline{\ell'}\ell',\, \ell'\overline{\ell''}\ell''$, 
	the double flavor changing transitions $\ell \to \ell'\overline{\ell''}\ell'$ \emph{only} 
	receive contributions at 1-loop from box diagrams while $Z$ and $\gamma$ 
	penguin diagrams (see Figure \ref{diagrams1}) do not contribute until 2-loops.\label{order}} 
On the other hand all three three-body decays receive contributions from four-point form factors which we can write for a generic 
decay $\ell \to \ell'\overline{\ell''}\ell'''$ as, 
\begin{align}
	{\cal{F}}_4 = \sum_{i,j} \chi^{\ell \ell' \ell'' \ell'''}_{ij} 
	{F}_4(m_{\ell_{H i}},m_{\ell_{H j}}, \dots ) 
	+ \sum_{i,j} \tilde{\chi}^{\ell \ell' \ell'' \ell'''}_{ij} 
	{G}_4(m_{\tilde{\nu}^c_i},m_{\tilde{\nu}^c_j}, \dots) \ ,
\end{align}
where we have defined the (flavor) mixing coefficients 
\begin{align}
	\label{chiij}
	\chi_{ij}^{\ell \ell' \ell'' \ell'''}= & \ V^\dagger_{\ell' i}V_{i \ell} \ V^\dagger_{\ell''' j} V_{j \ell''} + 
	(\ell' \leftrightarrow \ell''') \  , \\
	\label{tildechiij}
	\tilde{\chi}_{ij}^{\ell \ell' \ell'' \ell'''}= & 
	\sum_{k,n,r,s}  V^\dagger_{\ell' k} \frac{m_{\ell_{H k}}}{M_{W_H}} W^\dagger_{k i} 
	W_{i n} \frac{m_{\ell_{H n}}}{M_{W_H}} V_{n \ell} \ 
	V^\dagger_{\ell''' r} \frac{m_{\ell_{H r}}}{M_{W_H}} W^\dagger_{r j}
	W_{j s} \frac{m_{\ell_{H s}}}{M_{W_H}} V_{s \ell''} + 
	(\ell' \leftrightarrow \ell''') \ .
\end{align}
Again we have separated the contributions from the two sources of LFV.~Similarly, the amplitudes for $\mu \to e$ conversion 
in nuclei can be written in terms of analogous form factors with the appropriate insertion of lepton and quark (flavor) combinations 
(see below).~Obviously, the corresponding mixing coefficients do not have a crossed term, in contrast 
to the four-lepton mixing in Eqs. (\ref{chiij}) and (\ref{tildechiij}), because lepton and baryon number 
are preserved in the model.\footnote{SM neutrinos can be assumed to be massless throughout this work.} 
We perform various scans on these mixings and masses parametrizing the LHT and present results in 
Section~\ref{sec:ConfrontingLFVandLHT}.~Before doing so we first present the amplitudes for LFV processes mediated by gauge 
bosons including for the first time the full T--odd sector.

%----------- LFV PROCESSES IN THE LHT-----------%

\section{Lepton flavor violating processes in the LHT} 
\label{sec:LFVprocessesLHT}

The contributions to LFV processes of T--odd particles in the LHT, which can be mediated by gauge or Higgs bosons, have been discussed a number of times in the past.~In the former case only the contributions from the T--odd sector \emph{without} the heavy partner leptons have been considered~\cite{delAguila:2008zu,delAguila:2010nv} and furthermore, only at low $Q^2 \sim 0$.~This is justified because these contributions alone sum to a finite amplitude while the contributions from the partner leptons decouple when their masses are taken to be very large.~In contrast, for Higgs mediated processes, without including the partner leptons these amplitudes are not finite due to divergences introduced by the mirror leptons~\cite{delAguila:2017ugt}.~As a consequence, the partner leptons can not be considered infinitely heavy and with their presence in the spectrum an additional source of LFV is introduced.~In particular, the LFV processes mediated by gauge bosons at low $Q^2$ must be recalculated adding the contributions of the T--odd partner leptons which in turn introduces new contributions involving the T--odd electroweak triplet $\Phi$.~We recalculate here these amplitudes in the low $Q^2$ limit and in addition at $Q^2 = M_Z^2$ when we consider LFV on-shell $Z$ decays, which have not been previously examined. 

The LHT Lagrangian and our conventions are presented in~\cite{delAguila:2017ugt} to which we refer the reader for details.~The Feynman rules necessary to calculate LFV amplitudes mediated by gauge bosons are collected in Tables~\ref{Zg:VFF} and~\ref{Zg:SVV-SSV} and include the extra Feynman rules for the new partner lepton and $\Phi$ contributions which were not included in previous studies~\cite{delAguila:2008zu}. 
\begin{table}
	\begin{center}
		\begin{tabular}{c||c|c}
			[V$_\mu$FF] & $g_L$ & $g_R$ \\
%			&&\\[-0.6cm]
			\hline
			&&\\[-0.45cm]
			$\gamma\ \overline{\ell_{i}}\ \ell_j$ & $ \delta_{ij} e $ & $ \delta_{ij} e $ \\ 
%			&&\\[-0.6cm]
			$\gamma\ \overline{\ell_{H i}}\ \ell_{H j} $ & $ \delta_{ij} e $ & $ \delta_{ij} e $ \\
%			&&\\[-0.6cm]
			$\gamma\ \overline{\tilde{\ell^c_{i}}}\ \tilde{\ell^c_{j}} $ & $- \delta_{ij} e $ & $- \delta_{ij} e $ \\
%			&&\\[-0.6cm]
			$Z\ \overline{\ell_{i}}\ \ell_j$ & $ \delta_{ij} \frac{g}{2 c_W} (-1+2s_W^2) $ & $ \delta_{ij} g' s_W $ \\ 
%			&&\\[-0.6cm]
			$Z\ \overline{\nu_{H i}}\ \nu_{H j} $ & $ \delta_{ij} \frac{g}{2 c_W} $ & 
			$ \delta_{ij} \frac{g}{2 c_W} \left( 1 - \frac{v^2}{4 f^2} \right)$ \\
%			&&\\[-0.6cm]
			$Z\ \overline{\ell_{H i}}\ \ell_{H j}  $ & $\delta_{ij} \frac{g}{2 c_W} (-1+2s_W^2)$ & 
			$\delta_{ij} \frac{g}{2 c_W} (-1+2s_W^2)$ \\
%			&&\\[-0.6cm]
			$Z\ \overline{\tilde{\nu^c_{i}}}\ \tilde{\nu^c_{j}} $ & $- \delta_{ij} \frac{g}{2 c_W} \left( 1 - \frac{v^2}{4 f^2} \right)$ & 
			$- \delta_{ij} \frac{g}{2 c_W} \left( 1 - \frac{v^2}{4 f^2} \right)$ \\
%			&&\\[-0.6cm]
			$Z\ \overline{\tilde{\ell^c_{i}}}\ \tilde{\ell^c_{j}}  $ & $-\delta_{ij} \frac{g}{2 c_W} (-1+2s_W^2)$ & 
			$-\delta_{ij} \frac{g}{2 c_W} (-1+2s_W^2)$ \\
%			&&\\[-0.6cm]
			$A_H\ \overline{\ell_{H i}}\ \ell_{j}  $ & $\frac{g}{2} \left( \frac{t_W}{5} - x_H \frac{v^2}{f^2} \right) V_{ij}$ & 
			$0$ \\
%			&&\\[-0.6cm]
			$Z_H\ \overline{\ell_{H i}}\ \ell_{j} $ & $- \frac{g}{2} \left( 1 + x_H \frac{t_W}{5} \frac{v^2}{f^2} \right) V_{ij}$ & 
			$0$ \\
%			&&\\[-0.6cm]
			$W_H^+\ \overline{\nu_{H i}}\ \ell_{j}  $ & $\frac{g}{\sqrt 2} V_{ij}$ & 
			$0$ \\
		\end{tabular} 
	\end{center}
	\caption{Vector-Fermion-Fermion couplings at ${\cal O} (\frac{v^2}{f^2})$ 
		completing the Tables of Appendix B.2 in~\cite{delAguila:2008zu} and of \cite{delAguila:2017ugt} and 
		including the couplings of the partner lepton doublets $\tilde{l}^c_i = (\tilde{\nu}^c_i \ \tilde{\ell}_i^c)^T$.~Note the different vertex normalization used in~\cite{delAguila:2008zu}, where $e$ is factored out, and the opposite sign for charges of charge conjugate fields.~The couplings involving the neutral singlets $\chi_{i}$ needed to complete the $SO(5)$ multiplet vanish.}
	\label{Zg:VFF}
\end{table}
\begin{table}
	\begin{center}
		\begin{tabular}{c||ccc||c}
			[SV$_\mu$V$_\nu$] & $K$ & $\quad$ & [S$(p_1)$S$(p_2)$V$_\mu$] & $G$ \\
%			&&&&\\[-0.6cm]
			\cline{1-2} \cline{4-5}
			&&&&\\[-0.45cm]
			$\Phi^0\ Z_H\ Z$ & $- \frac{g^2}{\sqrt 2 c_W} \frac{v^2}{2 f}$ && 
			$\Phi^0\ \Phi^P\ Z$ & $i \frac{g}{c_W} (1-\frac{v^2}{4 f^2})$ \\
%			&&&&\\[-0.6cm]
			$\Phi^0\ A_H\ Z$ & $- \frac{g g'}{\sqrt 2 c_W} \frac{v^2}{2 f}$ && 
			$\Phi^0\ \omega^0\ Z$ & $- i \frac{g}{\sqrt 2 c_W} \frac{v^2}{4 f^2}$ \\
%			&&&&\\[-0.6cm]
			$\Phi^+\ W_H^-\ Z$ & $\frac{g^2}{c_W} \frac{v^2}{4 f}$ && 
			$\Phi^+\ \Phi^-\ Z$ & $- g' s_W (1-\frac{v^2}{8 s^2_W f^2})$ \\
%			&&&&\\[-0.6cm]
			$\omega^+\ W_H^-\ \gamma$ & $i g s_W M_{W_H}$ & &
			$\Phi^+\ \Phi^-\ \gamma$ & $-e$ \\ 
%			&&&&\\[-0.6cm]
			$\omega^+\ W_H^-\ Z$ & $-i g c_W M_{W_H} (1 - \frac{v^2}{4 c_W^2 f^2})$ & &
			$\Phi^+\ \omega^-\ Z$ & $i \frac{g}{c_W} \frac{v^2}{8 f^2}$ \\ 
%			\multicolumn{2}{c}{}&&&\\[-0.6cm]
			\multicolumn{2}{c}{}& &
			$\Phi^{++} \Phi^{--} Z$ & $\frac{g}{c_W} (1-2 s_W^2)$ \\ 
%			\multicolumn{2}{c}{}&&&\\[-0.6cm]
			\multicolumn{2}{c}{}& &
			$\Phi^{++} \Phi^{--} \gamma$ & $-2 e$ \\ 
%			\multicolumn{2}{c}{}&&&\\[-0.6cm]
			\multicolumn{2}{c}{}& &
			$\omega^{+}\ \omega^{-}\ \gamma$ & $-g s_W$ \\ 
%			\multicolumn{2}{c}{}&&&\\[-0.6cm]
			\multicolumn{2}{c}{}& &
			$\omega^{+}\ \omega^{-}\ Z$ & $g c_W(1 - \frac{v^2}{8 c_W^2 f^2})$ \\ 
		\end{tabular}
	\end{center}
	\caption{Non-vanishing Scalar-Vector-Vector and Scalar-Scalar-Vector couplings 
		at ${\cal O} (\frac{v^2}{f^2})$ completing the Tables of Appendix B.2 in~\cite{delAguila:2008zu} and of 
		\cite{delAguila:2017ugt} in order to include the couplings to the scalar triplet $\Phi$. 
		Note the different vertex normalization used in~\cite{delAguila:2008zu}, where $e$ is factored out.}
	\label{Zg:SVV-SSV}
\end{table}
They are given in terms of generic couplings defining the vertices for scalars (S), fermions (F) 
and/or gauge bosons (V): 
\bea
\label{couplingdefinition}
\mbox{[V$_\mu$FF]} &=& i \gamma^\mu(g_LP_L+g_RP_R)\ ,
\nonumber \\
\mbox{[SV$_\mu$V$_\nu$]} &=& i Kg^{\mu\nu}\ ,
\\ 
\mbox{[S$(p_1)$S$(p_2)$V$_\mu$]} &=& i G (p_1-p_2)^\mu\ ,
\nonumber 
\eea
where all momenta are assumed incoming.~The conjugate vertices are obtained replacing:
\bea
g_{L,R}\leftrightarrow g_{L,R}^*\ ,\; \; 
K\leftrightarrow K^*\ ,\; \; 
G\leftrightarrow G^*\ .
\eea
We use the conventions in \cite{delAguila:2017ugt} which differ from those in 
\cite{delAguila:2008zu} by the inclusion of the electromagnetic coupling constant $e$ in the generic coupling definition.\footnote{These conventions coincide with those in~\cite{Blanke:2006eb} up to a sign in the definition of the abelian gauge couplings.} 
The $g$ and $g'$ couplings are the $SU(2)_L$ and $U(1)_Y$ usual gauge couplings respectively and $e = g s_W = g' c_W$ (with 
$s_W$ and $c_W$ the sine and cosine of the electro--weak mixing or Weinberg angle, respectively, 
and $t_W = \frac{s_W}{c_W}, x_H=\frac{5t_W}{4(5-t_W^2)}$).~The Scalar-Fermion-Fermion vertices are given generically by 
\bea
\label{couplingdefinition1}
\mbox{[SFF]} &=& i (c_LP_L+c_RP_R)\, , 
\eea
where the couplings can be directly read off from Table 2 in~\cite{delAguila:2017ugt} with conjugate couplings $c_{L,R}\leftrightarrow c_{R,L}^*$ which also enter in the calculation.~There are also triple gauge boson vertices 
\bea
\label{couplingdefinition2} 
\mbox{[V$_\mu(p_1)$V$_\nu(p_2)$V$_\rho(p_3)$]} &=& i J \left[(g^{\mu \nu}(p_2-p_1)^\rho + 
g^{\nu \rho}(p_3-p_2)^\mu + g^{\rho \mu}(p_1-p_3)^\nu \right]\ ,  
\eea
with the various couplings given in Table~\ref{VVV}.
\begin{table}
	\begin{center}
		\begin{tabular}{c||c}
			[V$_\mu$V$_\nu$V$_\rho$] & \ J \\
%			&\\[-0.6cm]
			\hline
			&\\[-0.45cm]
			$\gamma \ W_H^+\ W_H^-$ & 
			\ $-e$ \\
%			&\\[-0.6cm]
			$Z\ W_H^+\ W_H^-$ & 
			\ $g c_W$ \\
		\end{tabular} 
	\end{center}
	\caption{Vector-Vector-Vector couplings at ${\cal O} (\frac{v^2}{f^2})$ 
		in the Table of Appendix B.2 in~\cite{delAguila:2008zu}.~Note the different vertex normalization used in Eq.\,(\ref{couplingdefinition}) as $e$ is not factorized.}
	\label{VVV}
\end{table}

Finally, for the calculation of muon to electron conversion in nuclei $\mu\ {\rm N} \rightarrow e\ {\rm N}$ the corresponding quark couplings to gauge and scalar bosons are also necessary.~The relevant Vector-Quark-Quark couplings are collected in Table~\ref{tab1}. 
\begin{table}
	\begin{center}
		\begin{tabular}{c||c|c}
			[V$_\mu$FF] & $g_L$ & \ $g_R$ \\
%			&&\\[-0.6cm]
			\hline
			&&\\[-0.45cm]
			$\gamma\ \overline{u_{i}}\ u_j$ & $ - \delta_{ij} \frac{2}{3} e $ & $ - \delta_{ij} \frac{2}{3} e $ \\ 
%			&&\\[-0.6cm]
			$\gamma\ \overline{d_{i}}\ d_j$ & $ \delta_{ij} \frac{1}{3} e $ & $ \delta_{ij} \frac{1}{3} e $ \\  
%			&&\\[-0.6cm]
			$Z\ \overline{u_{i}}\ u_j$ & $ \delta_{ij} \frac{g}{2 c_W} (1 - \frac{4}{3} s_W^2) $ & 
			 $ - \delta_{ij} \frac{2}{3} g' s_W $ \\
%			&&\\[-0.6cm]
			$Z\ \overline{d_{i}}\ d_j$ & $ \delta_{ij} \frac{g}{2 c_W} (-1+ \frac{2}{3} s_W^2) $ & 
			 $ \delta_{ij} \frac{1}{3} g' s_W $ \\
%			&&\\[-0.6cm]
			$A_H\ \overline{u_{H i}}\ u_j$ & $\frac{g}{2}(\frac{t_W}{5}+x_H\frac{v^2}{f^2}) V_{i j}^{u}$ & \ 0 \\ 
%			&&\\[-0.6cm]
			$A_H\ \overline{d_{H i}}\ d_j$ & $\frac{g}{2}(\frac{t_W}{5}-x_H\frac{v^2}{f^2}) V_{i j}^{d}$ & \ 0 \\ 
%			&&\\[-0.6cm]
			$Z_H\ \overline{u_{H i}}\ u_j$ & $\frac{g}{2}(1-x_H\frac{t_W}{5}\frac{v^2}{f^2}) V_{i j}^{u}$ & \ 0 \\ 
%			&&\\[-0.6cm]
			$Z_H\ \overline{d_{H i}}\ d_j$ & $-\frac{g}{2}(1+x_H\frac{t_W}{5}\frac{v^2}{f^2}) V_{i j}^{d}$ & \ 0 \\ 
%			&&\\[-0.6cm]
			$W_H^+\ \overline{u_{H i}}\ d_j$ & $\frac{g}{\sqrt{2}} V_{i j}^{d}$ & \ 0 \\ 
%			&&\\[-0.6cm]
			$W_H^-\ \overline{d_{H i}}\ u_j$ & $\frac{g}{\sqrt{2}} V_{i j}^{u}$ & \ 0 \\ 
		\end{tabular}
	\end{center}
	\caption{[V$_\mu$FF] vertices $i \gamma^\mu(g_LP_L+g_RP_R)$ for quarks at ${\cal O} (\frac{v^2}{f^2})$ 
		in the LHT.}
	\label{tab1}
\end{table}
We do the same for the Scalar-Quark-Quark couplings in Table~\ref{tab2} and include the partner quark couplings which are also needed to evaluate the $\mu\ {\rm N} \rightarrow e\ {\rm N}$ 
amplitude.\footnote{Note that the flavor index $i$ should read $j$ in the couplings 
	$c_R$ of Table 2 in~\cite{delAguila:2010nv}.} 
\begin{table}
	\begin{center}
		\begin{tabular}{c||c|c}
			[SFF] & $c_L$ & $c_R$ \\
%			&&\\[-0.6cm]
			\hline
			&&\\[-0.45cm]
			$\Phi^0 \ \overline{u_{H i}}\ u_{j} $ & $0$ 
			& $V^u_{ij} \frac{m_{u_j}}{\sqrt 2 f} \left( 1 + \frac{v^2}{4f^2} \right)$ \\
%			&&\\[-0.6cm]
			$\Phi^0 \ \overline{d_{H i}}\ d_{j} $ & $0$ 
			& $V^d_{ij} \frac{m_{d_j}}{\sqrt 2 f} \left( 1 + \frac{v^2}{4f^2} \right)$ \\
%			&&\\[-0.6cm]
			$\Phi^0 \ \overline{\tilde{u}_i}\ u_{j} $ & $-W_{ik}^q \frac{m_{d_{H k}}}{\sqrt{2} f} V_{kj}^u$ 
			& $0$ \\
%			&&\\[-0.6cm]
			$\Phi^P \ \overline{u_{H i}}\ u_{j} $ & $-i\frac{m_{d_{H i}}}{\sqrt{2} f} V_{ij}^u\frac{v^2}{4 f^2}$ 
			& $-iV^u_{ij} \frac{m_{u_j}}{\sqrt 2 f} \left( 1 - \frac{v^2}{4f^2} \right)$ \\
%			&&\\[-0.6cm]
			$\Phi^P \ \overline{d_{H i}}\ d_{j} $ & $0$ 
			& $iV^d_{ij} \frac{m_{d_j}}{\sqrt 2 f} \left( 1 + \frac{v^2}{4f^2} \right)$ \\
%			&&\\[-0.6cm]
			$\Phi^P \ \overline{\tilde{u}_i}\ u_{j}  $ & $-iW_{ik}^q \frac{m_{d_{H k}}}{\sqrt{2} f} V_{kj}^u$  
			& $0$ \\
%			&&\\[-0.6cm]
			$\Phi^+\ \overline{u_{H i}}\ d_{j} $ & $\frac{m_{d_{H i}}}{\sqrt 2 f} V^d_{ij} \frac{v^2}{8 f^2}$ & 
			$V^d_{ij} \frac{m_{d_j}}{\sqrt 2 f} \left( 1 - \frac{v^2}{8f^2} \right)$ \\
%			&&\\[-0.6cm]
			$\Phi^+\ \overline{\tilde{u}_i}\ d_{j} $ & 
			$W_{ik}^q \frac{m_{d_{H k}}}{\sqrt{2} f} V_{kj}^d$ 
			& $0$ \\
%			&&\\[-0.6cm]
			$\Phi^+\ \overline{\tilde{x}_i}\ u_{j} $ & 
			$W_{ik}^q \frac{m_{d_{H k}}}{\sqrt{2} f} V_{kj}^u$ 
			& $0$ \\
%			&&\\[-0.6cm]
			$\Phi^-\ \overline{d_{H i}}\ u_{j} $ & $-\frac{m_{d_{H i}}}{\sqrt 2 f} V^u_{ij} \frac{v^2}{8 f^2}$ & 
			$-V^u_{ij} \frac{m_{u_j}}{\sqrt 2 f} \left( 1 - \frac{v^2}{8f^2} \right)$ \\
%			&&\\[-0.6cm]
			$\Phi^{++}\ \overline{\tilde{x}_i}\ d_j$ & 
			$-W_{ik}^q \frac{m_{d_{H k}}}{f} V_{kj}^d$ 
			& $0$ \\
%			&&\\[-0.6cm]
			$\eta\ \overline{u_{H i}}\ u_j$ & $i\frac{m_{d_{H i}}}{2 \sqrt{5} f}\left[1+\frac{v^2}{f^2}
			\left(\frac{9}{8}+x_H\frac{s_W}{c_W}\right)\right]
			V_{i j}^{u}$ & 
			$-i V_{i j}^{u} \frac{m_{u_j}}{2 \sqrt{5} f} \left[1 + \frac{v^2}{f^2}(\frac{15}{8}+x_H\frac{s_W}{c_W})\right]$ \\ 
%			&&\\[-0.6cm]
			$\eta\ \overline{d_{H i}}\ d_j$ & $i\frac{m_{d_{H i}}}{2 \sqrt{5} f}\left[1-\frac{v^2}{f^2}
			\left(\frac{5}{8}+x_H\frac{s_W}{c_W}\right)\right]
			V_{i j}^{d}$ & $-i V_{i j}^{d} \frac{m_{d_j}}{2 \sqrt{5} f} \left[1 - \frac{v^2}{f^2}(\frac{5}{8}+x_H\frac{s_W}{c_W})\right]$\\ 
%			&&\\[-0.6cm]
			$\eta\ \overline{\tilde{u}_i}\ u_{j}$ & $-iW_{ik}^q \frac{m_{d_{H k}}}{2 \sqrt{5} f} V_{kj}^u\frac{3 v^2}{4f^2}$ 
			& $0$ \\ 
%			&&\\[-0.6cm]
			$\omega^0\ \overline{u_{H i}}\ u_j$ & $-i\frac{m_{d_{H i}}}{2 f}\left[1+\frac{v^2}{f^2}
			\left(\frac{1}{8}-x_H\frac{c_W}{s_W}\right)\right]
			V_{i j}^{u}$ & $i V_{i j}^{u} \frac{m_{u_j}}{2 f} \left[1+\frac{v^2}{f^2}(\frac{3}{8} - x_H\frac{c_W}{s_W})\right]$  \\ 
%			&&\\[-0.6cm]
			$\omega^0\ \overline{d_{H i}}\ d_j$ & $i\frac{m_{d_{H i}}}{2 f}\left[1+\frac{v^2}{f^2}
			\left(-\frac{1}{8}+x_H\frac{c_W}{s_W}\right)\right]
			V_{i j}^{d}$  & $-i V_{i j}^{d} \frac{m_{d_j}}{2 f} \left[1-\frac{v^2}{f^2}(\frac{1}{8} - x_H\frac{c_W}{s_W})\right]$ \\ 
%			&&\\[-0.6cm]
			$\omega^0\ \overline{\tilde{u}_i}\ u_{j}$ & $iW_{ik}^q \frac{m_{d_{H k}}}{2 f} V_{kj}^u \frac{v^2}{4f^2}$ 
			& $0$ \\ 
%			&&\\[-0.6cm]
			$\omega^+\ \overline{u_{H i}}\ d_j$ & $-i\frac{m_{d_{H i}}}{\sqrt{2} f}
			V_{i j}^{d}$ & $iV_{i j}^{d}\frac{m_{d_j}}{\sqrt{2} f} (1+\frac{v^2}{8 f^2})$ \\ 
%			&&\\[-0.6cm]
			$\omega^+\ \overline{\tilde{u}_i}\ d_{j} $ & 
			$iW_{ik}^q \frac{m_{d_{H k}}}{\sqrt{2} f} V_{kj}^d \frac{v^2}{8f^2}$ 
			& $0$ \\
%			&&\\[-0.6cm]
			$\omega^+\ \overline{\tilde{x}_i}\ u_{j} $ & 
			$iW_{ik}^q \frac{m_{d_{H k}}}{\sqrt{2} f} V_{kj}^u \frac{v^2}{8f^2}$ 
			& $0$ \\
%			&&\\[-0.6cm]
			$\omega^-\ \overline{d_{H i}}\ u_j$ & $-i \frac{m_{d_{H i}}}{\sqrt{2} f} 
			V_{i j}^{u}$ & $i V_{i j}^{u} \frac{m_{u_j}}{\sqrt{2} f} (1+\frac{v^2}{8 f^2})$ 
			\\ 
		\end{tabular}
	\end{center}
	\caption{[SFF] vertices $i(c_LP_L+c_RP_R)$ for quarks at ${\cal O} (\frac{v^2}{f^2})$ in the LHT.} 
	\label{tab2}
\end{table}
We provide further details elsewhere when discussing flavor changing top decays in the LHT.~Here we only note that the three generations of T--odd partner quarks in the right-handed $SO(5)$ multiplets are $SU(2)_L$ doublets $\tilde{q}_{R i} = (\tilde{x}_{R i}\ \tilde{u}_{R i})^T (i=1,2,3)$ with hypercharge 7/6 which get their vector-like masses by combining with three left-handed $SO(5)$ quark multiplets, analogously to the heavy lepton sector.~Also in analogy to the lepton sector, the misalignment between the partner quark mass eigenstates and the mirror quarks as well as those between the mirror and SM quarks are parametrized by the corresponding $3 \times 3$ unitary matrices,
\bea
\label{matrices} 
V^u = V^{q_H \dagger}_L V^u_L\ , \quad V^d = V^{q_H \dagger}_L V^d_L\ , \quad 
W^q = \tilde{V}^{\tilde{q} \dagger}_R V^{q_H}_R\, ,   
\label{quarkmixing}
\eea
where the products are of the unitary matrices rotating left and right handed fields in order to diagonalize the various mass matrices.\footnote{The mass matrices $M$ are in general diagonalized by 
	two $3 \times 3$ unitary matrices $V_{L, R}$ which we write as $M= V_L D V_R^\dagger$.~The indices $u, d, q_H, \tilde{q}$ stand for the SM up and down quarks and for the (heavy) mirror and partner quarks, respectively.~For charged leptons $\ell$ we omit above some of these indices, 
	$V = V^{H \dagger}_L V^\ell_L, W = \tilde{V}^{T}_L V^{H}_R$, following the conventions in Ref.~\cite{delAguila:2017ugt}. 
	Note also the different $W$ subscript convention for diagonalizing the partner leptons in order to take into 
	account that these lepton fields enter conjugated in the corresponding (right-handed) $SO(5)$ multiplets.
} 
The only physical combination is the Cabibbo-Kobayashi-Maskawa matrix $V_{\rm CKM} = V^{u \dagger} V^d$ relating the SM up and down quark sectors.~In the following we make use of these Feynman rules to evaluate various LFV processes. 

\subsection{Two or three-body lepton decays and $\mu \rightarrow e$ conversion in nuclei}
\label{muegamma}

As the contributions of mirror fermions and heavy gauge sector have been discussed in detail~\cite{delAguila:2008zu,delAguila:2010nv} in the low $Q^2$ limit, we will only quote their final expressions in the low $Q^2$ LFV processes studied in the following.\footnote{SM contributions mediated by $W$ bosons are negligible due to the tiny neutrino masses.}  

\subsubsection*{\underline{$\ell \to \ell' \gamma$}:}

\begin{figure}
	\begin{center}
		\begin{tabular}{cccc}
			\includegraphics[scale=0.7]{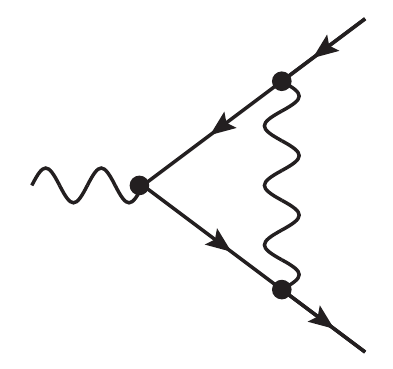} &
			\includegraphics[scale=0.7]{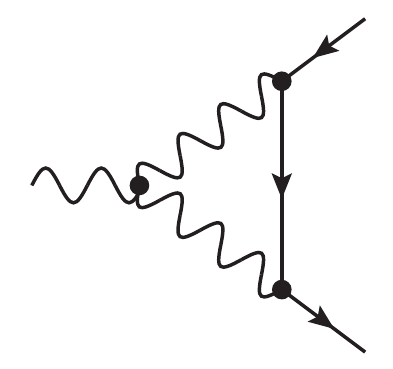} &
			\includegraphics[scale=0.7]{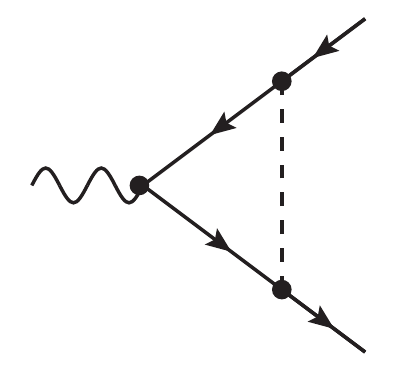} &
			\includegraphics[scale=0.7]{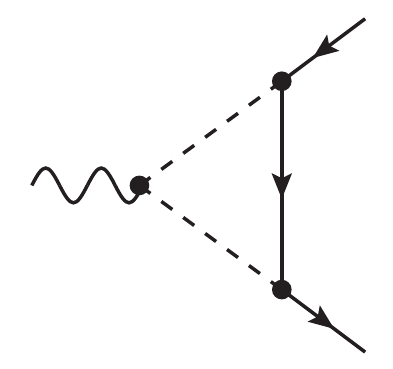} \\
			I & II & III & IV \\
			&
			\includegraphics[scale=0.7]{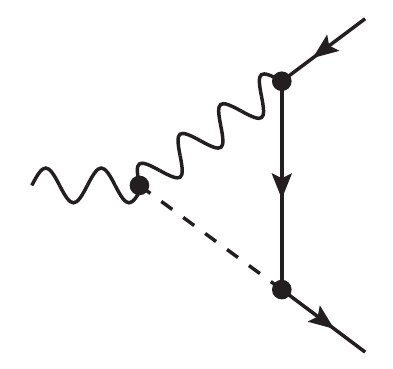} &
			\includegraphics[scale=0.7]{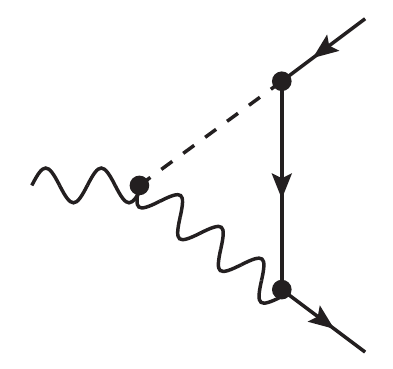} & 
			\\ 
			& V & VI & \\ 
			\includegraphics[scale=0.7]{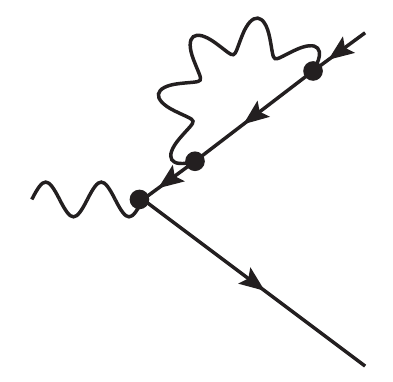} &
			\includegraphics[scale=0.7]{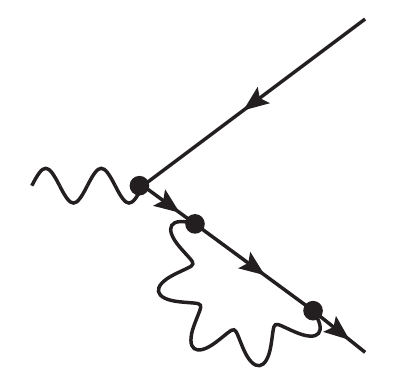} &
			\includegraphics[scale=0.7]{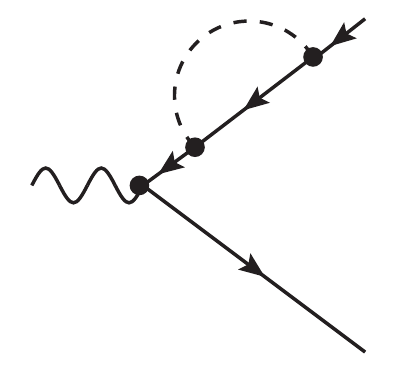} &
			\includegraphics[scale=0.7]{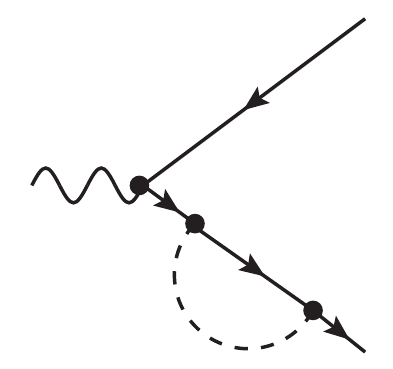} \\
			VII & VIII & IX & X 
		\end{tabular}
	\end{center}
	\caption{Topologies contributing to two and three-body lepton decays as well as $\gamma , Z \rightarrow \overline{\ell} \ell'$.}
	\label{diagrams}
\end{figure}

This process has garnered much attention in the past 
\cite{delAguila:1982yu,Hisano:1995cp,Illana:2000ic,Illana:2002tg,Arganda:2005ji} 
due to the stringent experimental bound on $\mu \to e \gamma$.~Gauge invariance reduces this vertex for an on-shell photon to a dipole transition,
\bea
\label{vertexmue} 
i\ \Gamma^\mu_\gamma (p_\ell, p_{\ell'}) = i\ e \left[ i F_M^\gamma (Q^2) + F_E^\gamma (Q^2) \gamma_5\right] 
\sigma^{\mu \nu}\ Q_\nu\ , 
\eea
where $Q_\nu = (p_{\ell'} - p_\ell)_\nu$, with decay width (neglecting $m_{\ell'}$) 
\bea
\label{decaywidthmue} 
\Gamma (\ell \to \ell' \gamma) = \frac{\alpha}{2}\ m_\ell^3\ ( |F_M^\gamma|^2 + |F_E^\gamma|^2 )\ ,  
\eea
where $\alpha = \frac{e^2}{4 \pi}$.~The contributions to these two dipole form factors from the T--odd mirror leptons are presented in detail in~\cite{Blanke:2007db,delAguila:2008zu} where the notation used here is also introduced (we rename some functions for easy comparison between processes).~The different one-loop topologies contributing to them in the 't Hooft-Feynman gauge are depicted in 
Figure~\ref{diagrams}.\footnote{There are two other new topologies with non-renormalizable 
	Scalar-Scalar-Fermion-Fermion couplings in the corresponding Higgs decays, $h \to \overline{\ell} \ell'$, compared with the analogous gauge transitions~\cite{delAguila:2017ugt}.}

The contributions from the partner leptons $\tilde{l^c} = (\tilde{\nu}^c\ \tilde{\ell}^c)$ only involve topologies III, IV, IX and X in Figure~\ref{diagrams} because they do not couple to one T--odd gauge boson and a SM charged lepton at the order considered (see Table \ref{Zg:VFF}).~Their total contribution is finite as expected and the cancellation of the infinities holds for the contributions exchanging $\tilde{\nu}^c$ and $\tilde{\ell}^c$ independently in $F_M^\gamma|_{\tilde{\nu}^c}$ and $F_M^\gamma|_{\tilde{\ell}^c}$ respectively.~In addition, these contributions decouple from the amplitude in the heavy $\tilde{l}^c$ limit.~In summary, neglecting $m_{\ell'} (\ll m_{\ell})$ and defining $\alpha_W = \alpha / s_W^2$ (with the five terms satisfying $F_M^\gamma|_a = -i F_E^\gamma|_a$),\footnote{
	We have validated our calculations with FeynCalc~\cite{Mertig:1990an,Shtabovenko:2016sxi} and utilized the scalar integrals in~\cite{Passarino:1978jh} using the conventions in Appendix C of~\cite{delAguila:2008zu}.}
\begin{align}
	\label{dipoleformfactormue} 
	F_M^\gamma & =  F_M^\gamma|_{W_H} + F_M^\gamma|_{A_H} + 
	F_M^\gamma|_{Z_H} + F_M^\gamma|_{\tilde{\nu}^c} + F_M^\gamma|_{\tilde{\ell}^c} \nonumber \\
	& = 
	\sum_i V_{\ell' i}^\dagger V_{i \ell}\ \frac{\alpha_W}{16\pi}\frac{m_{\ell}}{M_{W_H}^2} 
	\left[ F^W_{M}\left(\frac{m_{\ell_{H i}}^2}{M_{W_H}^2}\right) +  
	\frac{1}{5} F_M^{A/Z}\left(\frac{m_{\ell_{H i}}^2}{M_{A_H}^2}\right) 
	+ F_M^{A/Z}\left(\frac{m_{\ell_{H i}}^2}{M_{Z_H}^2}\right) \right]  \\ 
	& + \sum_{ijk} V_{\ell' i}^\dagger \frac{m_{\ell_{H i}}}{M_{W_H}} W_{ij}^\dagger 
	W_{jk} \frac{m_{\ell_{H k}}}{M_{W_H}} V_{k \ell} 
	\ \frac{\alpha_W}{16\pi}\frac{m_{\ell}}{M_{\Phi}^2} \left[ F_M^{\tilde{\nu}} \left(\frac{m_{\tilde{\nu}^c_j}^2}{M_{\Phi}^2}\right) + 
	F_M^{\tilde{\ell}} \left(\frac{m_{\tilde{\nu}^c_j}^2}{M_{\Phi}^2}\right) \right]\ ,  \nonumber 
\end{align}
with 
\begin{align}
	\label{dipoleformfactormue1} 
	F^W_M (x) & = \frac{5}{6}-\frac{3x-15x^2-6x^3}{12(1-x)^3}+\frac{3x^3}{2(1-x)^4}\ln x\ , \nonumber \\
	F_M^{A/Z} (x) & = -\frac{1}{3}+\frac{2x+5x^2-x^3}{8(1-x)^3}+\frac{3x^2}{4(1-x)^4}\ln x\, , \nonumber \\
	F_M^{\tilde{\nu}} (x) & = \frac{-1+5x+2x^2}{12(1-x)^3}+\frac{x^2}{2(1-x)^4}\ln x\ , \\
	F_M^{\tilde{\ell}} (x) &= \frac{-4+5x+5x^2}{6(1-x)^3}-\frac{x(1-2x)}{(1-x)^4}\ln x\ . \nonumber 
\end{align}
Note that these four loop functions $F_M^{W, A/Z, {\tilde{\nu}}, {\tilde{\ell}}}$ are finite and depend on the ratio of the particle masses circulating in the loop.~Once the global suppression factor $v^2 / f^2$ has been factored out in Eq.\,(\ref{dipoleformfactormue}) their own $v^2 / f^2$ corrections 
can be neglected as they are next order.~Thus, in general the (heavy) masses of the different components of the same $SU(2)_L$ multiplet can be taken to be degenerate when substituted in $F_M^{W, A/Z, {\tilde{\nu}}, {\tilde{\ell}}}$ .\footnote{
	This means that $m_{\nu_{H i}} = m_{\ell_{H i}} (1-v^2/8f^2) \simeq m_{\ell_{H i}},~M^2_{W_H} = M^2_{Z_H} = 5 M^2_{A_H} (1 + v^2 / f^2) / t_W^2 \simeq 5 M^2_{A_H} / t_W^2$ when used within these finite loop functions while the masses of ${\tilde{\nu}}_i$ and ${\tilde{\ell}}_i$ are the same.}
This is also the case for the components of the scalar electroweak triplet $\Phi$ within $F_M^{{\tilde{\nu}}, {\tilde{\ell}}}$ and in the denominator multiplying the last line in Eq.\,(\ref{dipoleformfactormue}). 

%%%%%%%%%%%%%%%%%
\subsubsection*{\underline{$\ell \to \ell'\overline{\ell'}\ell' , \ell'\overline{\ell''}\ell'' , \ell'\overline{\ell''}\ell'$}:}

\begin{figure}
	\begin{center}
		\begin{tabular}{ccc}
			\includegraphics[scale=0.65]{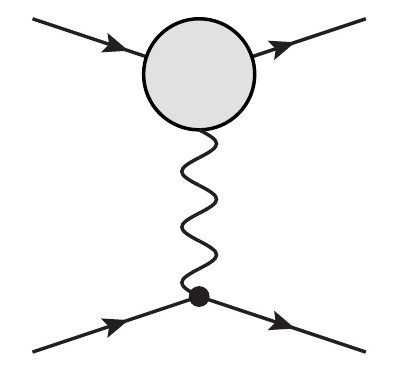} & 
			\quad \quad \quad \quad &
			\includegraphics[scale=0.65]{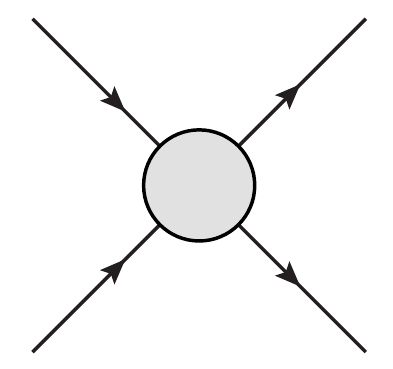} 
		\end{tabular}
	\end{center}
	\caption{Generic penguin (left) and box (right) amplitudes contributing to three-body $\ell$ decays and to 
		$\mu \rightarrow e$ conversion in nuclei.}
	\label{diagrams1}
\end{figure}

These processes involve photon and $Z$ penguin diagrams as well as box contributions which are depicted in Figure \ref{diagrams1}. 
\bea
\label{vertexmuegamma} 
i\ \Gamma^\mu_\gamma (p_\ell, p_{\ell'}) = i\ e \left\{\left[ i F_M^\gamma (Q^2) + F_E^\gamma (Q^2) \gamma_5\right] 
\sigma^{\mu \nu} Q_\nu + F_L^\gamma (Q^2) \gamma^\mu P_L \right\}\ , 
\eea
with $Q_\nu = (p_{\ell'} - p_\ell)_\nu$ and $P_{L, R} = \frac{1}{2} (1 \mp \gamma_5)$.~The corresponding right-handed vector form factor 
vanishes with $m_{\ell'}$ and is further suppressed as are the corresponding scalar form factors.~The left-handed vector form factor receives contributions from mirror and partner leptons in analogy with the dipole form factors above.~Keeping with the notation above, 
\begin{align}
	\label{gammapenguin} 
	F_L^\gamma & =  F_L^\gamma|_{W_H} + F_L^\gamma|_{A_H} + 
	F_L^\gamma|_{Z_H} + F_L^\gamma|_{\tilde{\nu}^c} + F_L^\gamma|_{\tilde{\ell}^c} \nonumber \\
	& = 
	\sum_i V_{\ell' i}^\dagger V_{i \ell}\ \frac{\alpha_W}{4\pi}\frac{Q^2}{M_{W_H}^2} 
	\left[ F^W_{L}\left(\frac{m_{\ell_{H i}}^2}{M_{W_H}^2}\right) +  
	\frac{1}{5} F_L^{A/Z}\left(\frac{m_{\ell_{H i}}^2}{M_{A_H}^2}\right) 
	+ F_L^{A/Z}\left(\frac{m_{\ell_{H i}}^2}{M_{Z_H}^2}\right) \right]  \\ 
	& + \sum_{ijk} V_{\ell' i}^\dagger \frac{m_{\ell_{H i}}}{M_{W_H}} W_{ij}^\dagger 
	W_{jk} \frac{m_{\ell_{H k}}}{M_{W_H}} V_{k \ell} 
	\ \frac{\alpha_W}{4\pi}\frac{Q^2}{M_{\Phi}^2} \left[ F_L^{\tilde{\nu}} \left(\frac{m_{\tilde{\nu}^c_j}^2}{M_{\Phi}^2}\right) + 
	F_L^{\tilde{\ell}} \left(\frac{m_{\tilde{\nu}^c_j}^2}{M_{\Phi}^2}\right) \right]\ ,  \nonumber
\end{align}
with the loop functions defined as
\begin{align}
	\label{gammapenguin1}
	F^W_L (x) & = -\frac{5}{18} + \frac{12x+x^2-7x^3}{24(1-x)^3} + \frac{12x^2-10x^3+x^4}{12(1-x)^4}\ln x\ , \nonumber \\
	F_L^{A/Z} (x) & = \frac{1}{36} + \frac{18x-11x^2-x^3}{48(1-x)^3} - \frac{4-16x+9x^2}{24(1-x)^4}\ln x\, , \nonumber \\
	F_L^{\tilde{\nu}} (x) & = \frac{2-7x+11x^2}{72(1-x)^3} + \frac{x^3}{12(1-x)^4}\ln x\ , \\
	F_L^{\tilde{\ell}} (x) &= \frac{20-43x+29x^2}{36(1-x)^3} + \frac{2-3x+2x^3}{6(1-x)^4}\ln x\ . \nonumber
\end{align}
The form factor $F_L^\gamma|_{W_H}$ also has a universal infinite contribution which cancels due to the unitarity of the mixing matrices multiplying it.~The new contributions proportional to $F_L^{\tilde{\nu}, \tilde{\ell}}$ decouple when the masses of the partner leptons $\tilde{l}^c_i$ are taken to infinity.~Summarizing, the form factors entering in photon penguin diagrams satisfy $F_M^\gamma = - i F_E^\gamma$ while current conservation implies that the vector form factors (in particular, $F_L^\gamma$) must be proportional to $Q^2$ and vanish for on-shell photons.~Of course they contribute to photon penguin diagrams which include a photon propagator proportional to $\sim Q^{-2}$.

In contrast, $Z$ penguin diagrams involve the $Z$ boson propagator which for small momentum transfer processes is proportional to $M_Z^{-2}$.~The dipole form factors, which flip chirality, are proportional to SM lepton masses and negligible when compared to the vector ones while
the scalar form factors are also negligible as they are proportional to SM lepton masses.~Thus, at leading order the $Z \overline{\ell} \ell'$ vertex reduces to,  
\bea
\label{vertexmueZ} 
i\ \Gamma^\mu_Z (p_\ell, p_{\ell'}) = i\ e\ F_L^Z (Q^2) \gamma^\mu P_L\  .
\eea
The corresponding right-handed vector form factor $F_R^Z$ is 
${\cal O} (m_\ell^2 / f^2)$ in the LHT and thus negligible at the order we work.~As in the case of the photon, the contributions of the T--odd fermions to $F_L^Z$ result from the running of the mirror and partner leptons inside the loops in Figure~\ref{diagrams}.~Using the Feynman rules introduced above and splitting these contributions as before, 
we obtain: 
\begin{align}
	\label{Zpenguin}
	F_L^Z & =  F_L^Z|_{W_H} + F_L^Z|_{A_H} + F_L^Z|_{Z_H} 
	+ F_L^Z|_{\tilde{\nu}^c} + F_L^Z|_{\tilde{\ell}^c} \nonumber \\
	& = \sum_i V_{\ell' i}^\dagger V_{i \ell}\ \frac{\alpha_W}{8\pi s_W c_W}  
	\left\{ \frac{v^2}{8 f^2} H^{W (0)}_{L}\left(\frac{m_{\ell_{H i}}^2}{M_{W_H}^2}\right) 
	+  \frac{Q^2}{M_{W_H}^2} H_L^{W}\left(\frac{m_{\ell_{H i}}^2}{M_{W_H}^2}\right) \right. \\
	& \quad \quad \quad \quad \quad \quad \quad \quad \quad \quad 
	\left. + (1-2c_W^2) \frac{Q^2}{M_{W_H}^2} \left[ \frac{1}{5} H_L^{A/Z}\left(\frac{m_{\ell_{H i}}^2}{M_{A_H}^2}\right) 
	+ H_L^{A/Z}\left(\frac{m_{\ell_{H i}}^2}{M_{Z_H}^2}\right) \right] \right\}  \nonumber \\ 
	& + \sum_{ijk} V_{\ell' i}^\dagger \frac{m_{\ell_{H i}}}{M_{W_H}} W_{ij}^\dagger 
	W_{jk} \frac{m_{\ell_{H k}}}{M_{W_H}}V_{k \ell} 
	\ \frac{\alpha_W}{8\pi s_W c_W}\frac{Q^2}{M_{\Phi}^2} \left[ H_L^{\tilde{\nu}} \left(\frac{m_{\tilde{\nu}^c_j}^2}{M_{\Phi}^2}\right) 
	+ (1-2c_W^2) H_L^{\tilde{\ell}} \left(\frac{m_{\tilde{\nu}^c_j}^2}{M_{\Phi}^2}\right) \right]\ ,  \nonumber
\end{align}
with the loop functions defined 
\begin{align}
	\label{Zpenguin1}
	H^{W (0)}_L (x) & = \frac{6x - x^2}{1-x} + \frac{2x+3x^2}{(1-x)^2}\ln x \ , \nonumber \\
	H^W_L (x) & = 2 F_L^{A/Z} (x) - 2c_W^2 F_L^W \ , \nonumber \\
	H_L^{A/Z} (x) & = F_L^{A/Z} (x) \ , \\
	H_L^{\tilde{\nu}} (x) & = \frac{1}{2} F_L^{\tilde{\ell}} (x) - 2c_W^2 F_L^{\tilde{\nu}} (x) \ , \nonumber \\
	H_L^{\tilde{\ell}} (x) & = F_L^{\tilde{\ell}} (x) \ . \nonumber 
\end{align}
Again $F_L^Z|_{W_H}$ has a universal infinite loop contribution 
which cancels due to the unitarity of the mixing matrices and there are similar relations among the finite loop functions $F_L^{W, A/Z, {\tilde{\nu}}, {\tilde{\ell}}}$ and $H_L^{W, A/Z, {\tilde{\nu}}, {\tilde{\ell}}}$ for the photon and left-handed $Z$ vector form factors in Eqs.\,(\ref{gammapenguin1}) and (\ref{Zpenguin1}).~The form factor $F_L^Z|_{W_H}$ has two different finite contributions, 
one which is proportional to $H^{W (0)}_L$ and independent of $Q^2$, so absent in the photon case, and another linear in $Q^2$ and proportional to $H^{W}_L$ (second line in Eq.\,(\ref{Zpenguin})).~Only the first contributes to three-body lepton decays and to $\mu \to e$ conversion in nuclei.~The second one as well as all the other contributions which are proportional to 
$Q^2$ in $F_L^Z$ (see Eq.\,(\ref{Zpenguin})) are negligible as long as $Q^2 \ll v^2$.~We give the complete result here because we will make use of it when 
discussing leptonic on-shell $Z$ decays below where we also provide further details of the calculation. 

\begin{figure}
	\begin{tabular}{cccc}
		\includegraphics[scale=0.7]{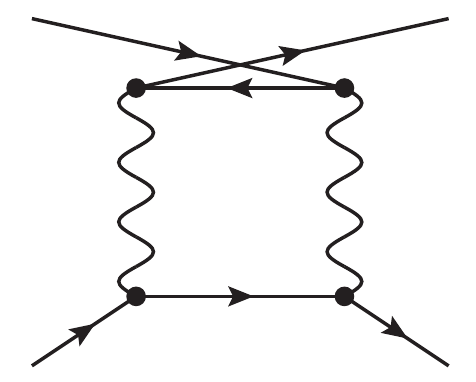} &
		\includegraphics[scale=0.7]{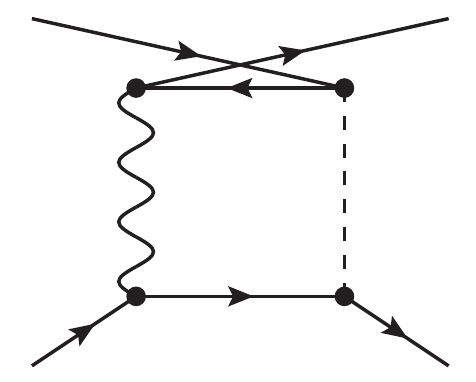} &
		\includegraphics[scale=0.7]{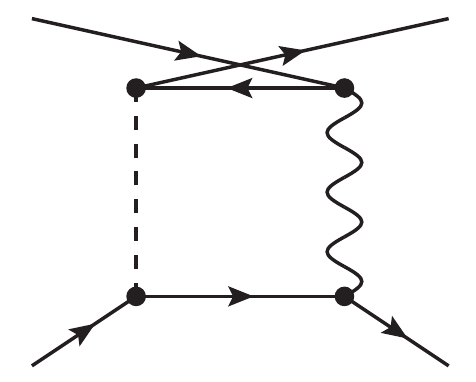} &
		\includegraphics[scale=0.7]{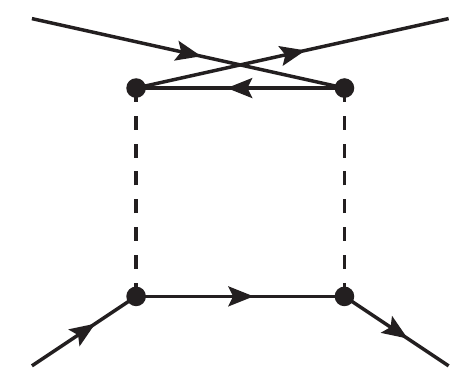} \\
		A1 & A2 & A3 & A4  \\
		\includegraphics[scale=0.7]{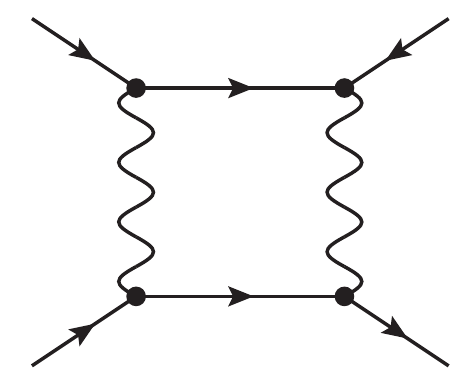} &
		\includegraphics[scale=0.7]{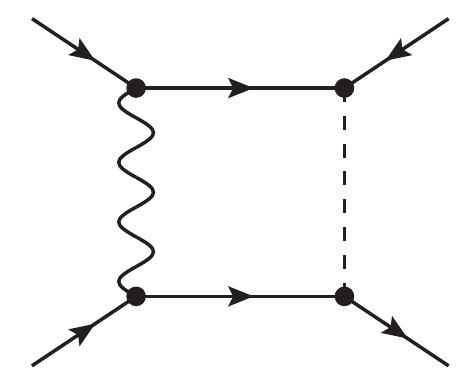} &
		\includegraphics[scale=0.7]{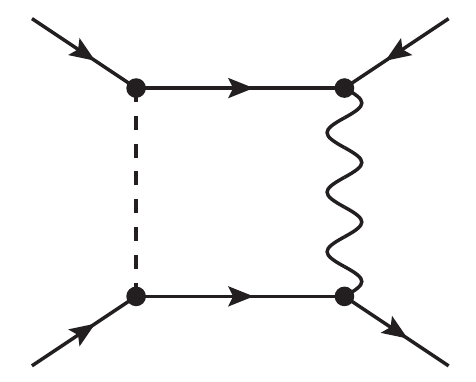} &
		\includegraphics[scale=0.7]{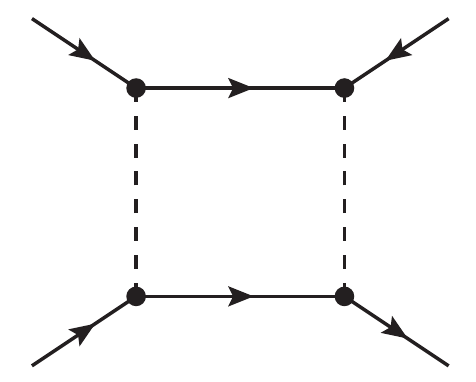} \\
		B1 & B2 & B3 & B4
	\end{tabular}
	\caption{Box diagram topologies contributing to three-body $\ell$ decays and to 
		$\mu \rightarrow e$ conversion in nuclei.~Crossed diagrams with the two outgoing leptons (whether they are identical or not) exchanged must be added in the three-body leptonic $\ell$ decays.}
	\label{diagrams2}
\end{figure}

The $\ell (p_\ell) \to \ell' (p_1) \overline{\ell'} (p_2) \ell' (p_3)$ amplitude 
${\cal M}^{\ell \to \ell'_1\overline{\ell'}_2\ell'_3}$ then gets contributions from the photon and $Z$ vertices in Eqs.\,(\ref{vertexmuegamma}) and (\ref{vertexmueZ}) after contracting them with the corresponding gauge boson propagators and the tree-level SM leptonic vertices.~It also receives contributions from the box diagrams shown in Figure~\ref{diagrams2}.  
Following the analysis in~\cite{delAguila:2008zu} (see also~\cite{Hisano:1995cp,Arganda:2005ji}) we write the full $\ell \to \ell'_1{\overline{\ell'}}_2\ell'_3$ amplitude as, 
\bea
\label{Amplitude}
{\cal M}^{\ell \to \ell'_1{\overline{\ell'_2}}\ell'_3} = 
{\cal M}^{\ell \to \ell'_1{\overline{\ell'_2}}\ell'_3}_\gamma + {\cal M}^{\ell \to \ell'_1{\overline{\ell'_2}}\ell'_3}_Z + 
{\cal M}^{\ell \to \ell'_1{\overline{\ell'_2}}\ell'_3}_{\rm Box}\ , 
\eea
where we have defined the individual amplitudes as,
\begin{align}
	\label{photonamplitude}
	{\cal M}^{\ell \to \ell'_1{\overline{\ell'_2}}\ell'_3}_\gamma = & \,\overline{u}(p_1)\, 
	e \left[ i\, F_M^\gamma (0)\, 2 P_R\, \sigma^{\mu \nu} (p_1 - p_\ell)_\nu + 
	F_L^\gamma ((p_1 - p_\ell)^2) \gamma^\mu P_L \right] u(p_\ell) \nonumber \\
	& \times \frac{1}{(p_1 - p_\ell)^2} \overline{u}(p_3) \gamma_\mu e\,v(p_2) 
	-  \left( p_1 \leftrightarrow p_3 \right) \ ,\\
	\label{Zamplitude}
	{\cal M}^{\ell \to \ell'_1{\overline{\ell'_2}}\ell'_3}_Z = & \,\overline{u}(p_1)\,(-e F_L^Z (0))\,\gamma^\mu P_L u(p_\ell) 
	\frac{1}{M_Z^2} \overline{u}(p_3) \gamma_\mu\,( g_L^Z P_L + g_R^Z P_R )\,v(p_2) \nonumber \\
	&\,- \left( p_1 \leftrightarrow p_3 \right) \ ,\\ 
	\label{Boxamplitude}
	{\cal M}^{\ell \to \ell'_1{\overline{\ell'_2}}\ell'_3}_{\rm Box} = & \,e^2 B_L (0)\ \overline{u}(p_1)\, 
	\gamma^\mu P_L u(p_\ell)\ \overline{u}(p_3) \gamma_\mu P_L\,v(p_2) \ , \\ 
	\nonumber 
\end{align}
with $F_E^\gamma = i\,F_M^\gamma$.~The photon magnetic and $Z$ left-handed vector form factors, $F_M^\gamma (0)$ and $F_L^Z (0)$ respectively, are evaluated at $Q^2 = (p_1 - p_\ell)^2 = 0$ because their leading terms are momentum independent for small momentum transfer $ Q^2 \sim m_\ell^2$ while the photon left-handed vector form factor, $F_L^\gamma ((p_1 - p_\ell)^2)$, is linear in $Q^2$.~The $g_L^Z$ and $g_R^Z$ couplings of the $Z$ boson to charged leptons are given in Table~\ref{Zg:VFF}.

Moving on to the box diagrams shown in Figure~\ref{diagrams2}, these all reduce to the same product of currents in Eq.\,(\ref{Boxamplitude}) in the limit of zero external momenta~\cite{delAguila:2008zu} (all internal masses are much larger than the external ones).~The box loop-integrals are finite by power counting and in this limit their contributions can all be absorbed into the form factor $e^2\,B_L(0)$.~In addition to the contributions from mirror leptons accounted for in~\cite{delAguila:2008zu}, those from partner leptons must also be included.~Bearing in mind the degenerate mass limit for different components of the heavy T--odd $SU(2)_L$ multiplets we can group them according to the bosonic particle masses running in the loop,\footnote{We use the Fierz identity $\langle 1 | \gamma^\mu P_L | \ell \rangle \langle 3 | \gamma_\mu P_L | 2 \rangle = - \langle 3 | \gamma^\mu P_L | \ell \rangle \langle 1 | \gamma_\mu P_L | 2 \rangle$ to relate crossed diagrams with $p_1\leftrightarrow p_3$.} 
\bea
\label{B}
B_L (0) = B_L^{W_H W_H+Z_H Z_H} + B_L^{A_H A_H} +  B_L^{A_H Z_H} + B_L^{\Phi \Phi} \ ,
\eea
where we have defined the functions,
\begin{align}
	\label{Bs}
	B_L^{W_H W_H+Z_H Z_H} & = \frac{\alpha_W}{32 \pi s_W^2} \frac{1}{M_{W_H}^2} 
	\times \nonumber \\
	\sum_{ij} \chi^{\ell \ell' \ell' \ell'}_{ij} & \left[ \frac{1}{2} \left(1+\frac{m_{\ell_{H i}}^2}{M_{W_H}^2}
	\frac{m_{\ell_{H j}}^2}{M_{W_H}^2}\right)
	\tilde d\left(\frac{m_{\ell_{H i}}^2}{M_{W_H}^2}, \frac{m_{\ell_{H j}}^2}{M_{W_H}^2}\right) 
	- 4\frac{m_{\ell_{H i}}^2}{M_{W_H}^2} \frac{m_{\ell_{H j}}^2}{M_{W_H}^2} 
	d\left(\frac{m_{\ell_{H i}}^2}{M_{W_H}^2}, \frac{m_{\ell_{H j}}^2}{M_{W_H}^2}\right) \right] \ ,\nonumber \\
	B_L^{A_H A_H} & = \frac{\alpha_W}{32 \pi s_W^2} \frac{1}{M_{W_H}^2} 
	\sum_{ij}\chi^{\ell \ell' \ell' \ell'}_{ij}\left[ -\frac{3}{50} \frac{M_{A_H}^2}{M_{W_H}^2}
	\tilde d\left(\frac{m_{\ell_{H i}}^2}{M_{A_H}^2}, \frac{m_{\ell_{H j}}^2}{M_{A_H}^2}\right) 
	\right] \ , \\
	B_L^{A_H Z_H} & = \frac{\alpha_W}{32 \pi s_W^2} \frac{1}{M_{W_H}^2} 
	\sum_{ij}\chi^{\ell \ell' \ell' \ell'}_{ij} \left[ -\frac{3}{5} 
	{\tilde{d}}^\prime \left(\frac{m_{\ell_{H i}}^2}{M_{A_H}^2}, 
	\frac{m_{\ell_{H j}}^2}{M_{A_H}^2}, \frac{M_{W_H}^2}{M_{A_H}^2} \right) 
	\right]\ ,\nonumber \\
	B_L^{\Phi \Phi} & = \frac{\alpha_W}{32 \pi s_W^2} \frac{1}{M_{\Phi}^2} 
	\sum_{ij} \tilde \chi^{\ell \ell' \ell' \ell'}_{ij} 
	\tilde d\left(\frac{m_{{\tilde{\nu}}^c_{i}}^2}{M_{\Phi}^2}, \frac{m_{{\tilde{\nu}}^c_{j}}^2}{M_{\Phi}^2}\right) 
	\ .\nonumber
\end{align}
The mixing coefficients are defined in 
Eqs.\,(\ref{chiij}) and (\ref{tildechiij}) with $\ell''' = \ell'' = \ell'$ respectively while the loop functions read, 
\begin{align}
	\label{Bs1}
	{\tilde{d}} (x,y) & = \frac{x^2 \ln x}{(1-x)^2 (y-x)} 
	+ \frac{y^2 \ln y}{(1-y)^2 (x-y)} - \frac{1}{(1-x) (1-y)} \ , \nonumber \\
	d (x,y) & = \frac{x \ln x}{(1-x)^2 (y-x)} + \frac{y \ln y}{(1-y)^2 (x-y)} - \frac{1}{(1-x) (1-y)} \ , \\
	{\tilde{d}}' (x,y,z) & = \frac{x^2 \ln x}{(1-x) (y-x) (z-x)} 
	+ \frac{y^2 \ln y}{(1-y) (x-y) (z-y)} + \frac{z^2 \ln z}{(1-z) (x-z) (y-z)} \ , \nonumber
\end{align}
with ${\tilde{d}} (x,y) = {\tilde{d}}' (x,y,1)$.~The new contributions from the partner leptons ${\tilde{\nu}}^c$ and ${\tilde{\ell}}^c$ are equal (neglecting small mass differences within the scalar triplet $\Phi$ components) and included in $B_L^{\Phi \Phi}$.~After integrating the three-body phase space the decay width reads:\footnote{
	The phase space factor for $|A_R|^2$ is singular when the second and third final lepton masses with $m_2=m_3$ vanish, and it must be carefully calculated.~We obtain the same result as in Ref.~\cite{Ilakovac:1994kj}, Eq. (C.4). 
	\label{integration}} 
\begin{align}
	\label{threeequal}
	\Gamma(\ell\to\ell'\overline{\ell'}\ell')&=
	\frac{\alpha^2m_{\ell}^5}{96\pi}\Big\{
	3|A_L|^2 + 2 |A_R|^2 \left(8\ln\frac{m_{\ell}}{m_{\ell'}}-13\right) 
	+ 2|F_{LL}|^2+|F_{LR}|^2 + \frac{1}{2}|B_L|^2 \nonumber \\
	& - \left[ 6 A_L A_R^{*} - ( A_L - 2 A_R )( 2 F_{LL}^* + F_{LR}^* + B_L^{*} )
	- F_{LL} B_L^{*} + \mbox{h.c.} \right]\Big\} \ ,
\end{align}
where we have defined in order to simplify the expression:
\bea
\label{simplification}
A_L = \frac{F_L^\gamma}{Q^2} , \ A_R = \frac{2 F_M^\gamma (0)}{m_\ell} , 
\ F_{LL} = - \frac{ g_L F_L^Z(0)}{e M_Z^2} , \ F_{LR} = - \frac{g_R F_L^Z(0)}{e M_Z^2} , 
\ B_L = B_L (0) , \,
\eea
with $g_{L,R}$ the corresponding $Z$ couplings to the charged lepton $\ell'$ (see Table \ref{Zg:VFF}).~With these results for $\ell \to \ell'\overline{\ell'}\ell'$, the other LFV three-body lepton decays are easily obtained. 

For the $\ell (p_\ell) \to \ell' (p_1) \overline{\ell''} (p_2) \ell'' (p_3)$ amplitude 
there are no crossed penguin diagram contributions due to swapping $\ell^{\prime}$ and $\ell^{\prime\prime}$ because two gauge boson LFV transitions would be needed.~This implies a higher order process both in loops and in $v^2 / f^2$.~This means that the $\ell (p_\ell) \to \ell' (p_1) \overline{\ell''} (p_2) \ell'' (p_3)$ amplitude ${\cal M}^{\ell \to \ell'_1\overline{\ell''}_2\ell''_3}$ has no $p_1 \leftrightarrow p_3$ term in Eqs.\,(\ref{photonamplitude}) and (\ref{Zamplitude}).~However, for the box amplitudes there are additional diagrams at this order for swapping $\ell^{\prime}$ and $\ell^{\prime\prime}$.~These are automatically taken into account by the mixing coefficient definitions in Eqs.\,(\ref{chiij}) and (\ref{tildechiij}) once the flavor factors with $\ell \ell' \ell' \ell'$ are replaced by the appropriate ones with $\ell \ell' \ell'' \ell''$.~Furthermore, now there is no symmetry factor of 1/2 in the phase space integration needed to obtain the decay width because all three final leptons are distinguishable.~The final decay width can be written as (see footnote~\ref{integration}),
\begin{align}
	\label{threedifferent}
	\Gamma(\ell\to\ell'\overline{\ell''}\ell'')&=
	\frac{\alpha^2m_{\ell}^5}{96\pi}\Big\{
	2 |A_L|^2  + 4 |A_R|^2 \left(4 \ln\frac{m_{\ell}}{m_{\ell''}} - 7\right) 
	+ |F_{LL}|^2+|F_{LR}|^2 + \left|B_L\right|^2 \nonumber \\ 
	& -  \left[  4 A_L A_R^* - ( A_L - 2 A_R ) \left( F_{LL}^* + F_{LR}^* + \frac{B_L^*}{2} \right) 
	- F_{LL} \frac{B_L^*}{2} + \mbox{h.c.}  
	\right] \Big\} \, ,
\end{align}
with the same simplifying definitions as in Eq.\,(\ref{simplification}) and corresponding changes to account for the $Z$ couplings to the charged lepton $\ell''$.

Finally, for the double flavor violating decay $\ell (p_\ell) \to \ell' (p_1) \overline{\ell''} (p_2) \ell' (p_3)$ the amplitude ${\cal M}^{\ell \to \ell'_1\overline{\ell''_2}\ell'_3}$ has no penguin contributions at the order we consider (see footnote \ref{order}).~The box contributions on the other hand are the same as for $\ell (p_\ell) \to \ell' (p_1) \overline{\ell'} (p_2) \ell' (p_3)$ but replacing the corresponding flavor coefficients in Eq. (\ref{Bs}) with those in this decay $\ell \ell' \ell'' \ell'$.~The decay width is also the same as in Eq. (\ref{threeequal}) but with the box loop form factor $|B_L|^2$ term only.~This gives for the total decay width (with same phase space as for $\ell\to\ell'\overline{\ell'}\ell'$),
\bea
\Gamma(\ell\to\ell'\overline{\ell''}\ell')=
\frac{\alpha^2m_{\ell}^5}{19 2\, \pi}
\left|B_L\right|^2\, .
\eea
%

%%%%%%%%
\subsubsection*{\underline{$\mu\ {\rm N} \rightarrow e\ {\rm N}$}:}

For this transition we follow Ref.~\cite{delAguila:2010nv} where the contributions 
of the (T--odd) mirror fermions to $\mu \rightarrow e$ conversion in nuclei are discussed.\footnote{
	Our definition of $Q_\nu = (p_e - p_\mu)_\nu$ has opposite sign to that in \cite{delAguila:2010nv}, 
	as well as our definition of $A_R$.
}
This process has penguin and box contributions as in Figure\,\ref{diagrams1}.~As for the leptonic decay $\ell \to \ell'\overline{\ell''}\ell''$,~it has no crossed penguin diagrams because the lower fermionic line where the gauge boson is attached is now a coherent sum of quarks composing the probed nucleus.~There is also no crossed box contributions due to the exchange of leptons.~Putting everything together, including the new partner fermion contributions, we can write the $\mu \to e$ conversion through the interaction with a quark $q$ equal to $u$ or $d$: 
\bea
{\cal M}^{\mu q \rightarrow e q} = 
{\cal M}^{\mu q \rightarrow e q}_\gamma + {\cal M}^{\mu q \rightarrow e q}_Z + 
{\cal M}^{\mu q \rightarrow e q}_{\rm Box}\ , 
\eea
with the amplitudes defined as,
\begin{align}
	\label{photonamplitudeN}
	{\cal M}^{\mu q \rightarrow e q}_\gamma = & \,\overline{u}(p_1)\, 
	e \left[ i\, F_M^\gamma (0)\, 2 P_R\, \sigma^{\mu \nu} (p_1 - p_\ell)_\nu + 
	F_L^\gamma ((p_1 - p_\ell)^2) \gamma^\mu P_L \right] u(p_\ell) \nonumber \\
	& \times \frac{1}{(p_1 - p_\ell)^2} 
	\overline{u}(p_3) \gamma_\mu\,( g_{L q}^\gamma P_L + g_{R q}^\gamma P_R )\,v(p_2) 
	\ ,\\
	\label{ZamplitudeN}
	{\cal M}^{\mu q \rightarrow e q}_Z = & \,\overline{u}(p_1)\,(-e F_L^Z (0))\,\gamma^\mu P_L u(p_\ell) 
	\frac{1}{M_Z^2} \overline{u}(p_3) \gamma_\mu\,( g_{L q}^Z P_L + g_{R q}^Z P_R )\,v(p_2) 
	\ ,\\ 
	\label{BoxamplitudeN}
	{\cal M}^{\mu q \rightarrow e q}_{\rm Box} = & \,e^2 B_{L}^q (0)\ \overline{u}(p_1)\, 
	\gamma^\mu P_L u(p_\ell)\ \overline{u}(p_3) \gamma_\mu P_L\,v(p_2) \ .
\end{align}
The form factors $F_M^\gamma (0)$, $F_L^\gamma$, and $F_L^Z (0)$ are given in Eqs.\,(\ref{dipoleformfactormue}), (\ref{gammapenguin}), and (\ref{Zpenguin}) respectively while the couplings $g_{L (R) q}^{\gamma (Z)}$ are gathered in Table \ref{tab1} for $q= u, d$.~The three form factors include the contributions from both mirror and partner leptons, the latter of which have not been previously computed.~Analogously, $B_{L}^q (0)$ can be read from Eqs.\,(\ref{B}) and (\ref{Bs}) and replacing the appropriate charges, masses, and mixings (and multiplying by a global factor of one-half to account for no crossed box diagrams for swapping leptons).~The mirror fermion contribution is detailed in Ref. \cite{delAguila:2010nv}, and is contained in the corresponding sums exchanging $W_H W_H$ (first two terms), $Z_H Z_H$ (third term), $A_H A_H$ (fourth term), and $A_H Z_H$ (fifth term) in Eq.\,(\ref{Bs}): 
\begin{align}
	B_{L u}^{W_H W_H+Z_H Z_H+A_H A_H+A_H Z_H} &=
	\frac{\alpha_W}{32\pi s_W^2} \frac{1}{M_{W_H}^2} \times 
	\nonumber \\ 
	\sum_{ij}\chi_{ij}^u\left[ -\left(8 +\frac{1}{2} \frac{m_{\ell_{H i}}^2}{M_{W_H}^2}\right.\right.&\left.
	\frac{m_{d_{H j}}^2}{M_{W_H}^2} \right)
	\tilde{d}\left(\frac{m_{\ell_{H i}}^2}{M_{W_H}^2} , \frac{m_{d_{H j}}^2}{M_{W_H}^2} \right) 
	+4 \frac{m_{\ell_{H i}}^2}{M_{W_H}^2} \frac{m_{d_{H j}}^2}{M_{W_H}^2} 
	d\left(\frac{m_{\ell_{H i}}^2}{M_{W_H}^2} , \frac{m_{d_{H j}}^2}{M_{W_H}^2}\right) 
	\nonumber \\ 
	-\frac{3}{2}\tilde{d}\left(\frac{m_{\ell_{H i}}^2}{M_{W_H}^2} , \frac{m_{d_{H j}}^2}{M_{W_H}^2}\right)
	&\left.-\frac{3}{50} \frac{M_{A_H}^2}{M_{W_H}^2}
	\tilde{d}\left(\frac{m_{\ell_{H i}}^2}{M_{A_H}^2} , \frac{m_{d_{H j}}^2}{M_{A_H}^2}\right)
	+\frac{3}{5}\tilde{d}' \left(\frac{m_{\ell_{H i}}^2}{M_{A_H}^2} , \frac{m_{d_{H j}}^2}{M_{A_H}^2} , 
	\frac{M_{W_H}^2}{M_{A_H}^2} \right)
	\right] \, , \nonumber \\
	B_{L d}^{W_H W_H+Z_H Z_H+A_H A_H+A_H Z_H} &=
	\frac{\alpha_W}{32\pi s_W^2} \frac{1}{M_{W_H}^2} \times 
	\\ 
	\sum_{ij}\chi_{ij}^d \left[ \left(2 +\frac{1}{2} \frac{m_{\ell_{H i}}^2}{M_{W_H}^2}\right.\right.&\left.
	\frac{m_{d_{H j}}^2}{M_{W_H}^2} \right)
	\tilde{d}\left(\frac{m_{\ell_{H i}}^2}{M_{W_H}^2} , \frac{m_{d_{H j}}^2}{M_{W_H}^2} \right) 
	-4 \frac{m_{\ell_{H i}}^2}{M_{W_H}^2} \frac{m_{d_{H j}}^2}{M_{W_H}^2} 
	d\left(\frac{m_{\ell_{H i}}^2}{M_{W_H}^2} , \frac{m_{d_{H j}}^2}{M_{W_H}^2}\right) 
	\nonumber \\ 
	-\frac{3}{2}\tilde{d}\left(\frac{m_{\ell_{H i}}^2}{M_{W_H}^2} , \frac{m_{d_{H j}}^2}{M_{W_H}^2}\right)
	&\left.-\frac{3}{50} \frac{M_{A_H}^2}{M_{W_H}^2}
	\tilde{d}\left(\frac{m_{\ell_{H i}}^2}{M_{A_H}^2} , \frac{m_{d_{H j}}^2}{M_{A_H}^2}\right)
	-\frac{3}{5}\tilde{d}' \left(\frac{m_{\ell_{H i}}^2}{M_{A_H}^2} , \frac{m_{d_{H j}}^2}{M_{A_H}^2} , 
	\frac{M_{W_H}^2}{M_{A_H}^2} \right)
	\right] \, .
	\nonumber 
\end{align}
Obviously, now the mixing coefficients involve mirror lepton mixing matrices 
as well as mirror quark ones, defined in Eq.\,(\ref{matrices}):
\bea
\chi_{ij}^u = V_{e i}^\dagger V_{i \mu} V_{u j}^{u \dagger} V_{j u}^u \,,\quad
\chi_{ij}^d = V_{e i}^\dagger V_{i \mu} V_{d j}^{d \dagger} V_{j d}^d \, .
\eea
The (new) partner fermion contribution can be also read from Eq.\,(\ref{Bs}).~In this case the exchanged bosons are the charged scalar triplet components with the contributions from the different field sets running in the box being equal up to mixing coefficients $\tilde{\chi}^{u, d}_{i j}$: 
\begin{align}
	B_{L u}^{\Phi \Phi} &= 
	\frac{\alpha_W}{64 \pi s_W^2} \frac{1}{M_\Phi^2}
	\sum_{i j} \tilde{\chi}^u_{i j} 
	\tilde{d} \left(\frac{m^2_{\tilde{\nu}^c_i}}{M_{\Phi}^2} , \frac{m^2_{\tilde{u}_j}}{M_{\Phi}^2} \right) \, , 
	\nonumber \\ 
	B_{L d}^{\Phi \Phi}  &= 5
	\frac{\alpha_W}{64 \pi s_W^2} \frac{1}{M_\Phi^2}
	\sum_{i j}\tilde{\chi}^d_{i j} 
	\tilde{d} \left(\frac{m^2_{\tilde{\nu}^c_i}}{M_{\Phi}^2} , \frac{m^2_{\tilde{u}_j}}{M_{\Phi}^2} \right) \, ,
\end{align}
where the mixing coefficients are defined in terms of the mixing matrices as, 
\begin{align}
	\tilde{\chi}^u_{i j} = \sum_{k,n,r,s} V_{e k}^\dagger \frac{m_{\ell_{H k}}}{M_{W_H}} 
	W_{k i}^\dagger W_{i n} \frac{m_{\ell_{H n}}}{M_{W_H}} V_{n \mu} V^{u\dagger}_{u r} 
	\frac{m_{d_{H r}}}{M_{W_H}} W_{r j}^{q \dagger} W_{j s}^q \frac{m_{d_{H s}}}{M_{W_H}} V^u_{s u}\, , 
	\nonumber \\
	\tilde{\chi}^d_{i j} = \sum_{k,n,r,s} V_{e k}^\dagger \frac{m_{\ell_{H k}}}{M_{W_H}} 
	W_{k i}^\dagger W_{i n} \frac{m_{\ell_{H n}}}{M_{W_H}} V_{n \mu} V^{d\dagger}_{d r} 
	\frac{m_{d_{H r}}}{M_{W_H}} W_{r j}^{q \dagger} W_{j s}^q \frac{m_{d_{H s}}}{M_{W_H}} V^d_{s d}\, . 
\end{align}
The extra factor of 5 for $B_{L d}^{\Phi \Phi}$ accounts for the fact that for down quarks we can have $\tilde{x}$ (of charge 5/3) and $\Phi^{++}$ as well as $\tilde{u}$ (of charge 2/3) and $\Phi^{+}$ circulating in the loop and keeping in mind the components of the heavy $SU(2)_L$ multiplets are degenerate at the order we work.~Summing the different box contributions, 
\bea
B_{L}^q (0) = B_{L q}^{W_H W_H+Z_H Z_H+A_H A_H+A_H Z_H} + B_{L q}^{\Phi \Phi} \, , \quad q = u , d \, .
\eea
This gives for the corresponding conversion width in a nucleus N with $Z$ protons and $N$ neutrons 
\cite{Hisano:1995cp} (with $Z$ not to be confused with the $Z$ gauge boson),
\begin{align}
	\Gamma (\mu\ {\rm N} \to e\ {\rm N}) = &\ \alpha^5 \frac{Z^4_{\rm eff}}{Z} F_p^2 m_\mu^5 
	\left| 2Z(A_L + A_R) - (2Z+N)(F_{LL}^u + F_{LR}^u + B_L^u) \right. \nonumber \\
	&\left. -\ (Z+2N)(F_{LL}^d + F_{LR}^d + B_L^d) \right|^2\ , 
\end{align}
where $Z_{\rm eff}$ is the nucleus effective charge for the muon and $F_p$ the associated form factor.~We also have $m_\ell = m_\mu$ in the definition of $A_R$ in Eq.\,(\ref{simplification}) while $g_{L(R) q}^Z$ are the $Z$ couplings to the quark $q = u, d$ given in Table \ref{tab1} and entering in the definition of $F_{LL(LR)}^q$ of the same equation. 
\begin{table}[t]
	\begin{center}
		\begin{tabular}{c||ccccc}
			Nucleus & $N$ & $Z$ & $Z_{\rm eff}$ & $F_p$  
			& $\Gamma_{\rm capture} [{\rm GeV}]$  \\
			\hline
			&&&&&\\[-0.45cm]
			$^{27}_{13}{\rm Al}$ & 14 & 13 & 11.5 & 0.64 & $4.6 \times 10^{-19}$ \\
			$^{48}_{22}{\rm Ti}$ & 26 & 22 & 17.6 & 0.54 & $1.7 \times 10^{-18}$ \\
			$^{197}_{\ 79}{\rm Au}$ & 118 & 79 & 33.5 & 0.16 & $8.6 \times 10^{-18}$ \\
		\end{tabular}
		\caption{Input parameters for the nuclei considered in our analysis (see Refs. \cite{Kitano:2002mt} 
			and \cite{Suzuki:1987jf}).}
		\label{Nuclei}
	\end{center}
\end{table}
In Table \ref{Nuclei} we gather the input parameters for Al and for Ti and Au, to be used below when 
comparing with future \cite{Abusalma:2018xem,Angelique:2018svf} and current limits, respectively. 

%%%%%%%%%%%%
\subsection{One-loop contributions to $Z \to \overline{\ell} \ell'$ decays}
\label{Zmue}

Let us finally evaluate the $Z$ boson decay into a pair of charged leptons of different flavor.~The corresponding $Z$ penguin contributing to rare processes with small transfer momentum $Q^2 \sim m_\ell^2$ has been discussed in the previous subsection (see Eqs.\,(\ref{vertexmueZ}), (\ref{Zpenguin}) and (\ref{Zpenguin1})).~In that case only the first contribution to $F_L^Z$ in Eq.\,(\ref{Zpenguin}), proportional to $v^2 / f^2$, is relevant because $Q^2 \ll v^2$.~However, in $Z$ decays with $Q^2 = M_Z^2 \sim v^2$ all the contributions to $F_L^Z$ in Eq.\,(\ref{Zpenguin}) are comparable and must be taken into account.~Overall the $Z \overline{\ell} \ell'$ vertex only receives significant contributions from the $F_L^Z$ form factor (see Eq.\,(\ref{vertexmueZ})), as all others are suppressed by (light) SM lepton masses.~Therefore, the corresponding amplitude for a $Z$ boson with polarization $s$ decaying into $\overline{\ell} \ell'$ can be written in our case 
\begin{eqnarray}
\label{amplitude}
{{\cal M}_s}(Z \rightarrow \overline{\ell}\ell') = i\,e\ 
\bar{u}(p_{\ell'},m_{\ell'})\ F^{Z}_L (M_Z^2)\ \gamma_\mu P_L\ v(p_\ell,m_\ell)\ \epsilon_s^\mu (p_{\ell'} - p_\ell) \ , 
\label{Zfiniteamplitude}
\end{eqnarray}
where $\epsilon_s^\mu (p_{\ell'} - p_\ell)$ is the $Z$ polarization vector, with $Q = p_{\ell'} - p_\ell$ and $Q^2 = M_Z^2$.~Hence, the $Z$ width reduces to 
\begin{eqnarray}
\label{ZBR}
\Gamma (Z \rightarrow \overline{\ell}\ell') = 
\frac{\alpha}{3} M_Z |F^{Z}_L(M_Z^2)|^2 \ .
\label{finiteamplitude}
\end{eqnarray}

In order to conclude this section we provide further details of the calculation of $F_L^Z$ in Eqs.\,(\ref{Zpenguin}) and (\ref{Zpenguin1}) where we obtained the leading contributions from the full T--odd spectrum in the LHT.~It is instructive to explicitly check the cancellation of the divergent terms in order to compare with the corresponding Higgs decay and previous calculations of LFV $Z$ decays which did not include the partner leptons and electroweak triplet $\Phi$. 
The unitarity of the mixing matrices $V$ and $W$ ensures that one is left only with divergent terms for a given topology (see Figure~\ref{diagrams}) and T--odd fields running in the loop which have at least 
two $m_{\ell_{H i}}$ insertions (see Eq.\,(\ref{Zpenguin})).~These terms can be ${\cal O} (1)$ or ${\cal O} (\frac{v^2}{f^2})$, but in both cases must sum to zero.~This can be ensured for the ${\cal O} (1)$ terms because they are generated by dimension four operators which can always be assumed (rotated) to be flavor diagonal.~For the higher dimensional operators this is not the case and a more subtle cancellation between different divergences is needed. 

These cancellations can be seen explicitly in Tables \ref{Z1infinities} and 
\ref{Zvfinfinities}.\footnote{
	We use the Feynman rules in the 't Hooft-Feynman gauge collected in the previous section and dimensional regularization.~Similarly as in the Higgs case, the divergent part of the amplitude (see Eq.\,(\ref{Zfiniteamplitude})) can be written as: ${\cal M}^\mu_{\rm div} (Z \rightarrow \overline{\ell}\ell') = i {1 \over 16\pi^2} {g\over 2 c_W} (C^{(1)}_{\rm UV} + {v^2 \over f^2} C_{\rm UV}^{(\frac{v^2}{f^2})}) \frac{1}{\epsilon} 
	\sum_{i=1}^3 V^\dagger_{\ell' i} V_{i \ell} \frac{m_{\ell_{H i}}^2}{f^2}
	\bar{u}(p_{\ell'},m_{\ell'}) \gamma^\mu P_L v(p_\ell,m_\ell)$.   
}%
\begin{table}
	\begin{center}
		\begin{tabular}{r||ccccccc|c}
			$C_{\rm UV}^{(1)}\ $ & {\small I} & {\small II} & {\small III} & 
			IV & {\small V+VI} & {\small VII+VIII} & {\small IX+X} &\ Sum \\
			\hline
			$W_H, \nu_H\ $ & 0 & 0 & $-$ & $-$ & $-$ & $0$ & $-$ & $0$ \\
			$W_H, \omega, \nu_H\ $ & $-$ & $-$ & $-$ & $-$ & $0$ & $-$ & $-$ & $0$ \\
			$\omega, \nu_H\ $ & $-$ & $-$ & $\frac{1}{2}$ & $-1 + s_W^2$ & $-$ & $-$ & $\frac{1}{2} - s_W^2$ & $\bullet$ \\
			\hline
			$A_H, \ell_H\ $ & 0 & $\bullet$ & $-$ & $-$ & $-$ & 0 & 
			$-$ & $\bullet$ \\
			$\eta, \ell_H\ $ & $-$ & $-$ & $-\frac{1}{20} + \frac{s^2_W}{10}$ & $\bullet$ & -- & $-$ & 
			$\frac{1}{20} - \frac{s^2_W}{10}$ & $\bullet$ \\
			\hline
			$Z_H, \ell_H\ $ & 0 & $\bullet$ & $-$ & $-$ & $-$ & 0 & 
			$-$ & $\bullet$ \\
			$\omega, \ell_H\ $ & $-$ & $-$ & $-\frac{1}{4} + \frac{s^2_W}{2}$ & $\bullet$ & -- & -- & 
			$\frac{1}{4} - \frac{s^2_W}{2}$ & $\bullet$ \\
			\hline
			$\Phi, \tilde{\nu}^c\ $ & -- & -- & $- \frac{1}{2}$ & $s_W^2$ & -- & -- & 
			$\frac{1}{2} - s_W^2$ & $\bullet$ \\
			\hline
			$\Phi, \tilde{\ell}^c\ $ & -- & -- & $1 - 2 s_W^2$ & $-2 + 4 s_W^2$ & -- & -- & 
			$1 - 2s_W^2$ & $\bullet$ \\
			\hline
			Total\ \ & 0 & 0 & $\frac{7}{10} - \frac{7 s_W^2}{5}$ & $-3 + 6 s_W^2$ & $0$ & 0 & 
			$\frac{23}{10} - \frac{23 s_W^2}{5}$ & $\bullet$ \\
		\end{tabular}
		\caption{The ${\cal O} (1)$ divergent contributions proportional to $\frac{1}{\epsilon}$ ($\epsilon = 4 - d$) of each particle set running in the loop and topology shown in Figure~\ref{diagrams}.~A dash means that the field set does not run in the diagram whereas a bullet indicates the infinite \emph{and} finite parts vanish.}
		\label{Z1infinities}
	\end{center}
\end{table}
\begin{table}
	\begin{center}
		\begin{tabular}{r||ccccccc|c}
			$C_{\rm UV}^{(\frac{v^2}{f^2})}$ & {\small I} & {\small II} & {\small III} & 
			 IV & {\small V+VI} & {\small VII+VIII} & {\small IX+X} & Sum \\
			\hline
			$W_H, \nu_H$ & 0 & 0 & $-$ & $-$ & $-$ & $0$ & $-$ & $0$ \\
			$W_H, \omega, \nu_H$ & $-$ & $-$ & $-$ & $-$ & $0$ & $-$ & $-$ & $0$ \\
			$\omega, \nu_H$ & $-$ & $-$ & $-\frac{1}{8}$ & $\frac{1}{8}$ & $-$ & $-$ & $0$ & $0$ \\
			\hline
			$A_H, \ell_H$ & 0 & $\bullet$ & $-$ & $-$ & $-$ & 0 & 
			$-$ & $0$ \\
			$\eta, \ell_H$ & $-$ & $-$ & $\frac{1}{16} - \frac{s^2_W}{8} + \frac{y_H s_W}{5 c_W}$ & $\bullet$ & -- 
			& $-$ & $-\frac{1}{16} + \frac{s^2_W}{8} - \frac{y_H s_W}{5 c_W}$ & 0 \\
			\hline
			$Z_H, \ell_H$ & 0 & $\bullet$ & $-$ & $-$ & $-$ & 0 & 
			$-$ & $0$ \\
			$\omega, \ell_H$ & $-$ & $-$ & $\frac{1}{16} - \frac{s^2_W}{8} - \frac{y_H c_W}{s_W}$ & $\bullet$ & -- 
			& $-$ & $-\frac{1}{16} + \frac{s^2_W}{8} + \frac{y_H c_W}{s_W}$ & 0 \\
			\hline
			$\Phi, \tilde{\nu}^c$ & -- & -- & $\frac{1}{8}$ & $-\frac{1}{8}$ & -- & -- & $\bullet$ & 0 \\
			\hline
			$\Phi, \tilde{\ell}^c$ & -- & -- & 0 & 0 & -- & -- & $\bullet$ & 0 \\
			\hline
			Total & 0 & 0 & 0 & 0 & 0 & 0 & 
			0 & 0 \\
		\end{tabular}
		\caption{As in Table \ref{Z1infinities} but to ${\cal O} (\frac{v^2}{f^2})$. 
			$y_H = \frac{5 s_W c_W (1-2 c_W^2)}{8(1-6 c_W^2)}$.}
		\label{Zvfinfinities}
	\end{center}
\end{table}
The columns label the contributing topologies listed in Figure~\ref{diagrams} while the rows label the different field sets running in the loop.~As pointed out, the leading divergences $C_{\rm UV}^{(1)}$ in Table \ref{Z1infinities} cancel, as do the corresponding finite parts, for each field set when adding all the diagrams.~This is represented by the bullets in the last column of this table.~The same happens for the next to leading ones $C_{\rm UV}^{(\frac{v^2}{f^2})}$ in Table \ref{Zvfinfinities}, but now the corresponding finite parts do not cancel.~This is indicated by the zeroes in the last column of this second table.~Hence, as noted in the previous subsection all the contributions to $F_L^Z$ in Eqs.\,(\ref{Zpenguin}) and (\ref{Zpenguin1}) are finite.~In particular, the contributions from the mirror leptons $F_L^Z|_{W_H, A_H, Z_H}$ are alone finite as are those from the partner leptons $F_L^Z|_{\tilde{\nu}^c, \tilde{\ell}^c}$ which decouple (go to zero) for large $m_{\tilde{l}}^2 / M_\Phi^2$.~In contrast, the mirror lepton contributions are proportional to the loop functions $H_L^{W (0)}$ and $H_L^{W, A/Z}$ in Eq.\,(\ref{Zpenguin1}) (or $F_L^{W, A/Z}$ in Eq.\,(\ref{gammapenguin1})) which do not vanish for $m_{\ell_H}^2 / M_{W_H}^2 \rightarrow \infty$, but instead $H_L^{W (0)}$ grows linearly with $m_{\ell_H}^2 / M_{W_H}^2$ (which scales with the Yukawa coupling $\kappa^2$) while $H_L^{W, A/Z}$ tends to a constant. 

Finally, we comment that the partner lepton contributions to the $Z$ left-handed vector form factor $F_L^Z$ cancel for $Q^2 \rightarrow 0$, but this is not the case for mirror lepton contributions.~Thus, although they are more restrictive, LFV processes with small momentum transfer ($Q^2 \sim m_{\ell}^2$) such as lepton decays and transitions, probe a different kinematic regime than the one probed by on-shell $Z$ decays which have $Q^2 = M_Z^2$.~This implies low energy LFV processes and LFV $Z$ (and Higgs) decays probe different form factor combinations making them sensitive to different regions of parameter space.~However, as we discuss in the next section, limits on LFV $Z$ decays are not yet sensitive to much of the presently allowed LHT parameter space.

%--CONFRONTING LFV WITH EXPERIMENT--%

\section{Confronting LFV processes with experiment}
\label{sec:ConfrontingLFVandLHT}

In this section we study the qualitative behavior of the different contributions to the LFV processes 
computed above and examine their dependence on the most relevant LHT parameters.~Although there 
are three families of light and heavy fermions, it is sufficient to consider just two and discuss the main 
implications of current experimental data.~We therefore concentrate on mixing in the $\mu - e$ or 
$\tau - \mu$ sectors.~Mixing in the $\tau - e$ sector is analogous to that in the $\tau - \mu$ sector while 
both are less constrained than the $\mu - e$ sector.~Allowing for three families all together would of course 
open potential cancellations which could restate the constraints on the LHT parameters in a different way.

The most restrictive constraints come from $\mu \to e \gamma$ and $\mu \to e$ conversion in nuclei.
~They require an effective alignment of the first two SM families with their mirror counterparts within 
$\sim 1 \%$ for $f \sim 1$ TeV.~In our studies of the parameter space below we will fix $f = 1.5$ TeV. 
We note that all predictions scale as  $f^{-4}$ and $f = 15$ TeV would give the same suppression for 
misalignment of order 1 between the SM and the T--odd leptons.~The mixing matrices are assumed 
to involve only two families and hence, for $\mu - e$ mixing 
\begin{equation}
V=\begin{bmatrix}
\cos\theta_V &  \sin\theta_V & 0 \\
-\sin\theta_V  & \cos\theta_V & 0 \\
0 & 0 & 1
\end{bmatrix}\ , \quad  
W=\begin{bmatrix}
\cos\theta_W &  \sin\theta_W & 0 \\
-\sin\theta_W  & \cos\theta_W  & 0 \\
0 & 0 & 1
\end{bmatrix} \; ,   
\label{mixingmatrix}
\end{equation}
where $\theta_W$ must not be confused with the electro--weak mixing or Weinberg angle. 
The physical range for the mixing angles is between $[0, \pi/ 2)$ since the amplitudes depend only on 
$\sin(2 \theta_{V, W})$.~Except otherwise stated, in the following we will use as default values 
$\theta_{V, W} = \pi / 4$ to maximize the LFV effects.~For the first two mirror lepton families we will take 
as default values $m_{\ell_{H 1}} m_{\ell_{H 2}} = 1$ TeV$^2$ and $m_{\ell_{H 2}}^2 - m_{\ell_{H 1}}^2  = 1$ TeV$^2$ 
and similarly for the partner lepton families $m_{\tilde{\nu}^c_{1}} m_{\tilde{\nu}^c_{2}} = 1$ TeV$^2$ and 
$m_{\tilde{\nu}^c_{2}}^2 - m_{\tilde{\nu}^c_{1}}^2 = 1$ TeV$^2$ 
while the mass of the third heavy lepton family is fixed to $m_{\ell_{H 3}} = m_{\tilde{\nu}^c_3} =1$ TeV. 
Summarizing the default choices for our input parameter point,
\bea
\label{defpt}
m_{\ell_{H 1}} m_{\ell_{H 2}} &\equiv& \tilde{x} = 1~\rm{TeV}^2\, , \nonumber \\
m_{\ell_{H 2}}^2 - m_{\ell_{H 1}}^2 &\equiv& \delta_{\ell_{H}} \tilde{x} = 1~\rm{TeV}^2\, , \nonumber \\
m_{\tilde{\nu}^c_{1}} m_{\tilde{\nu}^c_{2}} &\equiv& \tilde{y} = 1~\rm{TeV}^2 \, , \\
m_{\tilde{\nu}^c_{2}}^2 - m_{\tilde{\nu}^c_{1}}^2 &\equiv& \delta_{\tilde{\nu}^c} \tilde{y} = 1~\rm{TeV}^2 
\, , \nonumber \\
\theta_{V, W} &=& \pi / 4 \, , \nonumber 
\eea
where we have also defined the mass dimension squared variables $\tilde{x}, \tilde{y}$ parametrizing 
the product of heavy lepton masses and the dimensionless variables $\delta_{\ell_{H}}, \delta_{\tilde{\nu}^c}$ 
parametrizing the mass (squared) splittings between the heavy leptons.~The scalar triplet mass is related 
to the mass of the Higgs from the Coleman-Weinberg potential and fixed at leading order 
\cite{Coleman:1973jx,Han:2003wu} to be $M_\Phi = \sqrt{2} M_h f / v \approx f/\sqrt{2} \approx 1$~TeV for $f =1.5$~TeV. 
The evaluation of the $\mu \to e$ conversion rate also requires fixing the masses and mixings of T--odd quarks. 
We will assume no extra quark mixing and degenerate heavy quarks. 
Hence, $V_L^{q_H}$ and $W^q$ in Eq. (\ref{quarkmixing}) will be equal to the identity. 
We will use as default value $m_{d_{H i}} = m_{{\tilde{u}}_i} = 2$ TeV, fulfilling 
the current bound on pair--production of vector--like quarks \cite{Aaboud:2018pii} although it does 
not directly apply here because T--odd quark decays must involve lighter T--odd particles. 

For $\tau - \mu$ mixing we use the same notation but with the $2 \times 2$ rotation matrices in the 
bottom-right corner in Eq. (\ref{mixingmatrix}) (see Eq.\,(4.8) in Ref. \cite{delAguila:2017ugt}). We also use the same default 
masses and mixings with the understanding that the first and second T--odd lepton families, 1 and 2, stand for 
the second and third ones, 2 and 3, respectively. 
Similarly for the evaluation of processes with $\tau - e$ mixing but the $2 \times 2$ rotation matrices now 
involve the first and third T--odd lepton families, and 1 and 2 stand for 1 and 3. In both cases the mass of the 
remaining T--odd lepton family must be also fixed because it enters in the calculation of 
$\tau \to \mu \overline{e} e$ and $\tau \to e \overline{\mu} \mu$, respectively.  
In either case we will take this to be equal to the largest one of the other two T-odd lepton family masses.

%
%%%%%%%%LFV constraints Table%%%%%%%%%%%%%%%%%
\begin{table}
	\hspace*{-.1cm}
	\begin{center}
		\begin{tabular}{|c|c||c|c|}\hline
			%%%%%%%%%%%%%%%%%%%%%%%
			&~Branching Ratio~~&  
			& ~Branching Ratio~~\\\hline
			%%%%%%%%%%%%%%%%%%%%%%%
			$\mu \to e\ \gamma$ &~$4.3 \times 10^{-9}$~&
			$\mu \to e\ \overline{e}\ e$ &~$2.5\times 10^{-11}$~\\\hline
			%%%%%%%%%%%%%%%%%%%%%%%
			&~Conversion Rate~~&  &\\\hline
			$\ \mu \to e\ {\rm{(Au)}}\ $ &~$3.8\times 10^{-9}$~&&\\\hline
			$\ \mu \to e\ {\rm{(Ti)}}\ $ &~$3.3\times 10^{-9}$~&&\\\hline\hline     
			%%%%%%%%%%%%%%%%%%%%%%%
			&~Branching Ratio~~& ~~ &~~\\\hline
			$\tau \to e\ \gamma$ &~$7.3 \times 10^{-10}$~&
			~$\tau \to \mu\ \overline{e}\ \mu$ &~$0$~\\\hline
			$\tau \to \mu\ \gamma$ &~$7.3 \times 10^{-10}$~&
			$\tau \to e\ \overline{\mu}\ e$ &~$0$~\\\hline
			& &
			$\tau \to \mu\ \overline{e}\ e$ &~$8.2\times 10^{-12}$~\\\hline
			& &
			$\tau \to e\ \overline{\mu}\ \mu$ &~$ 2.2\times 10^{-12}$~\\\hline
			&&
			$\tau \to e\ \overline{e}\ e$ &~$7.4\times 10^{-12}$\\\hline
			&&
			$\tau \to \mu\ \overline{\mu}\ \mu$ &~$1.4\times 10^{-12}$~\\\hline\hline
			%%%%%%%%%%%%%%%%%%%%%%%
			$Z \to \mu\ e$ &~$2.7\times 10^{-12}$~&
			$h \to \mu\ e$ &~$1.2\times 10^{-15}$~\\\hline
			$Z \to \tau\ e$ &~$2.7\times 10^{-12}$~&
			$h \to \tau\ e$ &~$ 3.2\times 10^{-13}$~\\\hline
			$Z \to \tau\ \mu$ &~$2.7\times 10^{-12}$~&
			$h \to \tau\ \mu$ &~$3.2\times 10^{-13}$~\\\hline
		\end{tabular}
		\caption{LHT contributions mediated by T--odd (non-singlet) leptons to LFV processes for the default values 
			in~Eq.\,(\ref{defpt}) and the text.~The prediction for the (flavor conserving) muon magnetic moment 
			$a_\mu = a_\mu^{\rm SM} + \delta a_\mu^{\rm T-odd}$ is also included in the analysis with 
			$a_\mu^{\rm SM} = (116591823 \pm 43) \times 10^{-11}$ \cite{Tanabashi:2018oca} and 
			$\delta a_\mu^{\rm T-odd} = - 4.7 \times 10^{-13}$ obtained from Eqs. (\ref{App}) and 
			(\ref{dipoleformfactormue}) with $\ell' = \ell = \mu$.}
		\label{Values}
	\end{center}
\end{table}
%%%%%%%%%%%%%%%%%%%%%%%%%s
%
In Table \ref{Values} we collect the LHT contributions to the LFV processes in Table \ref{Limits} calculated in 
previous sections assuming the default values for the model parameters above. 
For processes involving $\tau$ leptons we assume $\tau - \mu$ or $\tau - e$ mixing depending 
on which final flavor enters an odd number of times. 
This is because the branching ratio vanishes when this flavor 
coincides with the unmixed one 
since unmixed fermions must be joined pairwise 
and there will always be one left unmatched. 
Moreover, the processes with double-flavor violation 
$\tau \to \mu \overline{e} \mu$ and $\tau \to e \overline{\mu} e$ 
also vanish when only $\tau - e$ or $\tau - \mu$ mixing is assumed. 
As is apparent comparing the two tables, the LHT predictions for the $\tau$ sector are similar to the $\mu$ sector, 
but do not constrain the model appreciably though they could do so in the future 
\cite{Kou:2018nap}.\footnote{The $\mu$ and $\tau$ sector predictions differ due to the different lepton masses involved in the process 
	as well as the different branching ratio normalization,
	\begin{equation}
	{\rm Br}(\ell \rightarrow \ell' \overline{\ell''} \ell''') = {\rm Br}(\ell \rightarrow \ell' \overline{\nu_{\ell'}} \nu_\ell) 
	\,  
	\frac{\Gamma (\ell \rightarrow \ell' \overline{\ell''} \ell''')}{\Gamma (\ell \rightarrow \ell' \overline{\nu_{\ell'}} \nu_\ell)} \, ,
	\nonumber
	\end{equation}
	where ${\rm Br}(\ell \rightarrow \ell' \overline{\nu_{\ell'}} \nu_\ell)$ 
	stands for the corresponding experimental value \cite{Tanabashi:2018oca} and 
	$\Gamma (\ell \rightarrow \ell' \overline{\nu_{\ell'}} \nu_\ell)$ for the SM prediction.~In particular, the prediction for 
	$\tau^- \to e^- e^+ e^-$ is larger 
	than for $\tau^- \to \mu^- \mu^+ \mu^-$ mainly due to the logarithmic mass dependent term in Eq. (\ref{threeequal}), 
	and similarly for $\tau^- \to \mu^- {e}^+ e^-$ relative to $\tau^- \to e^- {\mu}^+ \mu^-$. 
	The branching ratio for the radiative decay ${\rm Br}(\ell \rightarrow \ell' \gamma)$ is defined analogously but replacing 
	$\Gamma (\ell \rightarrow \ell' \overline{\ell''} \ell''')$ by $\Gamma (\ell \rightarrow \ell' \gamma)$ 
	while the prediction for the $\mu \to e$ conversion rate in nuclei is obtained dividing the conversion width by the 
	corresponding capture width in Table \ref{Nuclei}, 
	${\cal R} = \Gamma (\mu {\rm N} \rightarrow e {\rm N})  / \Gamma_{\rm capture}$. 
	Finally, the $Z$ and $h$ branching ratios into ${\overline{\ell}} \ell' + \ell {\overline{\ell'}}$ are normalized to the 
	SM total widths.}
On the other hand $\mu$ to $e$ transitions set stringent limits on the LHT. 

\begin{figure}{h}
	\begin{tabular}{cc}
		\includegraphics[scale=0.6]{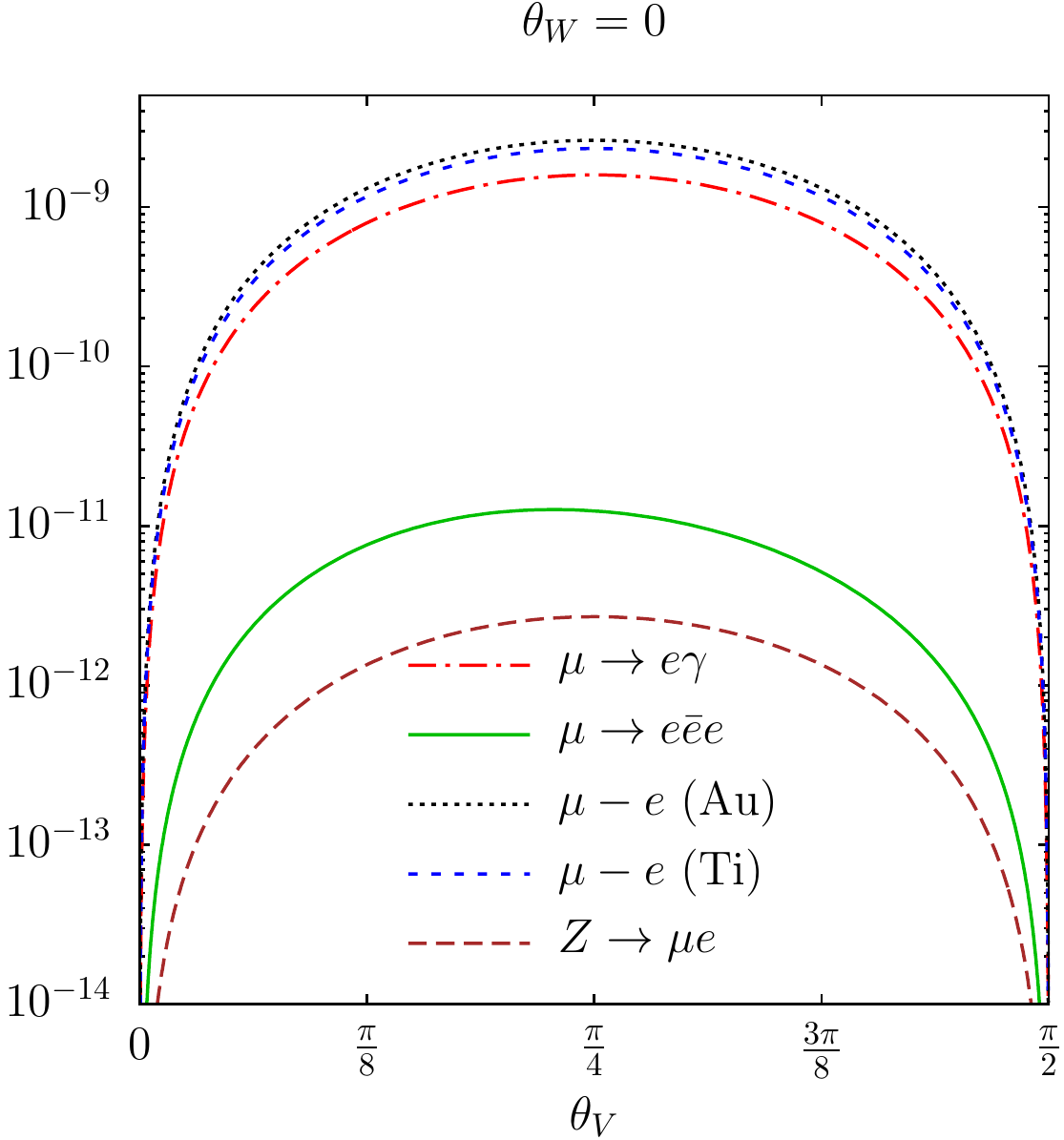} &
		\includegraphics[scale=0.6]{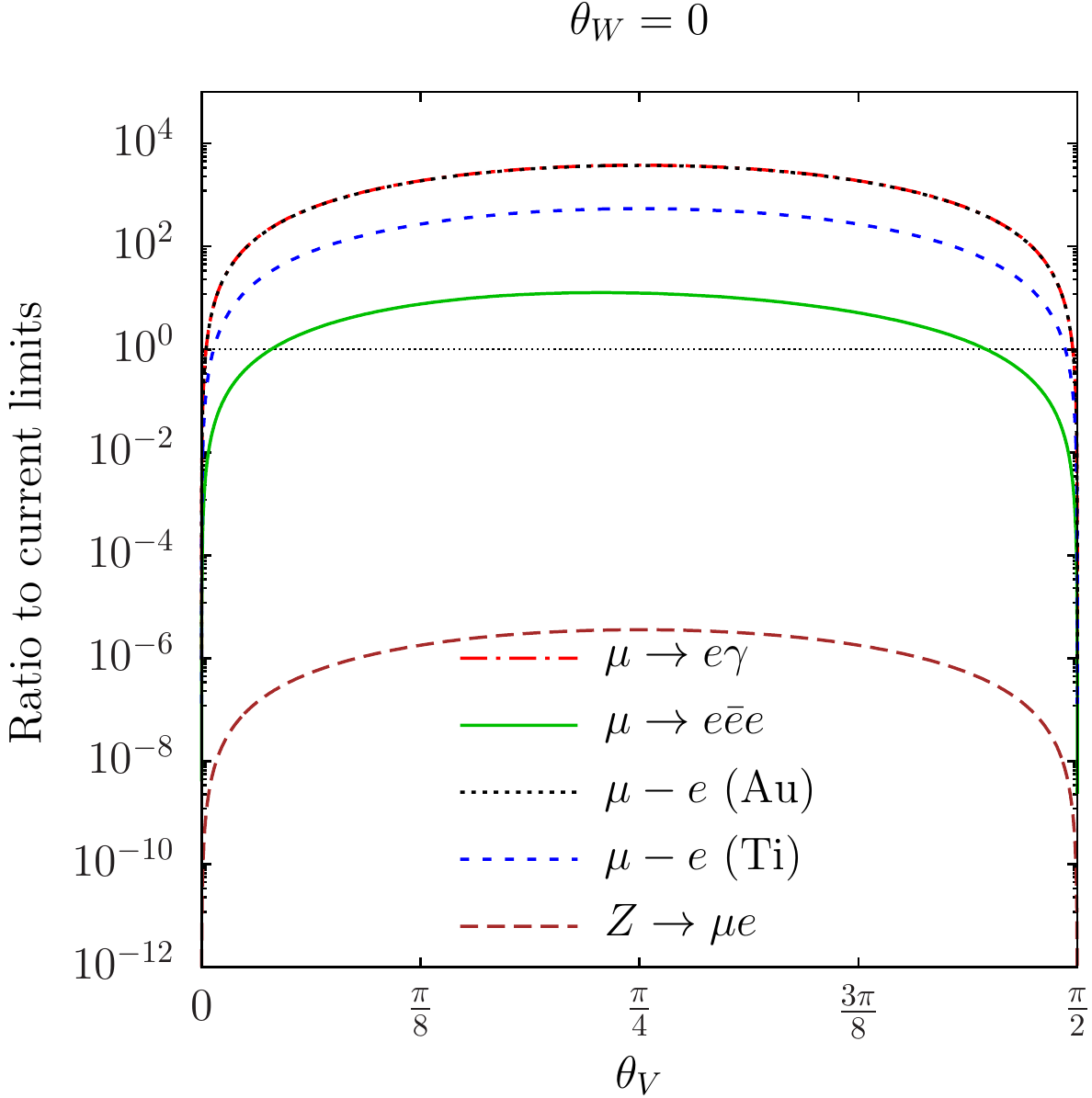} 
	\end{tabular}
	\caption{$\mu$ to $e$ transitions as a function of the $\theta_V$ mixing angle.~The other LHT parameters are fixed to 
		their default values in~Eq.\,(\ref{defpt}) and the text.~On the left we plot the branching ratios for the different 
		processes while on the right the branching ratios are normalized to the corresponding experimental limits. 
		Note that in the right-hand plot $\mu \to e \gamma$ and $\mu - e\, ({\rm Au})$ overlap.}
	\label{Panel1mu}
\end{figure}

\begin{figure}
	\begin{tabular}{cc}
		\includegraphics[scale=0.6]{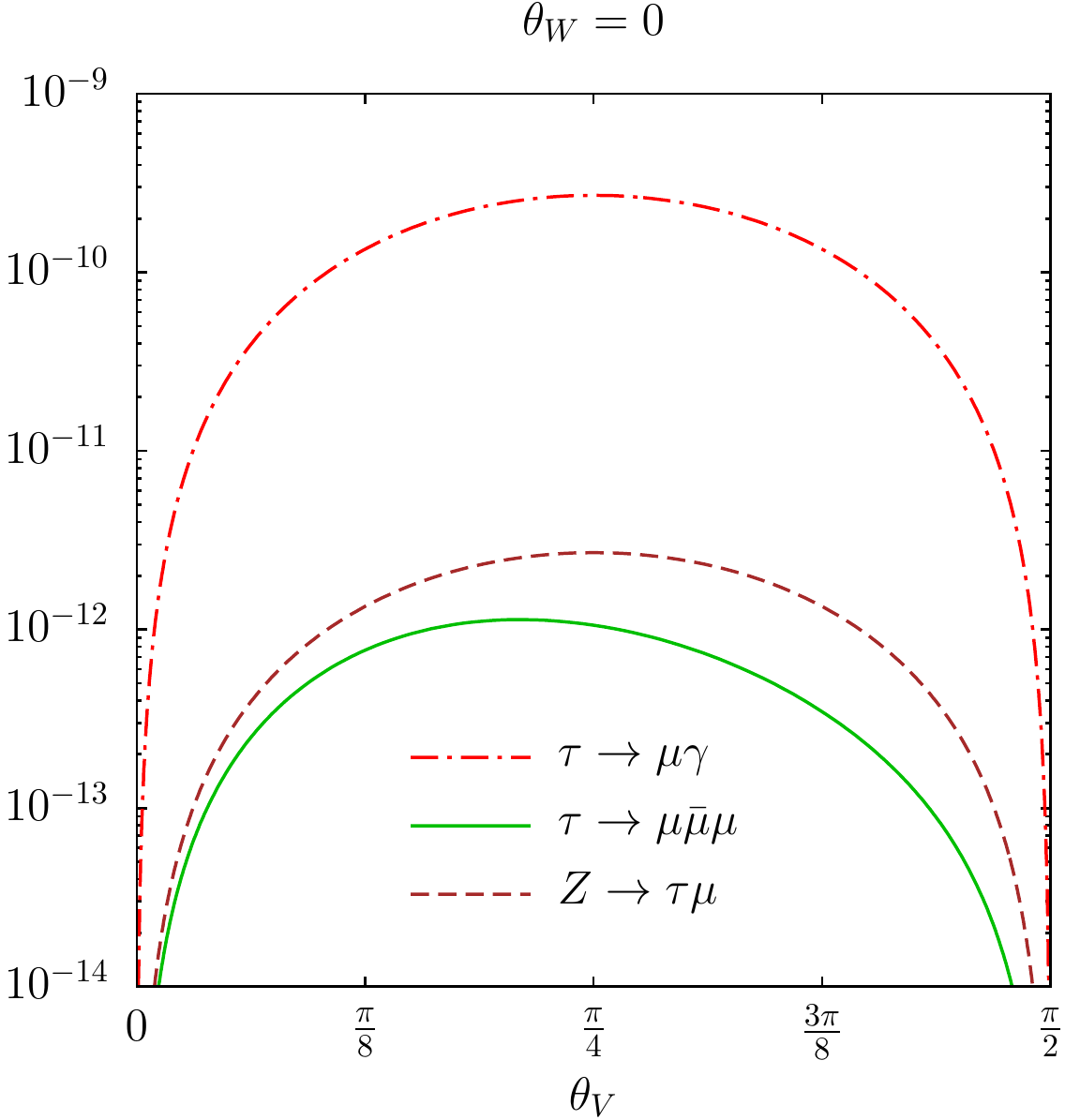} &
		\includegraphics[scale=0.6]{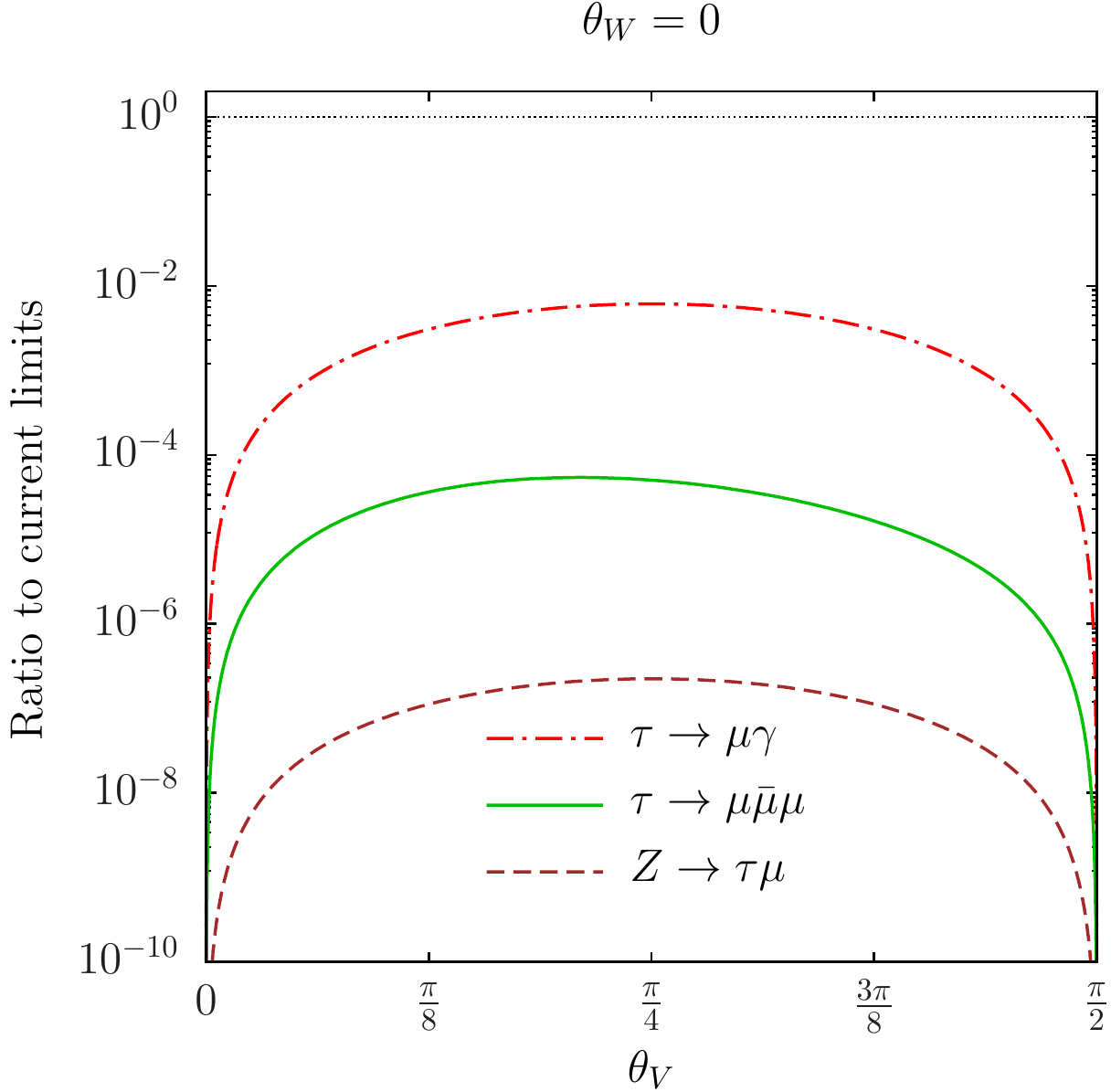} 
	\end{tabular}
	\caption{$\tau$ to $\mu$ transitions as a function of the $\theta_V$ mixing angle.~The other LHT parameters are fixed to 
		their default values in~Eq.\,(\ref{defpt}) and the text.~On the left we plot the branching ratios for the different processes 
		while on the right the branching ratios are normalized to the corresponding experimental limits.}
	\label{Panel1tau}
\end{figure}

In Figures~\ref{Panel1mu} and \ref{Panel1tau} we plot on the left panel the predictions for the corresponding LFV 
processes as a function of the mixing angle $\theta_V$ for $\theta_W = 0$ for $\mu - e$ and $\tau - \mu$ mixing, 
respectively. 
On the right panel we also show the branching ratios normalized to their current experimental limits 
in Table \ref{Limits} which illustrates the sensitivity of the different processes.~The first obvious observation about the behavior 
of the predictions of the LHT is that if the T--odd leptons are aligned with the SM ones, there is no LFV.~For vanishing mixing 
$\theta_{V, W} = 0, \pi / 2$, all of the LFV transitions go to zero.~We see in Figure \ref{Panel1mu} the largest branching ratios 
correspond to 
$\mu - e$ conversion in nuclei and to $\mu \to e \gamma$ (left panel) 
and at the same time they are also the best measured (right panel).~In contrast, limits from on-shell $Z$ and Higgs decays are 
much less restrictive.~In particular, Higgs decays are outside of the left panel and omitted in the following.~Furthermore, the 
anomalous magnetic moment $a_\mu$, which is flavor conserving and not sensitive to the mixing entering through $V$ and 
$W$, is more than two orders of magnitude below present sensitivity.~As can be observed in both panels, muon decay into 
three electrons shows an asymmetric dependence on the $\theta_V$ mixing angle.~This is in fact the generic behavior of all 
the observables when we vary the default values. 
In Figure \ref{Panel1tau} we plot the same decays as in Figure \ref{Panel1mu} but for $\tau - \mu$ (being similar the plots for 
processes involving $\tau - e$ mixing).~Comparing both figures, it is apparent how much less restrictive $\tau$ data is as all 
predictions for $\tau$ decays are below their current experimental limits (see right panel in Figure \ref{Panel1tau}).~Thus, in 
what follows we concentrate on $\mu - e$ transitions. 

\begin{figure}
	\begin{tabular}{cc}
		\includegraphics[scale=0.577]{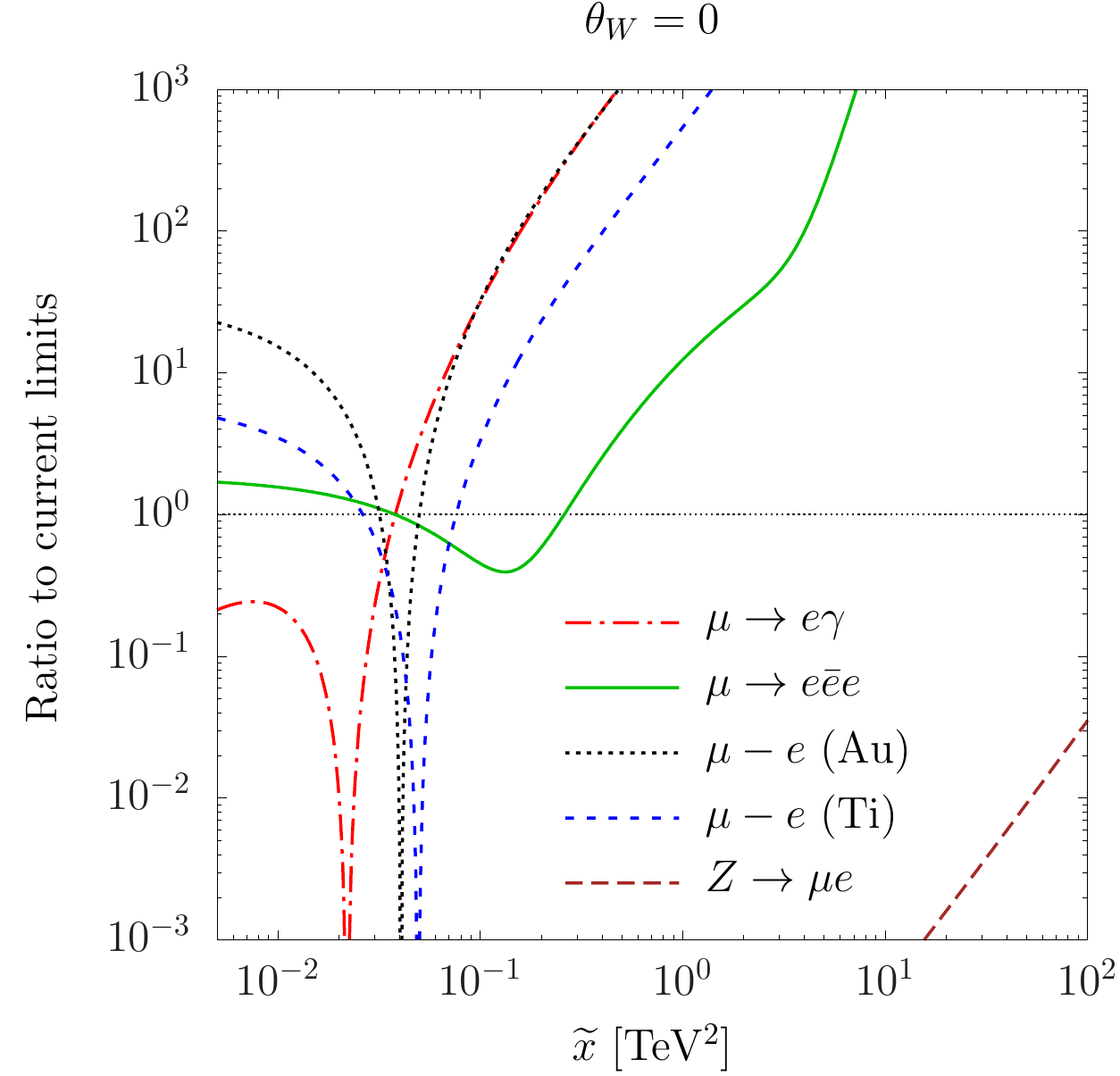} &
		\includegraphics[scale=0.577]{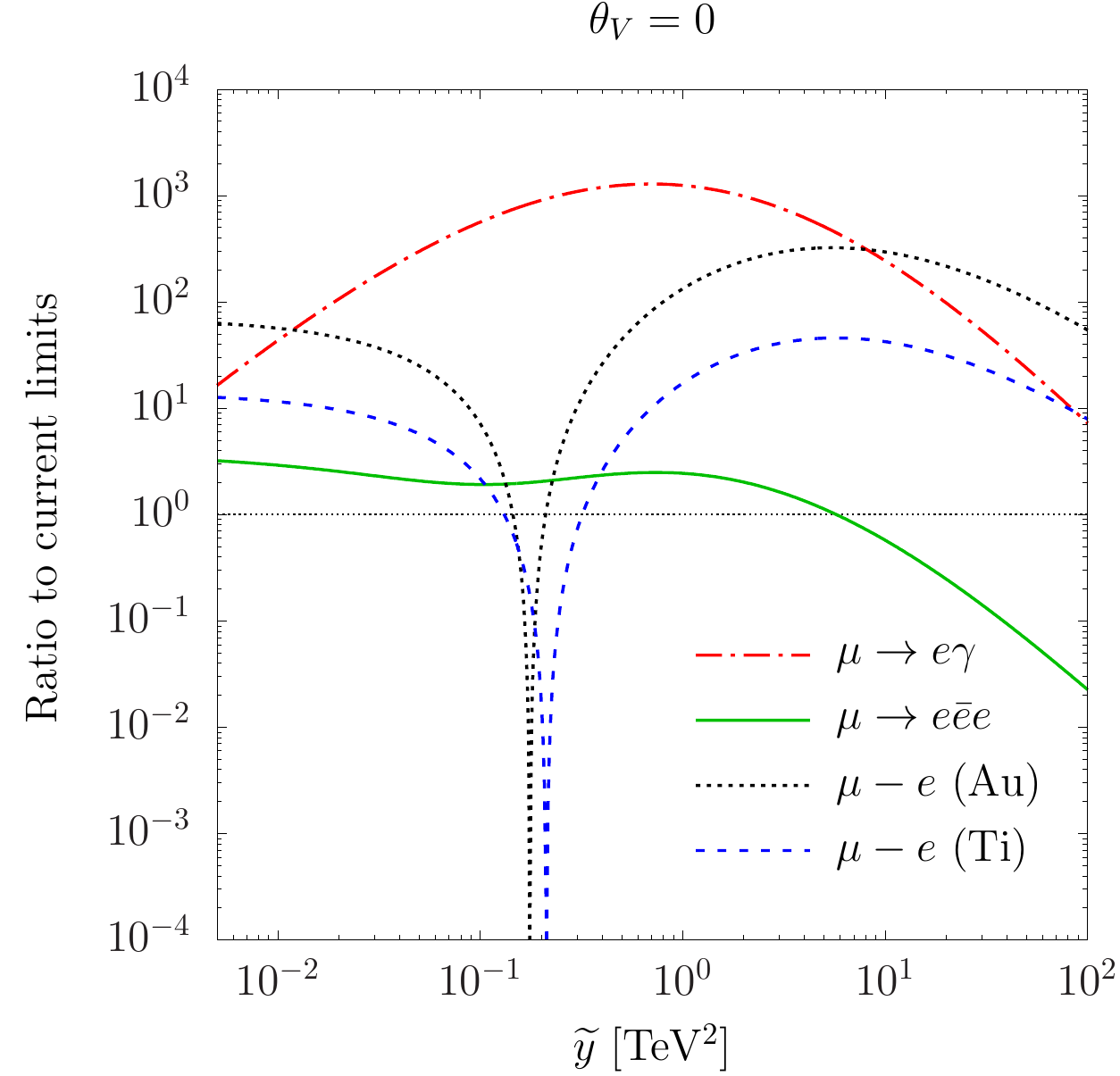} 
	\end{tabular}
	\caption{Ratios of the $\mu$ to $e$ transition branching ratios normalized to the corresponding experimental limits as a 
		function of $\tilde{x}$ and $\tilde{y}$ as defined in~Eq.\,(\ref{defpt}) for the mirror and partner leptons.~On the left (right) 
		we neglect the $\theta_{W (V)}$ mixing.~The other LHT parameters are fixed to their default values as defined 
		in~Eq.\,(\ref{defpt}).~On the right panel $Z \rightarrow \mu e$ lies outside the plot.}
	\label{Panel2mu}
\end{figure}

\begin{figure}
	\begin{tabular}{cc}
		\includegraphics[scale=0.577]{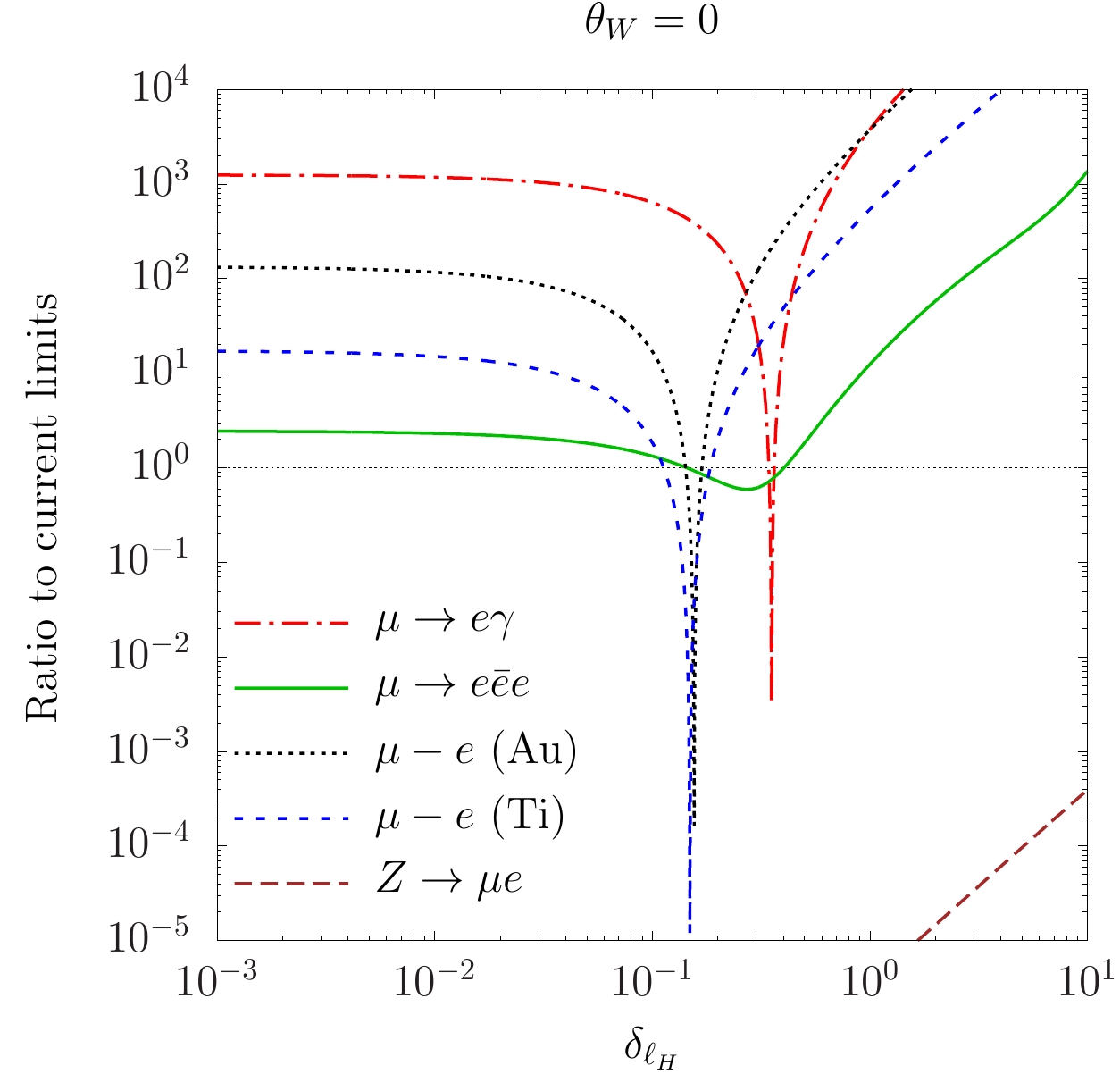} &
		\includegraphics[scale=0.577]{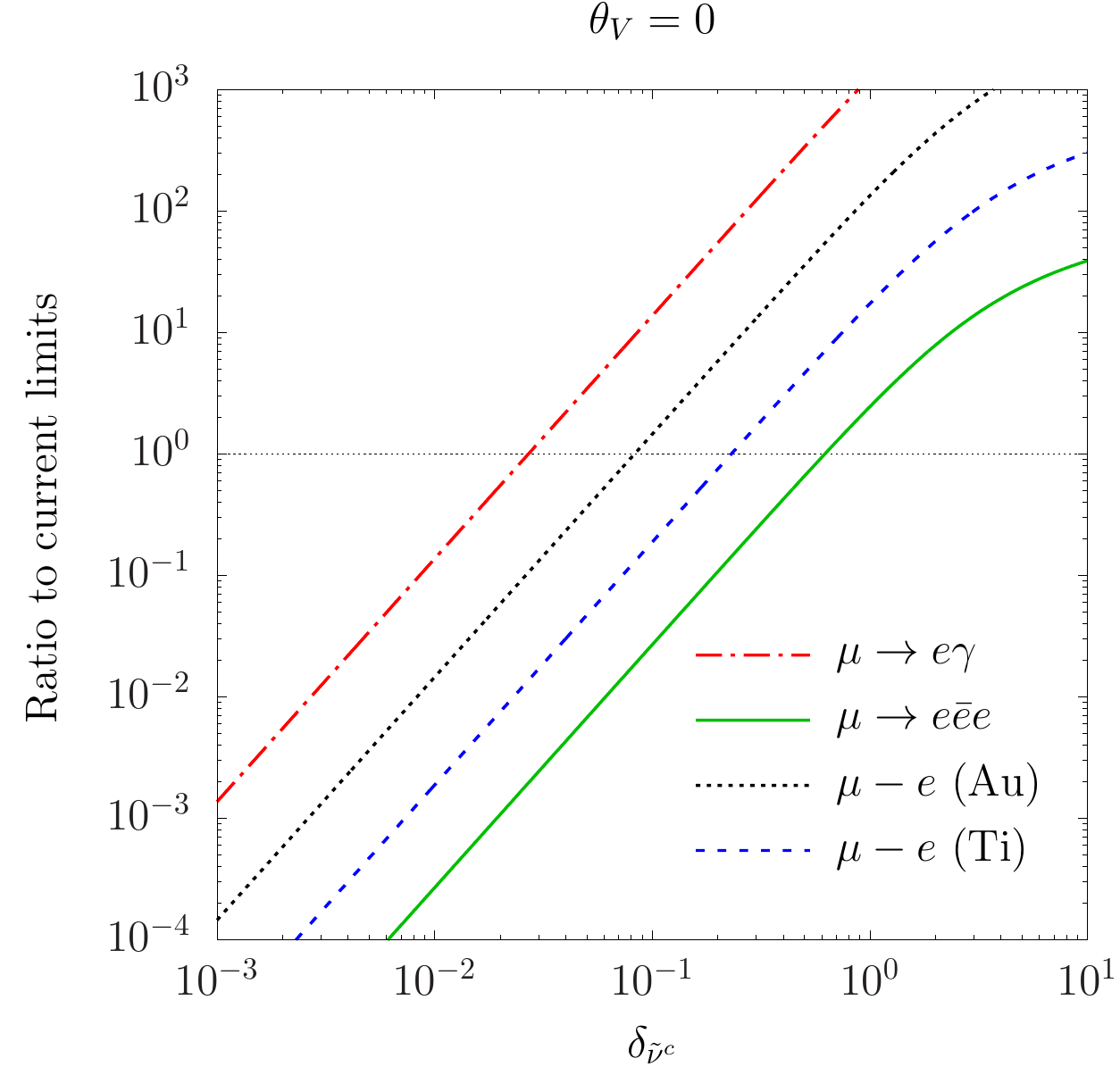} 
	\end{tabular}
	\caption{Ratios of the $\mu$ to $e$ transition branching ratios normalized to the corresponding experimental limits as a function 
		of $\delta_{\ell_H}$ (left) and $\delta_{\tilde{\nu}^c}$ (right) as defined in~Eq.\,(\ref{defpt}) for the mirror and partner leptons. 
		On the left (right) we neglect the $\theta_{W (V)}$ mixing while the other LHT parameters are fixed to the default values 
		in~Eq.\,(\ref{defpt}).}
	\label{Panel3mu}
\end{figure}

Once $f$ is fixed, the size of the LFV processes is largely determined by the masses of the mirror and partner mirror leptons. 
This is apparent from Figure \ref{Panel2mu} where we plot the branching ratios normalized to their current experimental limits 
for the different $\mu - e$ processes as a function of the product of mirror masses 
$\tilde{x}$ for vanishing $\theta_{W}$ in the left panel and of the product of partner mirror masses 
$\tilde{y}$ for vanishing $\theta_{V}$ in the right one. 
Comparing both panels one can also observe the different dependence on the mirror and partner mirror leptons which reflects 
the different dependence of the observables on the heavy fermion masses.~In the left (right) panel there is a clear structure 
depending on the mirror (partner mirror) lepton masses showing that the corresponding branching ratios can vanish if there 
is a parameter conspiracy leading to large cancellations.~These cancellations require a sharp correlation between the parameters 
as seen in the narrowness of the vanishing regions.~Similar comments apply to Figure~\ref{Panel3mu} although the dependence 
on $\delta_{\ell_H}$ is flat in 
the left panel as long as it is small while there is no available cancellation varying 
$\delta_{\tilde{\nu}^c}$ in the right panel for the values assumed for the other parameters. 
\begin{figure}
	\begin{tabular}{cc}
		\includegraphics[scale=0.59]{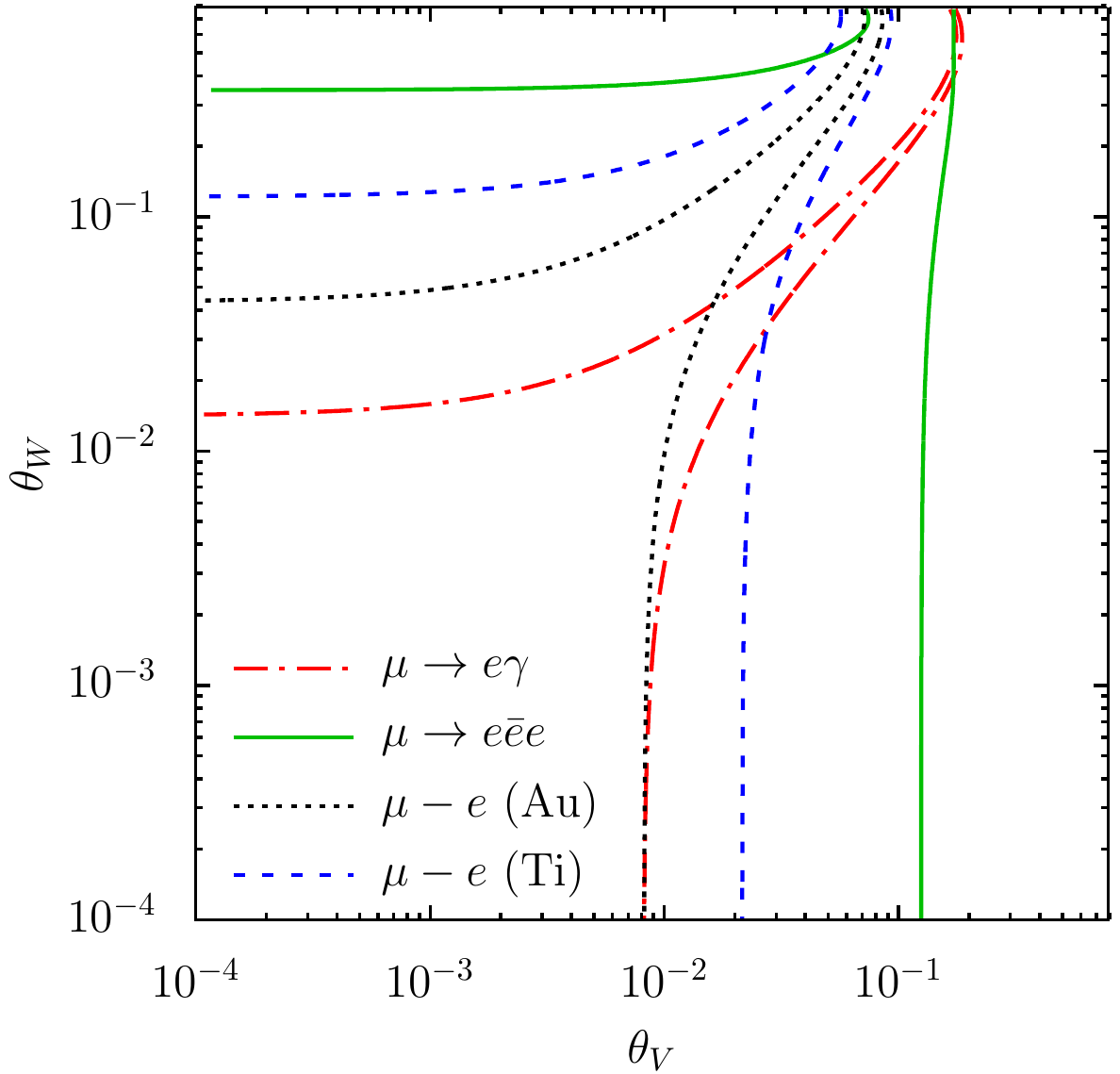} &
		\includegraphics[scale=0.59]{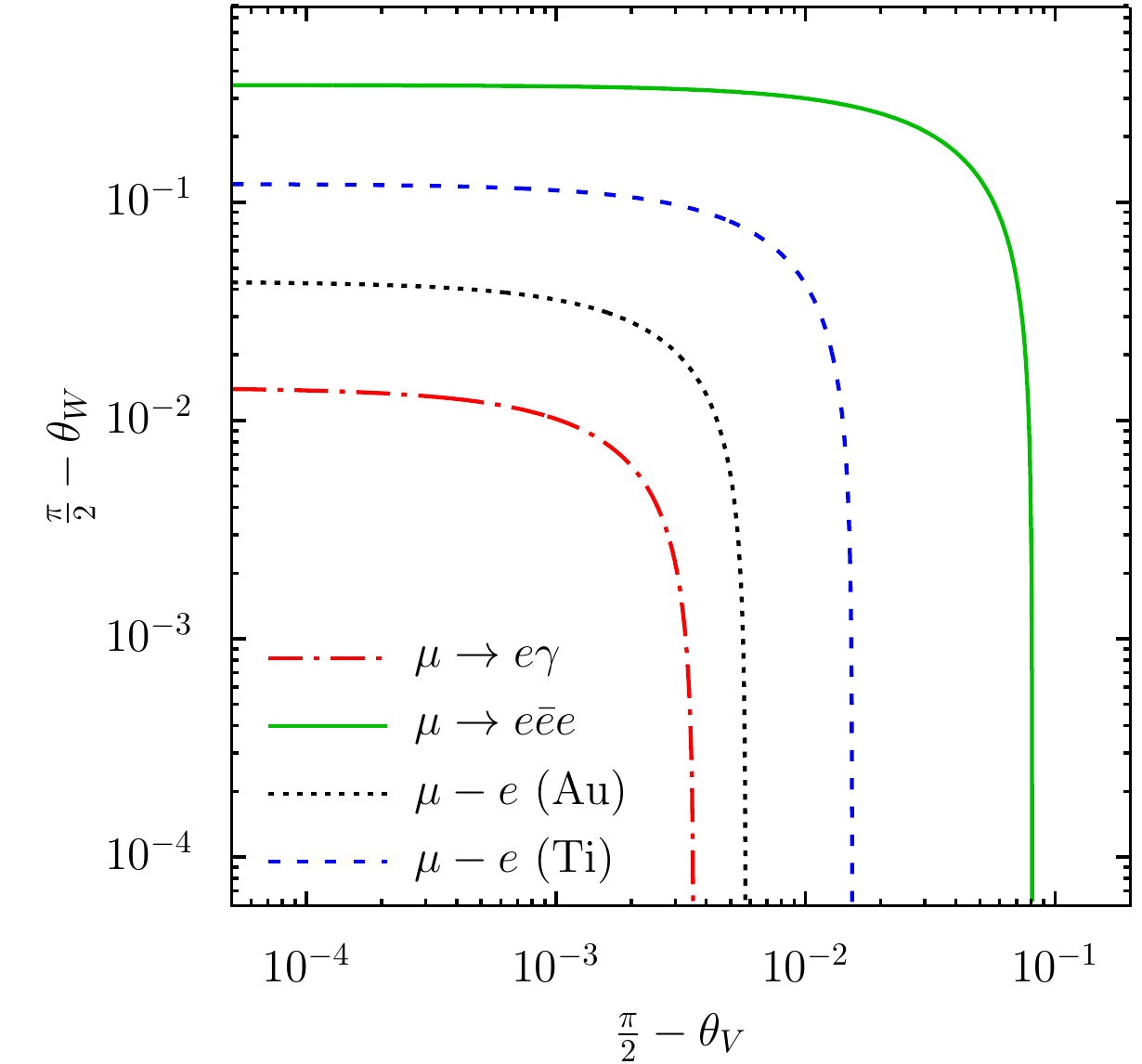} 
	\end{tabular}
	\caption{Mixing in the $\theta_V - \theta_W$ plane required to saturate the current bounds on $\mu$ to $e$ transition branching 
		ratios.~Different panels illustrate the different correlation between the two angles depending on where zero or 
		$(\theta_V, \theta_W) = (\pi / 2, \pi / 2)$ is chosen for expanding the misalignment.~The other LHT parameters are fixed to 
		their default values in~Eq.\,(\ref{defpt}).~The remaining processes are below present experimental limits for any mixing angle value.}
	\label{Panel4mu}
\end{figure}

Finally, in Figure \ref{Panel4mu} we plot contours in the mixing angle plane which saturate the current experimental limits for LFV processes 
having sufficient sensitivity.~We show two cases corresponding to expanding the misalignment around zero (left) or 
$(\theta_V, \theta_W) = (\pi / 2, \pi / 2)$ (right).~The other model parameters are fixed to the default values in Eq.\,(\ref{defpt}).~As can be 
seen, $\mu \to e \gamma$ decay and $\mu - e$ conversion in Au provide the most stringent constraints.~We also see that current limits 
require the misalignment between the SM and the mirror and partner mirror leptons to be $\sim 1$ \% as can be inferred from the 
$\theta_{V, W}$ values when the curves flatten near the axes.~For $\mu \to e$ conversion in nuclei we see similar shaped contours for 
Au and Ti, with Ti being less restrictive, while $\mu \to e \bar{e} e$ allows the misalignment to be generically $\sim 10$ times larger.~For 
misalignment around zero (left), the mixing angles can be larger if $\theta_{V, W}$ are correlated to equally high precision.~For misalignment 
around $(\theta_V, \theta_W) = (\pi / 2, \pi / 2)$ (right) the mixing angles are always constrained to be small, in particular by $\mu \to e \gamma$.
\begin{figure}
	\begin{tabular}{cc}
		\includegraphics[scale=0.6]{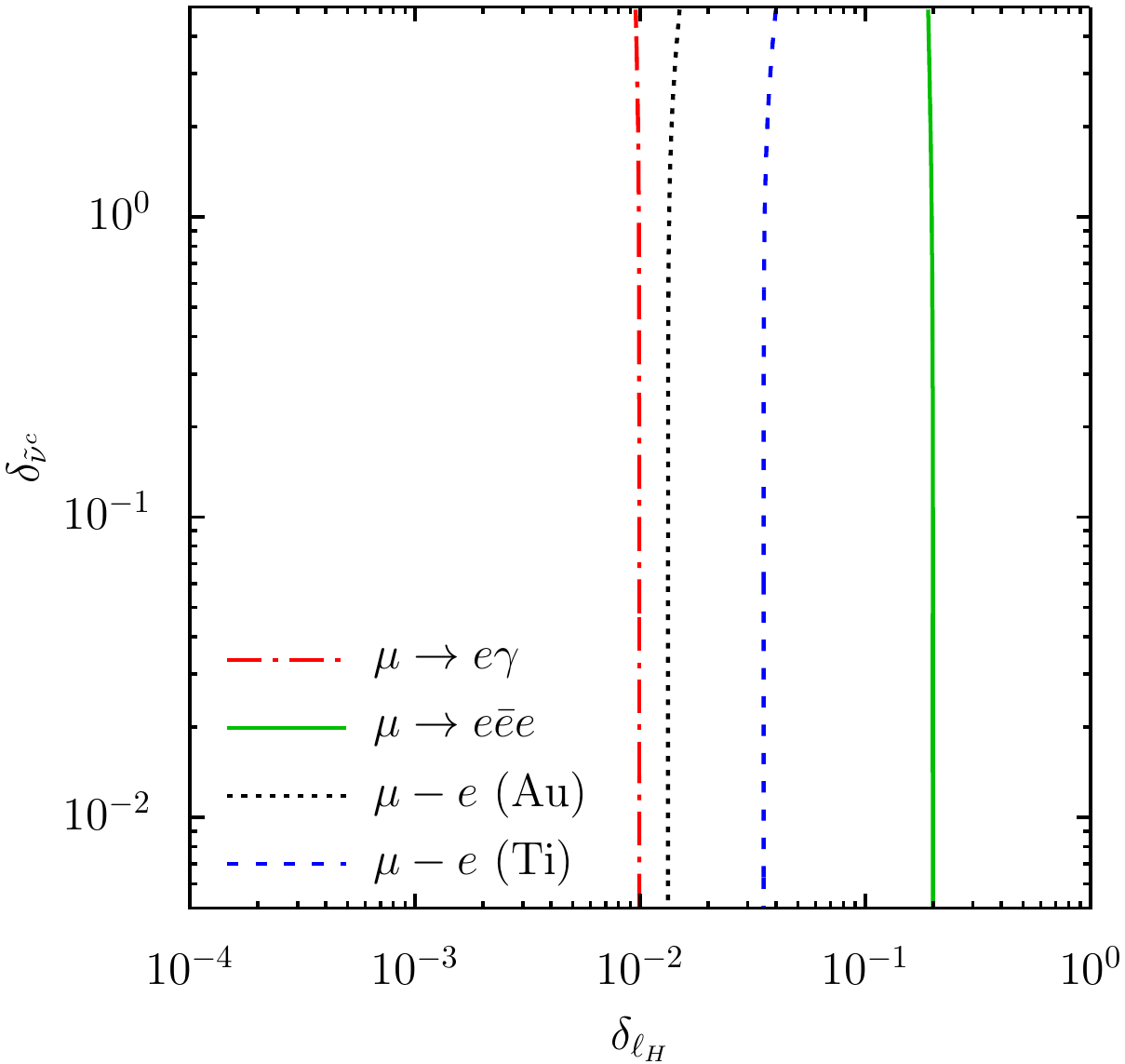} &
		\includegraphics[scale=0.6]{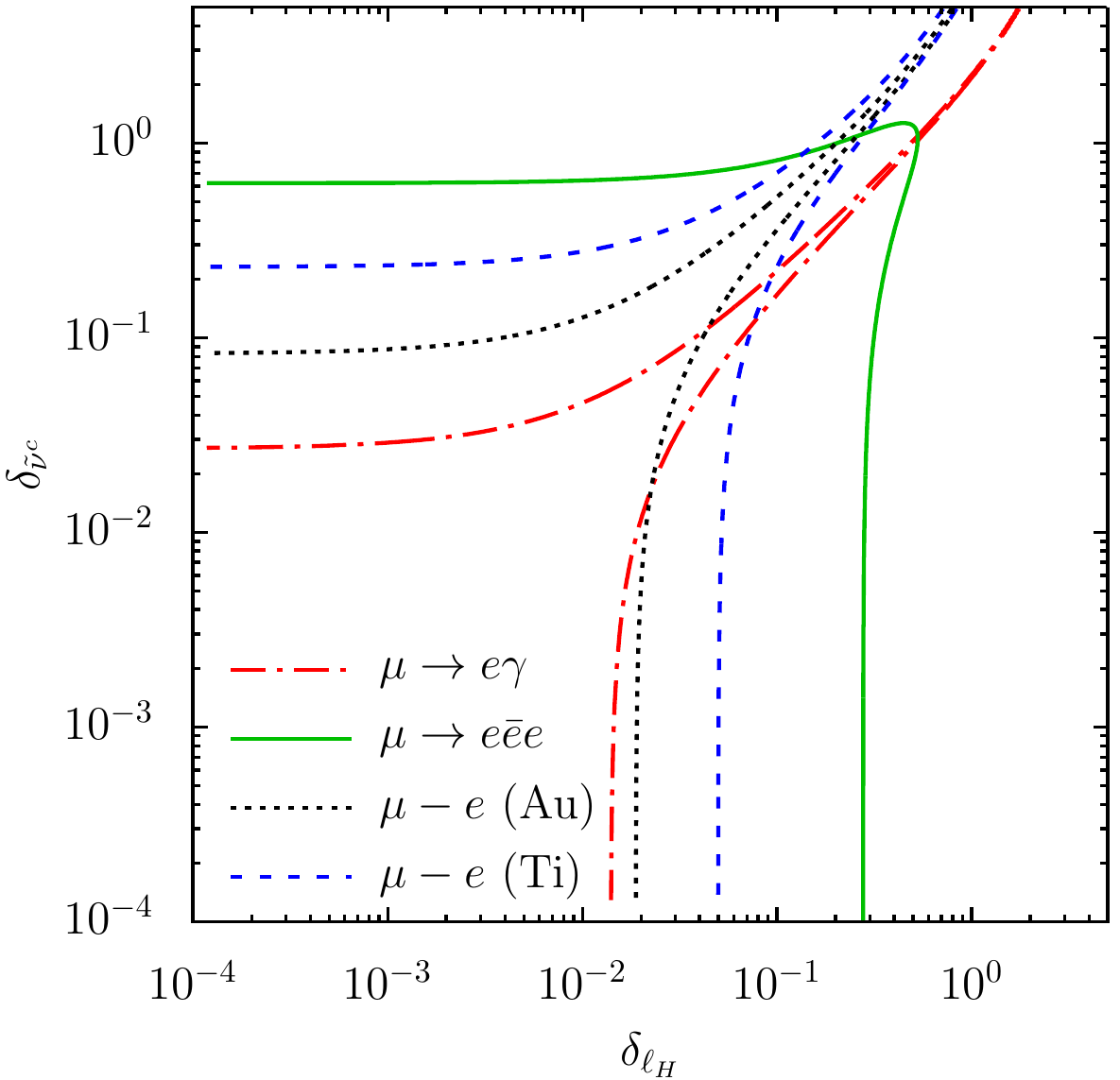} 
	\end{tabular}
	\caption{Contours in the $\delta_{\ell_H} - \delta_{\tilde\nu^c}$ plane which saturate the current bounds on $\mu$ to $e$ branching 
		ratios and conversion rates.~The left (right) panel corresponds to $\theta_{V, W} = \pi / 4\; ( \pi  / 8 )$ while the remaining LHT 
		parameters are fixed to the default values given in~Eq.\,(\ref{defpt}).~The remaining processes are below present experimental 
		limits for the range of mass (squared) splittings considered.}
	\label{Panel4tau}
\end{figure}

Similar comments apply when studying the dependence on the mirror and partner lepton masses and a similar precision is needed to fit 
current experimental data from LFV processes.~In Figure~\ref{Panel4tau} we show the corresponding contours in the 
$\delta_{\ell_H} - \delta_{\tilde\nu^c}$ (see~Eq.\,(\ref{defpt})) plane which saturate current experimental limits.~With the default values 
in Eq. (\ref{defpt}) chosen for the remaining parameters we see in the left hand panel that there is essentially no dependence on 
$\delta_{\tilde\nu^c}$ but $\delta_{\ell_H}$ must be tiny, $\sim 1$ \% to fulfill the bound from $\mu \to e \gamma$.~This asymmetric 
behavior is due to the choice of mixing angles $\theta_{V, W} = \pi / 4$ since for $\theta_{V} + \theta_{W} = \pi / 2$ the partner fermion 
contribution to the relevant LFV processes is much smaller than the mirror one.~This is made apparent in the right hand panel where we 
plot the corresponding contours for $\theta_{V, W} = \pi / 8$.~As we saw for the mixing angles, the $\mu \to e \gamma$ decay requires 
the mass (squared) splitting parameter to be tuned within 1 \% with analogous behavior in $\mu \to e$ conversion 
while again for $\mu \to e \bar{e} e$ limits are $\sim 10$ 
times weaker.

The tuning of the mass splittings and mixing angles required by $\mu$ to $e$ transitions can also be quantified using the measure 
defined in \cite{Barbieri:1987fn} in order to compare with other new physics scenarios.~Applying it to $\mu \rightarrow e \gamma$ in 
Figure \ref{Panel3mu},  
\bea
\Delta = \left| \frac{\partial \ln {\rm Br} (\mu \rightarrow e \gamma)}{\partial \ln \delta_{\ell_H}} \right| \, ,
\eea
we obtain $\Delta \sim 70$ which corresponds to a fine tuning of $\Delta^{-1} \sim 1.4$ \%  
for the two points around $\delta_{\ell_H} \sim 0.35$ where the branching ratio saturates the current experimental limit.~Though this can 
be accommodated in the LHT by fixing the parameters appropriately and there are large regions of parameter space in agreement with data, 
it would be more interesting to complete the LHT with a flavor symmetry which could align the first two heavy lepton families with the $e$ 
and $\mu$ families of the SM.~Explorations of this are left to future work.  

\begin{figure}
		\begin{tabular}{cc}
			\includegraphics[scale=0.38]{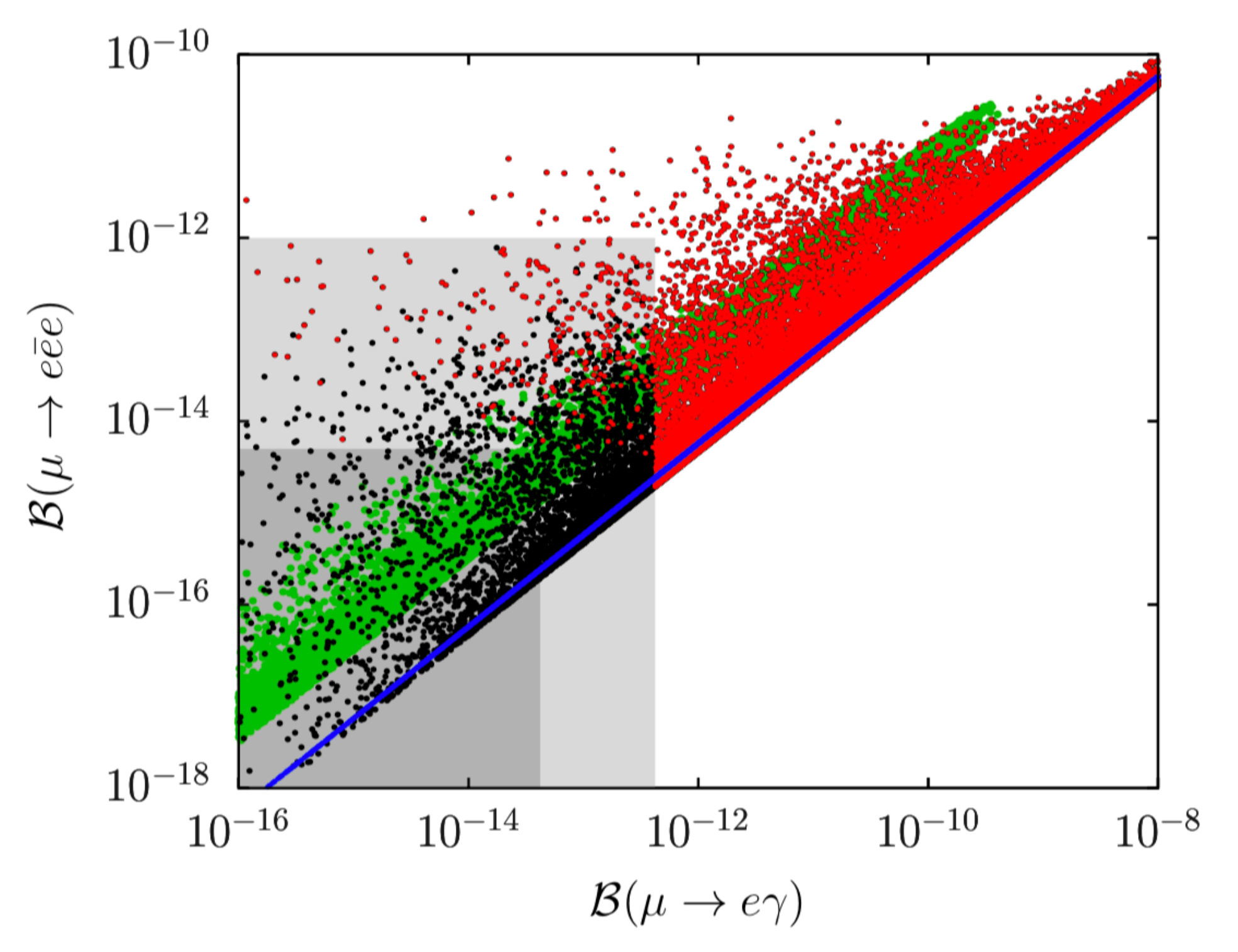} &
			\includegraphics[scale=0.38]{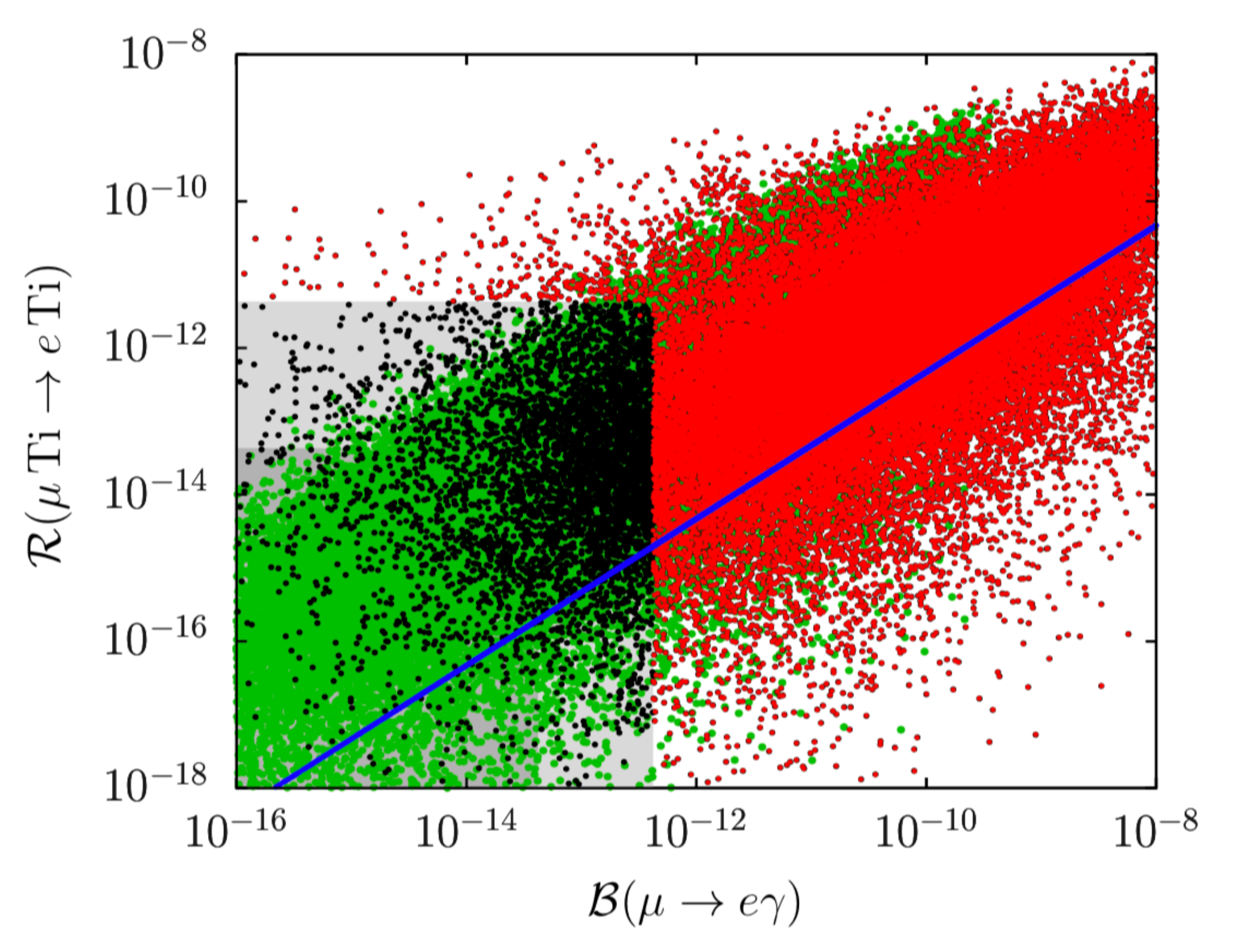} 
		\end{tabular}
		\caption{Scatter plots for the LHT contributions to ${\rm Br}(\mu \rightarrow e \bar{e} e)$ versus ${\rm Br}(\mu \rightarrow e \gamma)$ 
			(left) and ${\cal R}(\mu {\rm Ti} \to e {\rm Ti})$ versus ${\rm Br}(\mu \rightarrow e \gamma)$ (right). Green points result 
			from the mirror lepton contributions alone while red and black points result from the mirror and partner mirror contributions 
			together for the parameter values given in the text. But only black points correspond to LHT parameters satisfying the 
			experimental constraints (shaded regions) on the three LFV observables. See the text for further details.}
		\label{Scatter}
	\end{figure}

	As already emphasized and showed in previous figures, the inclusion of the partner mirror lepton contributions enlarge 
	the parameter space allowing for further cancellations but also for the mitigation of possible correlations between LFV observables implied 
	by the mirror lepton contributions previously considered. For illustration, we show in Figure \ref{Scatter} (left) how the scatter plot 
	of the LHT predictions for ${\rm Br}(\mu \rightarrow e \bar{e} e)$ versus ${\rm Br}(\mu \rightarrow e \gamma)$ 
	looks like when only the mirror leptons are 
	considered (green points, in agreement, for example, with \cite{Blanke:2009am,Blanke:2007db}) and when their partners are 
	also included (red and black points). 
	In the right panel we show the corresponding behavior of 
	${\cal R}(\mu {\rm Ti} \to e {\rm Ti})$ versus ${\rm Br}(\mu \rightarrow e \gamma)$. 
	Black points in both panels correspond to LHT parameter values satisfying the experimental constraints 
	on those three LFV observables. 
	As is apparent in the ${\rm Br}(\mu \rightarrow e \bar{e} e)$ versus ${\rm Br}(\mu \rightarrow e \gamma)$ panel, the existing correlation when only the mirror leptons are taken into account relaxes and the scatter region expands to fill in the experimentally allowed (shaded) area when the partner mirror leptons are also included. 
	Few comments are in order in this case: 
	$i$) The solid (blue) line corresponds to the mirror lepton dipole contribution alone. 
	This contribution is smaller than the mirror lepton $Z$ penguin and box contributions what explains the distance between 
	this solid line and the green region. 
	$ii$) In contrast, the red and black scatter region contains the solid (blue) line. This is because 
	when all the T--odd (non-singlet) lepton contributions are included the dipole term is typically larger, as the $Z$ penguin and 
	box terms, and they also in general interfere. 
	A consequence of this wider range of predictions is the difficulty of distinguishing between the LHT and 
	the corresponding supersymmetric predictions, in contrast with what was previously 
	argued in the absence of partner mirror leptons. 
	For comparison with Refs. \cite{Blanke:2009am,Blanke:2007db} 
	we have varied 
	\bea
	300\ {\rm GeV} \leq m_{\ell_{H1,2}}, m_{\tilde{\nu}^c_{1,2}},& m_{d_{H1}}, m_{\tilde{u}_1} 
    \leq & 1.5\ {\rm TeV}\ , \nonumber \\ 
    \theta_{V, W} \in& [0, \pi / 2) \; , \quad\quad & 
	\eea
	assuming no heavy quark mixing. 
	The scatter points are accumulated in the most probable observable region but 
	an unweighted scatter plot may be misleading because the less probable and then unpopulated region may not be 
	parametrically forbidden. 	
	As a matter of fact, in order to better illustrate what is going on we have weighted the scanned angle intervals 
	logarithmically. This populates the region for small observable values, which is the physically relevant one, and explains 
	why the green region penetrates the very low observable area in contrast with previous scatter plots \cite{Blanke:2009am,Blanke:2007db}). The darker shadow area defines the zone expected to be experimentally allowed in the 
	near future. Besides, we have fixed $f = 1.5$ TeV, contrary to the previous scatter plots, 
	where $f$ was assumed 
	to be equal to 1 TeV. Obviously, the LHT predictions can be made as small as needed not only increasing $f$ but 
	taking the mixing angles or the mass differences small enough (see Figs. \ref{Panel1mu} 
	and \ref{Panel4tau}). 
	This should be also interpreted as fine tuning, in agreement with our conclusion that the misalignment 
	between the T--odd and the SM leptons must be 
	$\sim 1$ \% when the other LHT parameters are fixed to their default values, as shown in Fig. \ref{Panel4mu}. 
    Finally, it is also worth noting that the bottom right region which is unpopulated in the left panel, corresponding to 
    ${\rm Br}(\mu \rightarrow e \gamma) / {\rm Br}(\mu \rightarrow e \bar{e} e) \geq 218$,   
    is inaccessible 
    for any value of the form factors involved in their decay widths in Eqs. (\ref{decaywidthmue}) and (\ref{threeequal}), respectively. The observation of these two processes in this region 
    could not be explained by the LHT contributions considered here nor by any model 
    which could be described by the form factors in Eqs. (\ref{vertexmue}) 
    and (\ref{Amplitude}--\ref{Boxamplitude}), respectively, at low energy.

    For completeness, we have also introduced and varied the CP violating phase 
	present in the general two family case analysed. While $V$ in Eq. (\ref{mixingmatrix}) can 
	be made real by a proper fermion field redefinition, $W$ is in general complex, 
	\begin{equation}
	W=\begin{bmatrix}
	\cos\theta_W &  \sin\theta_W e^{i\eta} \\
	-\sin\theta_W e^{-i\eta} & \cos\theta_W   
	\end{bmatrix} \, .
	\end{equation}
	No significant modification of Fig. \ref{Scatter} is found varying $\eta$, 
	although a given subset of predictions (and cancellations) can correspond to 
	different points in parameter space. 
	In any case we have not tried to characterize the full parameter space allowed by experiment, what is beyond the scope of the paper. 

%------- Future prospects --------------------------------%
\section{Future prospects}
\label{Futureprospects}

Future improvements of limits from LFV processes \cite{Baldini:2018uhj} 
will shed further light on the LHT model and to what degree it is natural. 
An improvement of sensitivity in measurements of $\mu \rightarrow e \gamma$ by an order of magnitude, as envisioned 
by the MEG Collaboration \cite{Baldini:2018nnn,Nakao:2018hip}, would increase the required tuning of the heavy lepton 
alignment by a factor of 1/3. 
Meanwhile an improvement of the sensitivity in $\mu \to e \bar{e} e$ by more than two orders of magnitude, as expected 
in Phase I of the Mu3e experiment \cite{Perrevoort:2018ttp}, would match the tuning currently required by 
$\mu \rightarrow e \gamma$.~Further improving it by almost two orders of magnitude in Phase II \cite{Blondel:2013ia} 
would result in a required alignment of the lepton sector at the per mille level or in a new physics scale $f  >  30$ TeV in 
the absence of any alignment or accidental cancellation.~Experiments on $\mu \to e$ conversion in nuclei are also expected 
to improve their sensitivity~\cite{Abusalma:2018xem,Angelique:2018svf,Teshima:2018msm}.~The Mu2e and COMET experiments 
aim to improve by two to more than four orders of magnitude in two phases.~We summarize these prospects for future 
LFV experiments using intense muon beams in Table \ref{Future}. 
%
%%%%%%%%LFV constraints Table%%%%%%%%%%%%%%%%%
\begin{table}
	\hspace*{-.1cm}
	\begin{center}
		\begin{tabular}{cc||cc|cccc}
			%%%%%%%%%%%%%%%%%%%%%%%
			Process & Experiment & $\begin{array}{c} {\rm Current} \\[-0.1cm]  {\rm precision} \end{array}$ 
			& $\begin{array}{c} {\rm Sensitivity} \\[-0.1cm]  {\rm improvement} \end{array}$ & 
			\multicolumn{2}{c}{$f$ [TeV] $>$}  
			& \multicolumn{2}{c}{
				$\begin{array}{c} {\rm Mixing\; angle} \\[-0.1cm] < \times 10^{-2} \end{array}$} \\ \hline
            &&&&&&& \\[-0.45cm]
			%%%%%%%%%%%%%%%%%%%%%%%
			$\mu \to e\ \gamma$ & [MEG] &
			$10^{-4}$ &  $10-500$ 
			&  $15$  & $27-71$ & \, $1$  & $0.3 - 0.04 $ \\ 
			%%%%%%%%%%%%%%%%%%%%%%%
			$\mu \to e\ \overline{e}\ e$ & [Mu3e] &
			$0.04$ &  $200-10^{4}$ 
			&  $3.4$  & $13-34$ & \, $20$  & $1-0.3 $ \\ 
			%%%%%%%%%%%%%%%%%%%%%%%
			$\mu \to e\ ({\rm Al})$ & [Mu2e] &
			$10^{-3}$ &  $10^{4}-10^{5}$ 
			&  $8.4$  & $84-150$ &\, $3$  & $0.03 - 0.01 $ \\   
			%%%%%%%%%%%%%%%%%%%%%%%
			$\mu \to e\ ({\rm Al})$ & [COMET] &
			$10^{-3}$ &  $10^{2}-10^{4}$ 
			&  $8.4$  & $27-84$ & \, $3$  & $0.3 - 0.03 $ \\  
			%%%%%%%%%%%%%%%%%%%%%%%}
		\end{tabular}
		\caption{Current precision and projected sensitivity improvement in Phase I--II for 
			different LFV experiments \cite{Baldini:2018uhj}. 
			In the absence of a signal the corresponding limits on the new physics scale $f$ 
			(for order 1 mixing) and on the mixing angle 
			of the heavy leptons with the first two families (for $f = 1.5$ TeV) are shown in the last two columns.}
		\label{Future}
	\end{center}
\end{table}
%%%%%%%%%%%%%%%%%%%%%%%%%s
%
The current precision is estimated by dividing the predictions in Table \ref{Limits} by the limits in Table \ref{Values},\footnote{
	The current precision estimate for $\mu \to e$ conversion in Al is 5 times larger than in Au because this is the ratio of their 
	rates in the LHT for the default values of the model parameters assumed in Table \ref{Values}, 
	${\cal R}(\mu {\rm Au} \to e {\rm Au}) / {\cal R}(\mu {\rm Al} \to e {\rm Al})  = 
	3.8 \times 10^{-9} / 8.4 \times 10^{-10}$.}
while limits on the new physics scale $f$ and alignment of the T--odd lepton with the first two SM families $\theta$ are obtained 
using the corresponding 
scaling dependence of the LFV processes for small mixing, $\theta^2 / f^4$.
\footnote{We use a loose meaning of mixing to 
	illustrate the stringent bounds on the first two lepton families in this summary 
	because although there are several mixing angles involved (see {\it e.g.} Eq. (\ref{mixingmatrix})), the corresponding estimate 
	roughly applies to all of them in large regions of parameter space. For instance, in Figure \ref{Panel4mu} the limit estimates 
	for $\theta_{V,W}$ can be directly read from the curves when they flatten.}

In the case of $\tau$ decays experimental bounds are less restrictive but constrain other mixings.~Belle II~\cite{Kou:2018nap} 
aims to improve the precision to $10^{-9}-10^{-10}$ in a variety of LFV processes~\cite{Liventsev:2018gin} while LHCb also 
expects to reach a similar sensitivity~\cite{Bediaga:2018lhg}.~The upper bound on $\tau \to \mu \overline{\mu} \mu$ from LHCb 
\cite{Aaij:2014azz} is already within a factor of 2 of its current best limit \cite{Tanabashi:2018oca} (see also \cite{DeBruyn:2017aqu} 
for the LHC collaborations). 
For the default values in Eq. (\ref{defpt}) we find that $\tau \rightarrow \mu \gamma$ could constrain the corresponding mixing 
angles in the LHT (see Table \ref{Values} for an estimate) 
but $\tau \to \mu \bar{\mu} \mu$ will likely not reach the necessary precision to test the LHT prediction of 
$\Gamma (\tau \to \mu \bar{\mu} \mu) / \Gamma (\tau \rightarrow \mu \gamma) \sim 2 \times 10^{-3}$ which 
we note is similar to predictions in supersymmetric models 
\cite{Ellis:2002fe,Babu:2002et,Brignole:2004ah} (see also \cite{Arganda:2005ji} and \cite{Dassinger:2007ru}).\footnote{We must 
	emphasize that the predictions in Table \ref{Values} for $\tau$ decays can be increased 
	by several orders of magnitude for larger $\tilde{x}$ values. Moreover, in the region where 
	$\tau \to \mu \gamma, e \gamma$ are suppressed three-body $\tau$ decays can easily saturate current 
	experimental bounds. \label{largetau}} 
This is in contrast to previous studies~\cite{Blanke:2009am,Blanke:2007db} which found that three-body $\tau$ decays could 
distinguish clearly between the LHT and supersymmetric models.~However, these studies only included effects from the mirror 
leptons as they worked in the limit of decoupled partner leptons.~When the contributions from both the mirror and partner leptons 
are included one finds that there are regions of parameter space where the LHT and supersymmetric predictions are similar.~Thus 
these decays cannot unambiguously distinguish between these two models.

\begin{figure}
	\begin{tabular}{ccc}
		\includegraphics[scale=0.6]{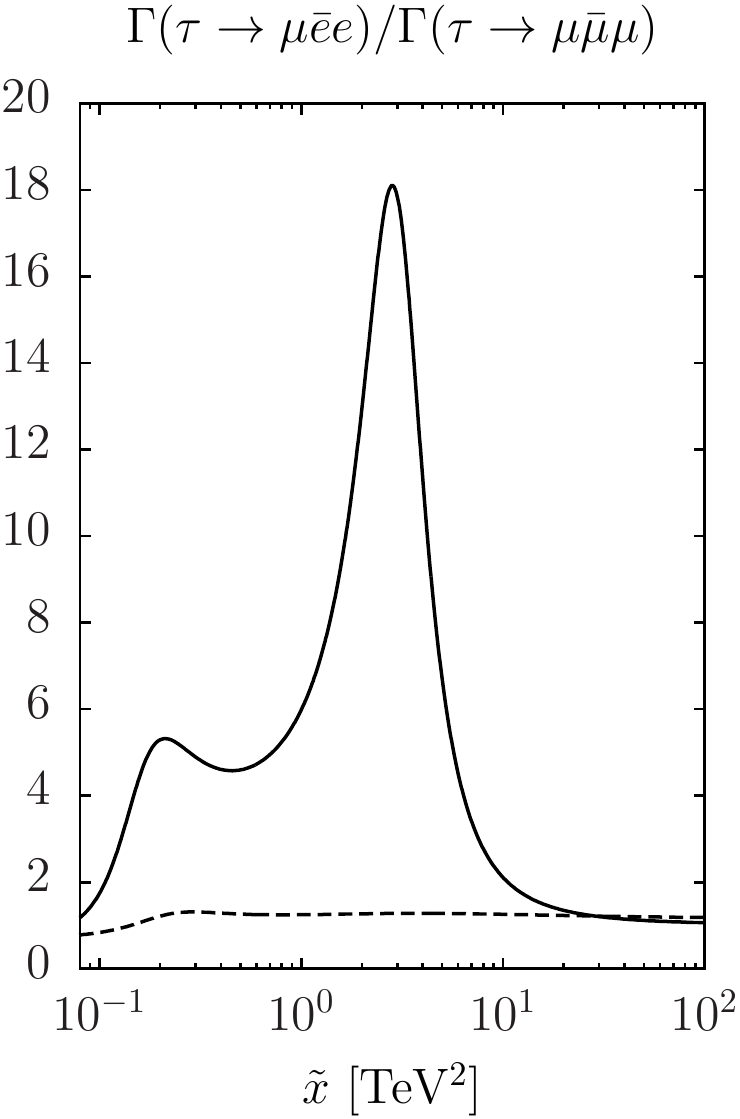} &
		\includegraphics[scale=0.6]{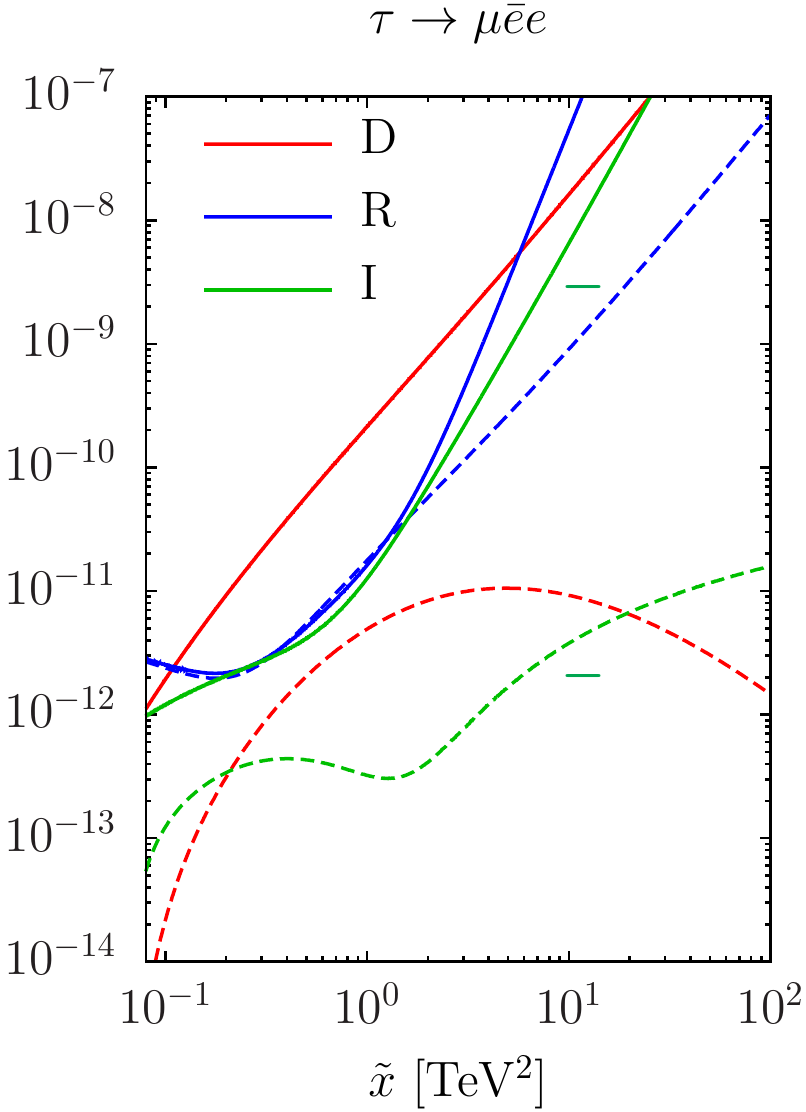} 
		\includegraphics[scale=0.6]{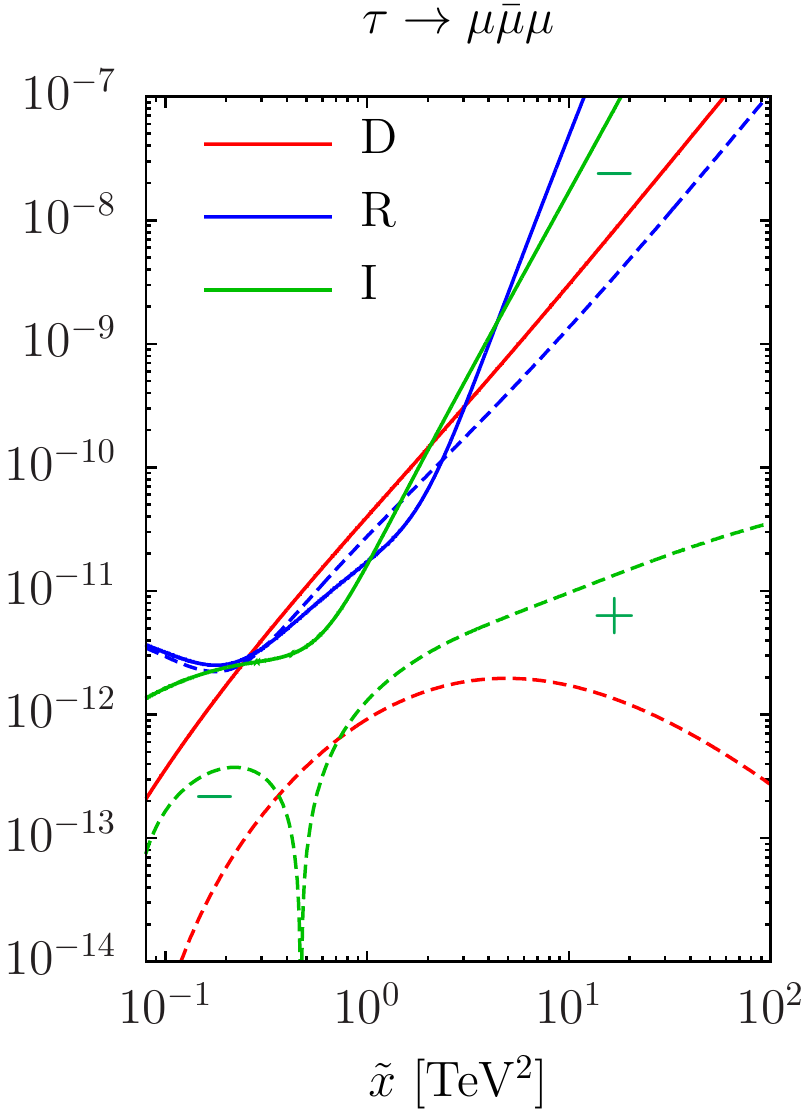}
	\end{tabular}
	\caption{Left: LHT prediction for the ratio ${\Gamma (\tau \to \mu \overline{e} e)}/{\Gamma (\tau \to \mu \overline{\mu} \mu)}$ 
		as a function of $\tilde{x} = m_{\ell_{H 2}} m_{\ell_{H 3}}$ for $\tau - \mu$ mixing, with the remaining parameters fixed to their 
		default values~Eq. (\ref{defpt}).~The solid line shows the prediction when both mirror and partner mirror leptons ($l_H$ and 
		$\tilde{l}$) are included while the dashed shows when only the mirror leptons ($l_H$) are taken into account.~Middle and 
		Right:~LHT model predictions for $ {\Gamma (\tau \to \mu \overline{e} e)} $ and $ {\Gamma (\tau \to \mu \overline{\mu} \mu)} $ 
		separating the photon dipole contribution (D), the other form factor contributions (R), and their interference (I).}
	\label{Behavior}
\end{figure}

To understand this further we recall that the supersymmetric prediction~\cite{Ellis:2002fe,Babu:2002et,Brignole:2004ah} relies on 
the observation that when the photon dipole contribution dominates the three-body $\tau$ decay, the ratio to the corresponding 
radiative two-body $\tau$ decay in Eq. (\ref{decaywidthmue}) is fixed by 
kinematics,\footnote{These relations, which follow from the corresponding LHT 
	expressions above when only $F^\gamma_M = - i F^\gamma_E$ is taken into account, differ slightly from those often given for 
	supersymmetric models  
	where the constant terms in parenthesis are $-11/4$ and $-8/3$, respectively.~These coefficients should be the same since they 
	are phase space factors, hence model independent, and must be carefully evaluated as emphasized in footnote \ref{integration}.} 
\bea
\label{logarithm}
\frac{\rm Br(\ell \to \ell' \overline{\ell'} \ell')}{\rm Br(\ell \to \ell' \gamma)} = 
\frac{\alpha}{3 \pi} \left(2 \ln \frac{m_\ell}{m_{\ell'}} - \frac{13}{4}\right)\,, \;\; 
\frac{\rm Br(\ell \to \ell' \overline{\ell''} \ell'')}{\rm Br(\ell \to \ell' \gamma)} = 
\frac{\alpha}{3 \pi} \left(2 \ln \frac{m_\ell}{m_{\ell''}} - \frac{7}{2}\right)\,. 
\eea
In the LHT the photon dipole term is the one proportional to $|A_R|^2$ in Eqs. (\ref{threeequal}) and (\ref{threedifferent}) so in 
regions of parameter space where this term dominates, the LHT prediction will also be fixed by kinematics and thus similar to the 
supersymmetric prediction.~In the left panel of Figure \ref{Behavior} we plot the ratio 
$\Gamma(\tau \to \mu \overline{e} e) / \Gamma(\tau \to \mu \overline{\mu} \mu)$ 
(see Eqs. (\ref{threedifferent}) and (\ref{threeequal}), respectively) as a function of the mass product parameter $\tilde{x}$ for 
the default values of the other LHT parameters fixed in~Eqs. (\ref{defpt}).~We show the prediction when all T--odd (non-singlet) 
leptons, $l_H$ and $\tilde{l}$, are included (solid line) and when only the mirror leptons, $l_H$, are taken into account (dashed line).
As can be seen, when only the mirror leptons are taken into account 
(dashed line) there is no logarithmic enhancement from the photon dipole term while when the partner leptons are also included 
(solid line) the logarithmic enhancement in Eq. (\ref{logarithm}) is significant.~How large it is of course depends on the particular 
choice of parameters.~To illustrate this we compare the predictions 
in Table \ref{Comparison} for $ {\rm Br(\tau \to \mu \overline{\mu} \mu)} / {\rm Br(\tau \to \mu \gamma)} $ and 
$ {\rm Br(\tau \to \mu \overline{e} e)} / {\rm Br(\tau \to \mu \gamma)} $ in supersymmetric models assuming the photon dipole 
dominates (see Eq. (\ref{logarithm})) and the LHT model for the default values of the parameters in~Eq. (\ref{defpt}) when all 
T--odd (non-singlet) leptons ($l_H$ and $\tilde{l}$) are included and when only the mirror leptons ($l_H$) are included with the 
partner leptons decoupled. 

\begin{table}
	\hspace*{-.1cm}
	\begin{center}
		\begin{tabular}{c||c|c|c}
			%%%%%%%%%%%%%%%%%%%%%%%
			Ratio & \quad\; SUSY \quad\; & \; LHT ($l_H + \tilde{l}$) \quad & 
			\quad LHT ($l_H$) \quad 
			\\ \hline
			&&&\\[-0.45cm]
			%%%%%%%%%%%%%%%%%%%%%%%
			\; $ \frac{\rm Br(\tau \to \mu \overline{\mu} \mu)}{\rm Br(\tau \to \mu \gamma)} $ \; 
			&  $1.9 \times 10^{-3}$ & $1.9 \times 10^{-3}$ 
			& \, $6.0 \times 10^{-2}$ \\ 
			%%%%%%%%%%%%%%%%%%%%%%%
			$ \frac{\rm Br(\tau \to \mu \overline{e} e)}{\rm Br(\tau \to \mu \gamma)} $ 
			& $9.9 \times 10^{-3}$ &  $1.1 \times 10^{-2}$ 
			& \, $7.5 \times 10^{-2}$ 
		\end{tabular}
		\caption{Predictions for ratios of $\tau$ decays in supersymmetric models when the photon dipole dominates and in the LHT model 
			assuming the default values for the parameters in~Eq. (\ref{defpt}) when including all T--odd (non-singlet) leptons ($l_H$ and 
			$\tilde{l}$) and when including only the mirror ones ($l_H$).}
		\label{Comparison}
	\end{center}
\end{table}
%%%%%%HERE%%%%
From Table \ref{Comparison} we see the prediction for the ratio of the two ratios is $\sim 5$ in both supersymmetric models and, for the parameter 
point with $\tilde{x} = 1\, {\rm TeV}^2$, in the LHT model when including $l_H$ and $\tilde{l}$ while the ratio of ratios is $\sim 1$ when only the $l_H$ 
are included.~Thus we see that for parameter points near $\tilde{x} = 1\, {\rm TeV}^2$, the LHT gives predictions similar to supersymmetric models 
when the mirror and partner leptons are included while they differ significantly when only mirror leptons are taken into account.~As can be deduced 
from Figure \ref{Behavior} (left panel), when both $l_H$ and $\tilde{l}$ are included, this ratio of ratios decreases rapidly starting from 
$\tilde{x} \sim {\rm TeV}^2$ and converges to $\sim 1$ as $\tilde{x}$ becomes large.~The behavior around $\tilde{x} = 1\, {\rm TeV}^2$ can be better 
understood by examining $\Gamma (\tau \to \mu \overline{e} e)$ and $\Gamma (\tau \to \mu \overline{\mu} \mu)$ and separating it into the photon 
dipole contribution (D) the contribution from other form factors (R) and the contribution from their interference (I) which we show in Figure \ref{Behavior} 
(middle and right panels, respectively). 
When only the mirror leptons are included (dashed lines) the photon dipole contribution is rather small and $\Gamma (\tau \to \mu \overline{e} e)$ and 
$\Gamma (\tau \to \mu \overline{\mu} \mu)$ are almost equal (the $\pm$ on the interference term (dashed line) indicates the overall sign).~When the 
partner leptons are also taken into account (solid lines) the photon dipole contribution dominates in the region around $\tilde{x} \sim 1\, {\rm TeV}^2$. 
The strong dependence on $\tilde{x}$ seen in Figure \ref{Behavior} in this region can be traced back to the behavior of the loop functions in 
Eq. (\ref{dipoleformfactormue}) and, in particular, the charged partner lepton loop function $F^{\tilde{\ell}}_M (x)$.~Thus we see how when all the 
heavy T--odd leptons are included, supersymmetric models and the LHT model cannot be unambiguously distinguished utilizing three-body $\tau$ decays.

\begin{figure}
	\begin{tabular}{cc}
		\includegraphics[scale=0.6]{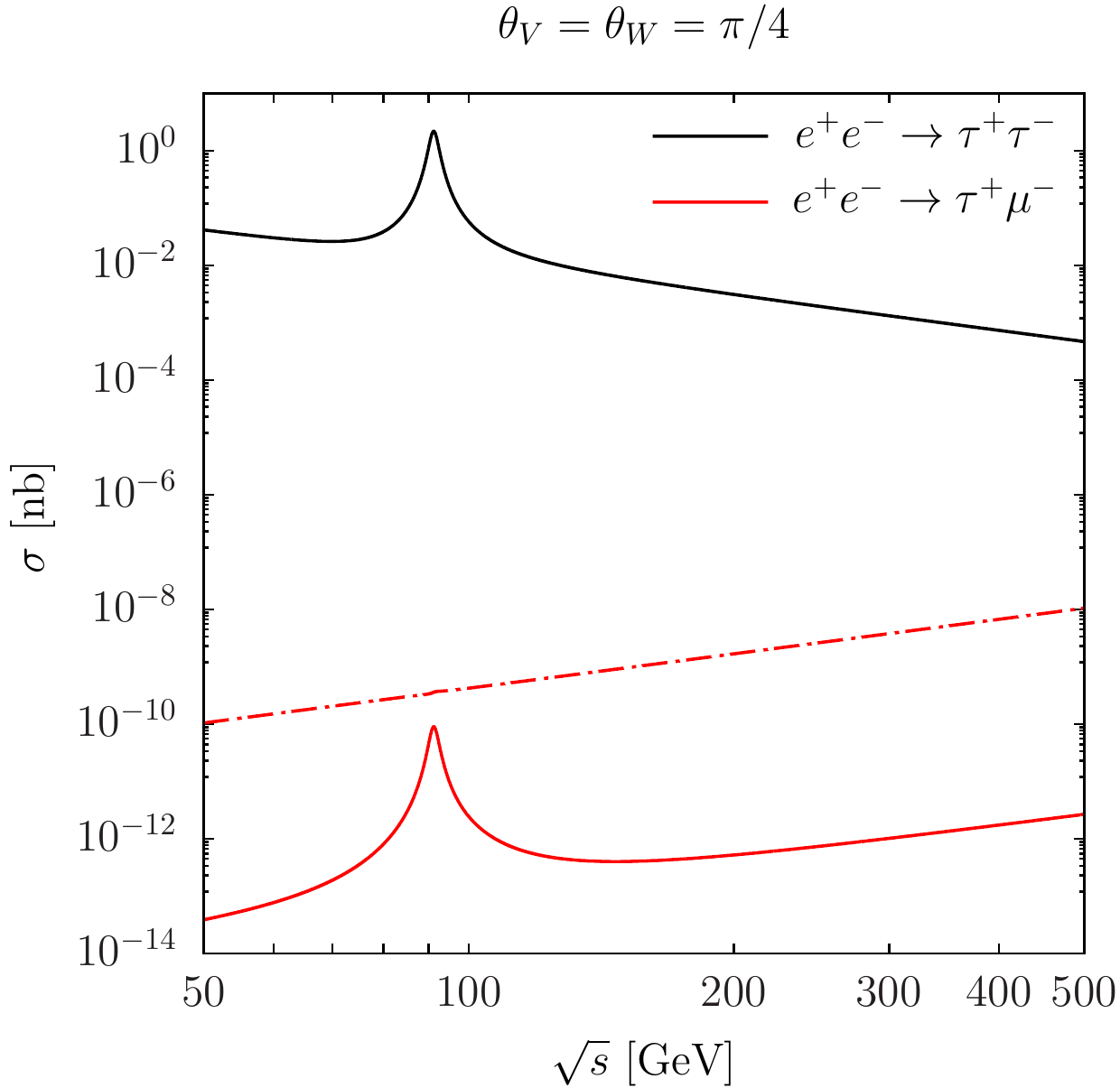} &
		\includegraphics[scale=0.6]{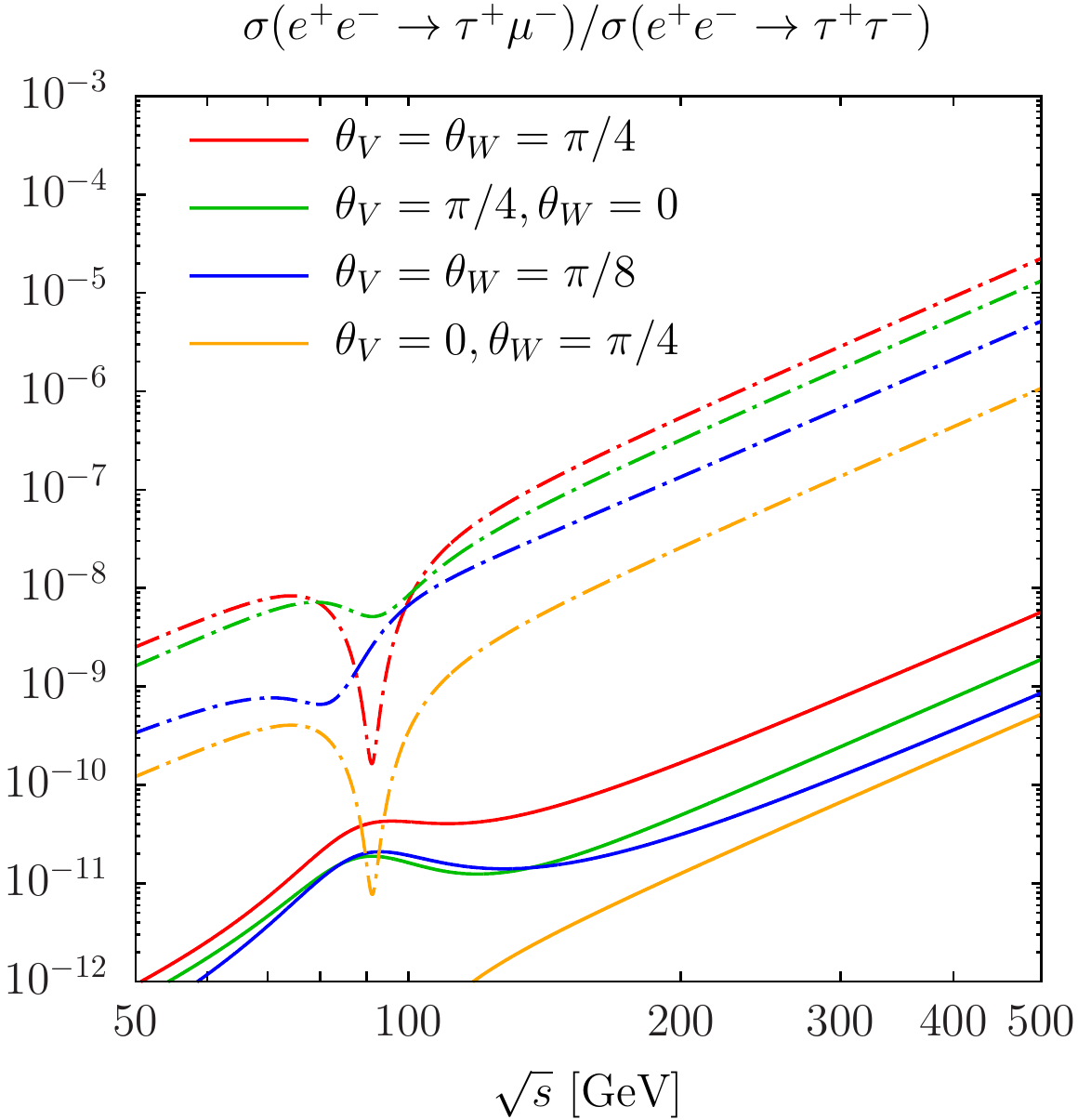} 
	\end{tabular}
	\caption{LHT predictions for $e \overline{e} \to \overline{\tau} \mu$ (left panel) 
		and its ratio to $e \overline{e} \to \overline{\tau} \tau$ (right panel) 
		as a function of the center of mass energy squared ($s = Q^2$) for different 
		mixing angles $\theta_{V,W}$ and for the default values in Eqs. (\ref{defpt}) for the other LHT parameters. 
		Solid (dashed-dotted) lines correspond to $\tilde{x} = 1$ TeV$^{2}$ (10 TeV$^{2}$).}
	\label{Colliders}
\end{figure}

Finally, it is also important to emphasize that future leptonic colliders 
will produce a large number of Drell--Yan lepton pairs and in particular $\overline{\tau} \tau$.\footnote{We thank A. Blondel for stressing to us the 
	potential of the FCC-ee \cite{Benedikt:2018qee}.\label{Alain}}  
A potentially promising probe of LFV in the LHT at $e \overline{e}$ colliders 
is searching for $\overline{\ell'} \ell''$ production (for a recent review of LFV $\tau$ decays at the FCC-ee see \cite{Dam:2018rfz}).~For illustration, in 
Fig. \ref{Colliders} we show the prediction for $e \overline{e} \to \overline{\tau} \mu$ (left panel) and its ratio to $e \overline{e} \to \overline{\tau} \tau$ 
(right panel) as a function of the center of mass energy squared ($s = Q^2$) for different mixing angles and for the default values of the remaining 
LHT parameters in Eqs. (\ref{defpt}). 
For our estimate we use the amplitude in Eq. (\ref{Amplitude}) which after crossing reads:\footnote{The dipole term is 
	suppressed because the total momentum 
	factor $Q_\nu = ( p_{\ell} + p_{\overline{\ell}} )_\nu = ( p_{\overline{\ell'}} + p_{\ell''} )_\nu$ results in light lepton masses 
	$m_{\ell', \ell''}$ ( $\ll \sqrt{s} $ ) when acting on the external legs.~There are no cross-terms either because the final 
	leptons are assumed to be different from the initial ones.~The flavor dependence is taken into account by the mixing 
	coefficients within the form factors.}
\begin{align}
	{\cal M}^{\ell \overline{\ell} \to \overline{\ell'} \ell''}  & = \,\overline{v}(p_{\overline{\ell}})
	\,\gamma_\mu\, e\,u(p_\ell)\, 
	\frac{1}{Q^2}\,\overline{u}(p_{\ell''})\,e\, F_L^\gamma (Q^2)\,\gamma^\mu P_L\,v(p_{\overline{\ell'}}) \nonumber \\
	+ & \,\overline{v}(p_{\overline{\ell}})\, \gamma_\mu\,( g_L^Z P_L + g_R^Z P_R )\,u(p_\ell)\, 
	\frac{1}{Q^2 - M_Z^2 +i M_Z \Gamma_Z}\, 
	\overline{u}(p_{\ell''})\,e\, F_L^Z (Q^2)\,\gamma^\mu P_L\,v(p_{\overline{\ell'}}) \nonumber \\
	+ & \, e^2 B_L (0)\,\overline{v}(p_{\overline{\ell}})\, 
	\gamma^\mu P_L\, u(p_\ell)\, \overline{u}(p_{\ell''})\, \gamma_\mu P_L\,v(p_{\overline{\ell'}}) \ , 
\end{align}
resulting in a cross section 
\begin{align}
\label{cross-section}
	\sigma (e \overline{e} \to \overline{\tau} \mu) &= \frac{\pi \alpha^2}{3 s} 
	\left\{ 2\, |F_L^\gamma(s)|^2 + \frac{(g_L^Z)^2+(g_R^Z)^2}{e^2} |F_L^Z(s)|^2 |\chi_Z(s)|^2  
	+ {s^2} |B_L|^2  \right.  \\ 
	+  & \left. \left[ F_L^\gamma(s) \left( \frac{g_L^Z + g_R^Z}{e} F_L^{Z *}(s) \chi^*_Z(s) 
	+  s B_L^* \right) + \frac{g_L^Z}{e} F_L^Z(s) \chi_Z(s) s B_L^* +\, {\rm h.c.} \right] \right\}\, , \nonumber
\end{align}
with $\chi_Z(s) = s (s-M_Z^2+iM_Z\Gamma_Z)^{-1}$. 

As is apparent from Fig. \ref{Colliders} (left panel) where we plot the total cross-section for 
$e \overline{e} \to \overline{\tau} \mu$ and $e \overline{e} \to \overline{\tau} \tau$, the LFV cross-section grows with 
energy.~When the box diagram contributions are large, no $Z$ peak emerges (dashed-dotted line corresponding to 
$\tilde{x} = 10\, {\rm TeV}^2$) in the invariant mass spectrum which opens up the possibility that a large enough LFV 
production cross-section can be observable without being excluded by $Z$-pole measurements.~On the right hand 
panel we show the variation of this cross-section with the mixing angles which we see can be almost two orders of 
magnitude 
for ${\cal O} (1)$ mixing. 
    The growth with $s$ of the cross-section 
	$e \overline{e} \to \overline{\tau} \mu$ below the production threshold of new (T--odd) particles reflects the fact that, in the presence of a mass gap, any extension of the SM 
	can be described at low energy by effective operators of dimension 6 
	at leading order. Then, as they are suppressed by the large new physics scale 
	squared ($f^{-2}$), their contributions must grow with the low energy scale $s$ (see Eq. (\ref{cross-section})). 
	This is a sensible approximation in our case (with the T--odd particles near the TeV) for 
	$\sqrt{s} \leq 500$ GeV. 
	Above the resonance region, for much larger $s$ values, the LHT penguin and box diagram 
	contributions scale like $s^{-1}$ and hence, the cross-section also decreases as $s^{-1}$,  satisfying the unitarity bound. 
To summarize, although for the default values of the parameters it may be difficult to observe 
LFV events, in other regions of parameter space it will be possible to constrain the $\tau$ sector in the LHT at a future 
FCC-ee \cite{Benedikt:2018qee} or ILC \cite{Baer:2013cma} machine assuming $\mathcal{O}(10^{11})$ 
$\overline{\tau} \tau$ pairs are collected. 
As noted in footnote \ref{largetau} these regions can be compatible with current constraints on LFV $\tau$ decays. 
Thus, considering the huge statistics available at future lepton colliders, it will be worthwhile to study the detectability 
of such signals in detail. 

%------- CONCLUSIONS --------------------------------%
\section{Conclusions}
\label{Conclusions}

The LHT is an elegant and phenomenologically viable example of a composite Higgs model with improved radiative behavior 
due to its T--parity which helps mitigate the flavor hierarchy problem significantly.~This discrete symmetry also requires the 
heavy T--odd spectrum to be pair-produced allowing their masses to be $\mathcal{O} (\rm{TeV})$ without violating present 
direct and indirect constraints.~In this study we have revised the constraints on the T--odd spectrum of the LHT implied by 
present bounds on LFV processes.~We have completed previous phenomenological analyses in two ways:~First, we have 
included for the first time all contributions from the full electroweak charged T--odd sector and in particular, the T--odd 
partner leptons previously neglected in phenomenological studies of low momentum transfer LFV processes.~As we have 
discussed, lepton electroweak singlets can be safely omitted in this phenomenological analysis.~Second, we have computed 
on-shell LFV $Z$ decays which, along with calculations of on-shell LFV Higgs decays~\cite{delAguila:2017ugt}, completes 
the set of observables at both small and large momentum transfers needed to analyze LFV in the LHT.~We have compared 
these predictions with experiment for all available LFV processes in Table 
\ref{Limits}.~The $\mu - e$ and $\tau - \mu$ and $\tau - e$ sectors are studied separately assuming a two-family description 
for each.~The general case can be obtained combining the parameters of the three analyses, 
noting that the restrictions on the $\tau - e$ sector are virtually identical to those on $\tau - \mu$ (see Table \ref{Limits}).

For the $\tau-\mu$ LFV observables we find that low energy observables, namely 
$\tau \to \mu \gamma$ and $\tau \to \mu\overline{\mu} \mu$, are much more restrictive than on-shell LFV decays of the $Z$ 
or Higgs bosons 
which are typically orders of magnitude below the current limits. 
We also find that even when allowing low energy observables to saturate their current experimental limits, on-shell LFV $Z$ 
and Higgs decays can be at most $\mathcal{O}(10^{-7})$.~The pattern of LFV is quite similar for $\tau-e$ transitions with 
small differences due to the slightly different experimental bounds.

On the other hand, $\mu-e$ transitions present significant
differences.~First, Higgs decays are proportional to the masses of the
SM leptons in the final state and are therefore strongly suppressed
for decays not involving the $\tau$ lepton.~In general constraints on $\mu-e$ transitions from low energy LFV observables 
are very strong and render on-shell $Z$ or Higgs decays to $\mu e$ too small to be observable even in future experiments. 
Furthermore, there is very little correlation between the different low-energy LFV observables so the only way of generically 
satisfying $\mu-e$ constraints is by reducing the global suppression factor by means of a relatively large $f$ and/or small 
mixing angles since even with the new source of flavor violation, cancellations are far from generic.~This implies large global 
symmetry breaking scales $f > 10$ TeV or small mixing angles $\theta_{V, W} < 10^{-2}$. 

Motivated by considerations of naturalness, we have focused our phenomenological studies on the case of small 
mixing angles and/or accidental cancellations with $f = 1.5$~TeV. We have explicitly shown that although all 
current bounds can be satisfied with small mixing angles, some of them can be the result of accidental 
cancellations between the new contributions to these processes. 
Furthermore, we have quantified the amount of fine tuning necessary and found it to be $\sim 1$ \% for particular parameter points. 
While the LHT can accommodate limits from searches of FCNC, a flavor completion of the model with a natural 
suppression of the leptonic FCNC would be welcome.

Future LFV experiments using intense muon beams will typically improve the current sensitivity by two to four 
orders of magnitude, constraining the effective $\mu - e$ mixing to 1 per mille for $f  \gtrsim $ TeV in the 
absence of a signal. In this case, future LFV experiments will strongly challenge the LHT scenario probing 
scales $f  \sim 10$ TeV even for mixings of order 1~\%, thus pushing the concept of naturalness in the LHT to its 
limits. We have also shown, in contrast to previous studies, that in certain regions of parameter space the LHT 
model can give similar predictions to supersymmetric models for LFV processes. 
In other LHT parameter regions $\tau$ LFV transitions can saturate current experimental bounds 
and hence, can be observed at Belle II as well as the LHC and/or future 
lepton colliders. 
As the current sensitivity on $\tau$ LFV processes does not favor any definite pattern of 
$\tau$ mixing, a systematic study of $\tau \mu$ and $\tau e$ production is worthwhile.

%%%%%%%%%%%%%%
%%%%%%%%%%%%%%

\section*{Acknowledgments}
We thank useful discussions and comments by A. Blondel, A. Falkowski, G. Hern\'andez-Tom\'e, J. Hubisz, I. Low and J.M. P\'erez-Poyatos. 
This work has been supported in part by the Ministry of Science, Innovation and Universities, under grant numbers FPA2016-78220-C3-1,2,3-P 
(fondos FEDER), and by the Junta de Andaluc{\'\i}a grant FQM 101 as well as by the Juan de la Cierva program.~J.S.~and R.V.M.~thank the Mainz 
Institute for Theoretical Physics (MITP) for its hospitality and partial support during the completion of this work.~R.V.M.~would also like to thank 
the Fermilab National Accelerator Laboratory and Northwestern University for their hospitality.

\appendix

%%%%%%%%%%%%%%
%%%%%%%%%%%%%%

\section{Flavor conserving observables}

For $\ell' = \ell$ the vertex in Eq.\,(\ref{vertexmue}) allows to define the 
anomalous magnetic and the electric dipole moments of the $\ell$ lepton, 
\bea
a_\ell \equiv \frac{(g-2)_\ell}{2} = 2\, m_\ell F_M^\gamma (0)\ \quad {\rm and}\ \quad 
d_\ell = - e F_E^\gamma (0) \ ,
\label{App}
\eea
respectively, where both moments are real if the interaction is to be Hermitian (see, for instance,~\cite{Branco:1999fs}).~The contribution of the T--odd spectrum in the LHT to $F_M^\gamma (0)$ and hence, its contribution to $a_\ell$ just defined, can be read from Eq.\,(\ref{dipoleformfactormue}).~It is real by inspection:~the loop functions are real because only heavy particles far away from any physical threshold run in the loops, and the mixing matrix elements as well as their complex conjugates enter symmetrically. 

On the other hand the contribution of these T--odd particles to the real part of $F_E^\gamma (0)$ (and then to $d_\ell$) vanishes because the form factors satisfy $F_E^\gamma (0) = i F_M^\gamma (0)$ and thus, are purely imaginary. 
    Indeed, from Eq.\,(\ref{dipoleformfactormue}) 
	\bea
	d_e = e\ {\rm Im} F_M^\gamma (0) = e \sum_{i,j,k}
	\frac{{\rm Im} (V^*_{ie}W^*_{ji}W_{jk}V_{ke}+V^*_{ke}W^*_{jk}W_{ji}V_{ie} )}{2}\ 
	G_{ijk} = 0\; ,
	\label{AppEDM}
	\eea
	where the first term of that equation combining the real products $V^*_{ie}V_{ie}$ is also real and the second term involving $V^*_{ie}W^*_{ji}W_{jk}V_{ke}$ products multiplied 
	by a real function $G_{ijk}$ symmetric under the exchange of the mirror leptons $i$ and $k$ 
	can be reordered as above to make the sum explicitly real. The current experimental precision $|d_e| < 1.1 \times 10^{-29}\ e \cdot$cm 
	at 90\% C.L. \cite{Andreev:2018ayy} deserves a full two-loop calculation \cite{Barr:1990vd}, what is beyond the scope of this paper. 
	It is worth emphasizing, however, that: 
	$i$) One-loop dimension 6 operators contributing to $d_e$ via mixing \cite{Panico:2018hal} 
	can be generated exchanging T--odd (non-singlet) leptons, but only through triangular diagrams. In particular, no box diagram can be closed to generate 
	the dimension 6 operator $|\phi|^2 V^{\mu \nu} \tilde{V}_{\mu \nu}$, 
	with $\phi$ the SM Higgs doublet (of hypercharge 1/2), $V^{\mu \nu}$ an electro-weak gauge boson field strength 
	and $\tilde{V}_{\mu \nu}$ its dual tensor, because there is no trilinear coupling of the Higgs doublet to two T--odd 
	lepton doublets. On the other hand there is a quartic coupling with two Higgs doublets to two T--odd lepton doublets, 
	$(\overline{l_{HL}} \tilde{\phi}) (\tilde{\phi}^\dagger l_{HR}) + {\rm h.c.} \supset \frac{1}{4} h^2 \overline{\nu_{H}} \nu_{H}$, 
	with $\tilde{\phi} = i\sigma^2 \phi^*$, \cite{delAguila:2017ugt}  
	which can be inserted in a (triangular) fermionic loop emitting two 
	SM gauge bosons, none of them being a photon. $ii$) In contrast, we expect non-vanishing box (and triangular) contributions 
	when considering the quark sector, not studied here, because this also includes T--odd electro-weak quark singlets. In any case, we can 
	use the rough estimate in \cite{Panico:2018hal}, 
	\bea
	\frac{d_e}{e} \simeq \left( \frac{g^2}{16 \pi^2} \right)^2 \frac{m_e}{f^2} \sin \vartheta \; ,
	\eea
	to obtain a bound on the effective CP violating phase $| \sin \vartheta | < 0.34$ for $f = 1.5$ TeV (and $g = 0.65$). What is a milder fine tuning compared to the 1 \% alignment of the T--odd (non-singlet) leptons with the SM ones required by current limits on LFV processes involving the first two families. $iii$) A discussion of the sub-leading contributions from dimension 8 operators is also necessary 
	\cite{Panico:2018hal}.


\begin{thebibliography}{10}
	
	

%\cite{Chatrchyan:2012xdj}
\bibitem{Chatrchyan:2012xdj}
S.~Chatrchyan {\it et al.} [CMS Collaboration],
%``Observation of a new boson at a mass of 125 GeV with the CMS experiment at the LHC,''
Phys.\ Lett.\ B {\bf 716} (2012) 30
doi:10.1016/j.physletb.2012.08.021
[arXiv:1207.7235 [hep-ex]].
%%CITATION = doi:10.1016/j.physletb.2012.08.021;%%
%7096 citations counted in INSPIRE as of 12 May 2017

%\cite{Aad:2012tfa}
\bibitem{Aad:2012tfa}
G.~Aad {\it et al.} [ATLAS Collaboration],
%``Observation of a new particle in the search for the Standard Model Higgs boson with the ATLAS detector at the LHC,''
Phys.\ Lett.\ B {\bf 716} (2012) 1
doi:10.1016/j.physletb.2012.08.020
[arXiv:1207.7214 [hep-ex]].
%%CITATION = doi:10.1016/j.physletb.2012.08.020;%%
%7233 citations counted in INSPIRE as of 12 May 2017

%\cite{Falkowski:2013dza}
\bibitem{Falkowski:2013dza}
A.~Falkowski, F.~Riva and A.~Urbano,
%``Higgs at last,''
JHEP {\bf 1311} (2013) 111
doi:10.1007/JHEP11(2013)111
[arXiv:1303.1812 [hep-ph]].
%%CITATION = doi:10.1007/JHEP11(2013)111;%%
%204 citations counted in INSPIRE as of 12 May 2017

%\cite{Khachatryan:2016vau}
\bibitem{Khachatryan:2016vau}
G.~Aad {\it et al.} [ATLAS and CMS Collaborations],
%``Measurements of the Higgs boson production and decay rates and constraints on its couplings from a combined ATLAS and CMS analysis of the LHC pp collision data at $ \sqrt{s}=7 $ and 8 TeV,''
JHEP {\bf 1608} (2016) 045
doi:10.1007/JHEP08(2016)045
[arXiv:1606.02266 [hep-ex]].
%%CITATION = doi:10.1007/JHEP08(2016)045;%%
%846 citations counted in INSPIRE as of 26 Dec 2018

%\cite{Graham:2015cka}
\bibitem{Graham:2015cka} 
P.~W.~Graham, D.~E.~Kaplan and S.~Rajendran,
%``Cosmological Relaxation of the Electroweak Scale,''
Phys.\ Rev.\ Lett.\  {\bf 115}, no. 22, 221801 (2015)
doi:10.1103/PhysRevLett.115.221801
[arXiv:1504.07551 [hep-ph]].
%%CITATION = doi:10.1103/PhysRevLett.115.221801;%%
%142 citations counted in INSPIRE as of 30 Aug 2017

%%%%%%%% LHC constraints %%%%%%%%%

%\cite{Falkowski:2015fla}
\bibitem{Falkowski:2015fla} 
A.~Falkowski,
%``Effective field theory approach to LHC Higgs data,''
Pramana {\bf 87}, no. 3, 39 (2016)
doi:10.1007/s12043-016-1251-5
[arXiv:1505.00046 [hep-ph]].
%%CITATION = doi:10.1007/s12043-016-1251-5;%%
%44 citations counted in INSPIRE as of 30 Aug 2017

%\cite{Englert:2015hrx}
\bibitem{Englert:2015hrx} 
C.~Englert, R.~Kogler, H.~Schulz and M.~Spannowsky,
%``Higgs coupling measurements at the LHC,''
Eur.\ Phys.\ J.\ C {\bf 76}, no. 7, 393 (2016)
doi:10.1140/epjc/s10052-016-4227-1
[arXiv:1511.05170 [hep-ph]].
%%CITATION = doi:10.1140/epjc/s10052-016-4227-1;%%
%29 citations counted in INSPIRE as of 30 Aug 2017

%\cite{Sirunyan:2017nrt}
\bibitem{Sirunyan:2017nrt}
A.~M.~Sirunyan {\it et al.} [CMS Collaboration],
%``Combination of searches for heavy resonances decaying to WW, WZ, ZZ, WH, and ZH boson pairs in protonÐproton collisions at $\sqrt{s}=8$ and 13 TeV,''
Phys.\ Lett.\ B {\bf 774} (2017) 533
doi:10.1016/j.physletb.2017.09.083
[arXiv:1705.09171 [hep-ex]].
%%CITATION = doi:10.1016/j.physletb.2017.09.083;%%
%27 citations counted in INSPIRE as of 26 Dec 2018 

%\cite{Aaboud:2018bun}
\bibitem{Aaboud:2018bun}
M.~Aaboud {\it et al.} [ATLAS Collaboration],
%``Combination of searches for heavy resonances decaying into bosonic and leptonic final states using 36 fb$^{-1}$ of proton-proton collision data at $\sqrt{s} = 13$ TeV with the ATLAS detector,''
Phys.\ Rev.\ D {\bf 98} (2018) no.5,  052008
doi:10.1103/PhysRevD.98.052008
[arXiv:1808.02380 [hep-ex]].
%%CITATION = doi:10.1103/PhysRevD.98.052008;%%
%8 citations counted in INSPIRE as of 26 Dec 2018


%%%%%%%%%%% EW Fits %%%%%%%%%%

%\cite{deBlas:2013qqa}
\bibitem{deBlas:2013qqa}
J.~de Blas, M.~Chala and J.~Santiago,
%``Global Constraints on Lepton-Quark Contact Interactions,''
Phys.\ Rev.\ D {\bf 88} (2013) 095011
doi:10.1103/PhysRevD.88.095011
[arXiv:1307.5068 [hep-ph]].
%%CITATION = doi:10.1103/PhysRevD.88.095011;%%
%45 citations counted in INSPIRE as of 18 Dec 2018

%\cite{Falkowski:2014tna}
\bibitem{Falkowski:2014tna} 
A.~Falkowski and F.~Riva,
%``Model-independent precision constraints on dimension-6 operators,''
JHEP {\bf 1502}, 039 (2015)
doi:10.1007/JHEP02(2015)039
[arXiv:1411.0669 [hep-ph]].
%%CITATION = doi:10.1007/JHEP02(2015)039;%%
%91 citations counted in INSPIRE as of 30 Aug 2017

%\cite{deBlas:2015aea}
\bibitem{deBlas:2015aea}
J.~de Blas, M.~Chala and J.~Santiago,
%``Renormalization Group Constraints on New Top Interactions from Electroweak Precision Data,''
JHEP {\bf 1509} (2015) 189
doi:10.1007/JHEP09(2015)189
[arXiv:1507.00757 [hep-ph]].
%%CITATION = doi:10.1007/JHEP09(2015)189;%%
%46 citations counted in INSPIRE as of 18 Dec 2018

%\cite{Berthier:2015gja}
\bibitem{Berthier:2015gja} 
L.~Berthier and M.~Trott,
%``Consistent constraints on the Standard Model Effective Field Theory,''
JHEP {\bf 1602}, 069 (2016)
doi:10.1007/JHEP02(2016)069
[arXiv:1508.05060 [hep-ph]].
%%CITATION = doi:10.1007/JHEP02(2016)069;%%
%50 citations counted in INSPIRE as of 30 Aug 2017

%\cite{Buckley:2015lku}
\bibitem{Buckley:2015lku}
A.~Buckley, C.~Englert, J.~Ferrando, D. J.~Miller, L.~Moore, M.~Russell and C.~D.~White,
%``Constraining top quark effective theory in the LHC Run II era,''
JHEP {\bf 1604} (2016) 015
doi:10.1007/JHEP04(2016)015
[arXiv:1512.03360 [hep-ph]].
%%CITATION = doi:10.1007/JHEP04(2016)015;%%
%68 citations counted in INSPIRE as of 18 Dec 2018

%\cite{deBlas:2016ojx}
\bibitem{deBlas:2016ojx}
J.~de Blas, M.~Ciuchini, E.~Franco, S.~Mishima, M.~Pierini, L.~Reina and L.~Silvestrini,
%``Electroweak precision observables and Higgs-boson signal strengths in the Standard Model and beyond: present and future,''
JHEP {\bf 1612} (2016) 135
doi:10.1007/JHEP12(2016)135
[arXiv:1608.01509 [hep-ph]].
%%CITATION = doi:10.1007/JHEP12(2016)135;%%
%61 citations counted in INSPIRE as of 18 Dec 2018

%\cite{Falkowski:2017pss}
\bibitem{Falkowski:2017pss}
A.~Falkowski, M.~Gonz\'alez-Alonso and K.~Mimouni,
%``Compilation of low-energy constraints on 4-fermion operators in the SMEFT,''
JHEP {\bf 1708} (2017) 123
doi:10.1007/JHEP08(2017)123
[arXiv:1706.03783 [hep-ph]].
%%CITATION = doi:10.1007/JHEP08(2017)123;%%
%37 citations counted in INSPIRE as of 25 Dec 2018

%\cite{Ellis:2018gqa}
\bibitem{Ellis:2018gqa}
J.~Ellis, C. W.~Murphy, V.~Sanz and T.~You,
%``Updated Global SMEFT Fit to Higgs, Diboson and Electroweak Data,''
JHEP {\bf 1806} (2018) 146
doi:10.1007/JHEP06(2018)146
[arXiv:1803.03252 [hep-ph]].
%%CITATION = doi:10.1007/JHEP06(2018)146;%%
%31 citations counted in INSPIRE as of 18 Dec 2018

%\cite{Grojean:2018dqj}
\bibitem{Grojean:2018dqj}
C.~Grojean, M.~Montull and M.~Riembau,
%``Diboson at the LHC vs LEP,''
arXiv:1810.05149 [hep-ph].
%%CITATION = ARXIV:1810.05149;%%
%4 citations counted in INSPIRE as of 20 Dec 2018

%\cite{Almeida:2018cld}
\bibitem{Almeida:2018cld}
E. d. S.~Almeida, A.~Alves, N. R.~Agostinho, O. J. P.~\'Eboli and M. C.~Gonzalez-Garcia,
%``Electroweak Sector Under Scrutiny: A Combined Analysis of LHC and Electroweak Precision Data,''
arXiv:1812.01009 [hep-ph].
%%CITATION = ARXIV:1812.01009;%%

%%%%%%%%%% LH(T) %%%%%%%%%%

%\cite{ArkaniHamed:2001nc}
\bibitem{ArkaniHamed:2001nc}
N.~Arkani-Hamed, A.~G.~Cohen and H.~Georgi,
%``Electroweak symmetry breaking from dimensional deconstruction,''
Phys.\ Lett.\ B {\bf 513} (2001) 232
doi:10.1016/S0370-2693(01)00741-9
[hep-ph/0105239].
%%CITATION = doi:10.1016/S0370-2693(01)00741-9;%%
%1210 citations counted in INSPIRE as of 12 May 2017

%\cite{ArkaniHamed:2001ca}
\bibitem{ArkaniHamed:2001ca} 
N.~Arkani-Hamed, A.~G.~Cohen and H.~Georgi,
%``(De)constructing dimensions,''
Phys.\ Rev.\ Lett.\  {\bf 86}, 4757 (2001)
doi:10.1103/PhysRevLett.86.4757
[hep-th/0104005].
%%CITATION = doi:10.1103/PhysRevLett.86.4757;%%
%672 citations counted in INSPIRE as of 31 Aug 2017

%\cite{ArkaniHamed:2002qy}
\bibitem{ArkaniHamed:2002qy} 
N.~Arkani-Hamed, A.~G.~Cohen, E.~Katz and A.~E.~Nelson,
%``The Littlest Higgs,''
JHEP {\bf 0207}, 034 (2002)
doi:10.1088/1126-6708/2002/07/034
[hep-ph/0206021].
%%CITATION = doi:10.1088/1126-6708/2002/07/034;%%
%1063 citations counted in INSPIRE as of 31 Aug 2017

%\cite{Cheng:2003ju}
\bibitem{Cheng:2003ju}
H.~C.~Cheng and I.~Low,
%``TeV symmetry and the little hierarchy problem,''
JHEP {\bf 0309} (2003) 051
doi:10.1088/1126-6708/2003/09/051
[hep-ph/0308199].
%%CITATION = doi:10.1088/1126-6708/2003/09/051;%%
%557 citations counted in INSPIRE as of 12 May 2017

%\cite{Cheng:2004yc}
\bibitem{Cheng:2004yc}
H.~C.~Cheng and I.~Low,
%``Little hierarchy, little Higgses, and a little symmetry,''
JHEP {\bf 0408} (2004) 061
doi:10.1088/1126-6708/2004/08/061
[hep-ph/0405243].
%%CITATION = doi:10.1088/1126-6708/2004/08/061;%%
%496 citations counted in INSPIRE as of 12 May 2017

%\cite{Cheng:2005as}
\bibitem{Cheng:2005as}
H.~C.~Cheng, I.~Low and L.~T.~Wang,
%``Top partners in little Higgs theories with T-parity,''
Phys.\ Rev.\ D {\bf 74} (2006) 055001
doi:10.1103/PhysRevD.74.055001
[hep-ph/0510225].
%%CITATION = doi:10.1103/PhysRevD.74.055001;%%
%130 citations counted in INSPIRE as of 22 Jul 

%\cite{Hubisz:2005tx}
\bibitem{Hubisz:2005tx} 
J.~Hubisz, P.~Meade, A.~Noble and M.~Perelstein,
%``Electroweak precision constraints on the littlest Higgs model with T parity,''
JHEP {\bf 0601}, 135 (2006)
doi:10.1088/1126-6708/2006/01/135
[hep-ph/0506042].
%%CITATION = doi:10.1088/1126-6708/2006/01/135;%%
%262 citations counted in INSPIRE as of 31 Aug 2017

%\cite{Low:2004xc}
\bibitem{Low:2004xc}
I.~Low,
%``T parity and the littlest Higgs,''
JHEP {\bf 0410} (2004) 067
doi:10.1088/1126-6708/2004/10/067
[hep-ph/0409025].
%%CITATION = doi:10.1088/1126-6708/2004/10/067;%%
%309 citations counted in INSPIRE as of 12 May 2017

%\cite{Hubisz:2004ft}  
\bibitem{Hubisz:2004ft}
J.~Hubisz and P.~Meade,
%``Phenomenology of the littlest Higgs with T--parity,''
Phys.\ Rev.\ D {\bf 71} (2005) 035016
doi:10.1103/PhysRevD.71.035016
[hep-ph/0411264].
%%CITATION = doi:10.1103/PhysRevD.71.035016;%%
%296 citations counted in INSPIRE as of 12 May 2017

%\cite{Hubisz:2005bd}
\bibitem{Hubisz:2005bd}
J.~Hubisz, S.~J.~Lee and G.~Paz,
%``The Flavor of a little Higgs with T--parity,''
JHEP {\bf 0606} (2006) 041
doi:10.1088/1126-6708/2006/06/041
[hep-ph/0512169].
%%CITATION = doi:10.1088/1126-6708/2006/06/041;%%
%132 citations counted in INSPIRE as of 12 May 2017

%\cite{Chen:2006cs}
\bibitem{Chen:2006cs}
C.~R.~Chen, K.~Tobe and C.-P.~Yuan,
%``Higgs boson production and decay in little Higgs models with T--parity,''
Phys.\ Lett.\ B {\bf 640} (2006) 263
doi:10.1016/j.physletb.2006.07.053
[hep-ph/0602211].
%%CITATION = doi:10.1016/j.physletb.2006.07.053;%%
%100 citations counted in INSPIRE as of 12 May 2017

%\cite{Blanke:2006sb}
\bibitem{Blanke:2006sb}
M.~Blanke, A.~J.~Buras, A.~Poschenrieder, C.~Tarantino, S.~Uhlig and A.~Weiler,
%``Particle-Antiparticle Mixing, epsilon(K), Delta Gamma(q), A**q(SL), A(CP) (B(d) ---> psi K(S)), A(CP) (B(s) ---> psi phi) and B ---> X(s,d gamma) in the Littlest Higgs Model with T-Parity,''
JHEP {\bf 0612} (2006) 003
doi:10.1088/1126-6708/2006/12/003 
[hep-ph/0605214].  %%CITATION = doi:10.1088/1126-6708/2006/12/003;%%
%164 citations counted in INSPIRE as of 12 May 2017

%\cite{Buras:2006wk}
\bibitem{Buras:2006wk}
A.~J.~Buras, A.~Poschenrieder, S.~Uhlig and W.~A.~Bardeen,
%``Rare $K$ and $B$ Decays in the Littlest Higgs Model without $T^-$ Parity,''
JHEP {\bf 0611} (2006) 062
doi:10.1088/1126-6708/2006/11/062
[hep-ph/0607189].
%%CITATION = doi:10.1088/1126-6708/2006/11/062;%%
%65 citations counted in INSPIRE as of 12 May 2017

%\cite{Belyaev:2006jh}
\bibitem{Belyaev:2006jh}
A.~Belyaev, C.~R.~Chen, K.~Tobe and C.-P.~Yuan,
%``Phenomenology of littlest Higgs model with $T^-$ parity: including effects of $T^-$ odd fermions,''
Phys.\ Rev.\ D {\bf 74} (2006) 115020
doi:10.1103/PhysRevD.74.115020
[hep-ph/0609179].
%%CITATION = doi:10.1103/PhysRevD.74.115020;%%
%115 citations counted in INSPIRE as of 12 May 2017

%\cite{Blanke:2006eb}
\bibitem{Blanke:2006eb}
M.~Blanke, A.~J.~Buras, A.~Poschenrieder, S.~Recksiegel, C.~Tarantino, S.~Uhlig and A.~Weiler,
%``Rare and CP-Violating $K$ and $B$ Decays in the Littlest Higgs Model with $T^-$ Parity,''
JHEP {\bf 0701} (2007) 066
doi:10.1088/1126-6708/2007/01/066
[hep-ph/0610298].
%%CITATION = doi:10.1088/1126-6708/2007/01/066;%%
%220 citations counted in INSPIRE as of 12 May 2017

%\cite{Hill:2007nz}
\bibitem{Hill:2007nz}
C.~T.~Hill and R.~J.~Hill,
%``Topological Physics of Little Higgs Bosons,''
Phys.\ Rev.\ D {\bf 75} (2007) 115009
doi:10.1103/PhysRevD.75.115009
[hep-ph/0701044].
%%CITATION = doi:10.1103/PhysRevD.75.115009;%%
%54 citations counted in INSPIRE as of 12 May 2017

%\cite{Hill:2007zv}
\bibitem{Hill:2007zv}
C.~T.~Hill and R.~J.~Hill,
%``$T^-$ parity violation by anomalies,''
Phys.\ Rev.\ D {\bf 76} (2007) 115014
doi:10.1103/PhysRevD.76.115014
[arXiv:0705.0697 [hep-ph]].
%%CITATION = doi:10.1103/PhysRevD.76.115014;%%
%84 citations counted in INSPIRE as of 12 May 2017

%\cite{Han:2008wb}
\bibitem{Han:2008wb}
X.~F.~Han, L.~Wang and J.~M.~Yang,
%``Higgs and Z-boson FCNC decays correlated with B-meson decays in littlest Higgs model with T--parity,''
Phys.\ Rev.\ D {\bf 78} (2008) 075017
doi:10.1103/PhysRevD.78.075017
[arXiv:0807.4480 [hep-ph]].
%%CITATION = doi:10.1103/PhysRevD.78.075017;%%
%23 citations counted in INSPIRE as of 12 May 2017

%\cite{Goto:2008fj}
\bibitem{Goto:2008fj}
T.~Goto, Y.~Okada and Y.~Yamamoto,
%``Ultraviolet divergences of flavor changing amplitudes in the littlest Higgs model with T--parity,''
Phys.\ Lett.\ B {\bf 670} (2009) 378
doi:10.1016/j.physletb.2008.11.022
[arXiv:0809.4753 [hep-ph]].
%%CITATION = doi:10.1016/j.physletb.2008.11.022;%%
%54 citations counted in INSPIRE as of 12 May 2017

%\cite{delAguila:2008zu}
\bibitem{delAguila:2008zu}
F.~del Aguila, J.~I.~Illana and M.~D.~Jenkins,
%``Precise limits from lepton flavour violating processes on the Littlest Higgs model with T--parity,''
JHEP {\bf 0901} (2009) 080
doi:10.1088/1126-6708/2009/01/080
[arXiv:0811.2891 [hep-ph]].
%%CITATION = doi:10.1088/1126-6708/2009/01/080;%%
%61 citations counted in INSPIRE as of 12 May 2017

%\cite{Blanke:2009am}
\bibitem{Blanke:2009am}
M.~Blanke, A.~J.~Buras, B.~Duling, S.~Recksiegel and C.~Tarantino,
%``FCNC Processes in the Littlest Higgs Model with T-Parity: a 2009 Look,''
Acta Phys.\ Polon.\ B {\bf 41} (2010) 657
[arXiv:0906.5454 [hep-ph]].
%%CITATION = ARXIV:0906.5454;%%
%138 citations counted in INSPIRE as of 12 May 2017

%\cite{delAguila:2010nv}
\bibitem{delAguila:2010nv}
F.~del Aguila, J.~I.~Illana and M.~D.~Jenkins,
%``Muon to electron conversion in the Littlest Higgs model with T--parity,''
JHEP {\bf 1009} (2010) 040
doi:10.1007/JHEP09(2010)040
[arXiv:1006.5914 [hep-ph]].
%%CITATION = doi:10.1007/JHEP09(2010)040;%%
%9 citations counted in INSPIRE as of 12 May 2017

%\cite{Goto:2010sn}
\bibitem{Goto:2010sn}
T.~Goto, Y.~Okada and Y.~Yamamoto,
%``Tau and muon lepton flavor violations in the littlest Higgs model with T--parity,''
Phys.\ Rev.\ D {\bf 83} (2011) 053011
doi:10.1103/PhysRevD.83.053011
[arXiv:1012.4385 [hep-ph]].
%%CITATION = doi:10.1103/PhysRevD.83.053011;%%
%18 citations counted in INSPIRE as of 13 May 2017 

%\cite{Zhou:2012cja}
\bibitem{Zhou:2012cja}
H.~S.~Hou, H.~Sun and Y.~J.~Zhou,
%``Flavor changing top quark decay and bottom-strange production in the littlest Higgs model with T--parity,''
Commun.\ Theor.\ Phys.\  {\bf 59} (2013) 443
doi:10.1088/0253-6102/59/4/10
[arXiv:1210.3904 [hep-ph]].
%%CITATION = doi:10.1088/0253-6102/59/4/10;%%
%3 citations counted in INSPIRE as of 12 May 2017

%\cite{Reuter:2013iya}
\bibitem{Reuter:2013iya}
J.~Reuter, M.~Tonini and M.~de Vries,
%``Littlest Higgs with T--parity: Status and Prospects,''
JHEP {\bf 1402} (2014) 053
doi:10.1007/JHEP02(2014)053
[arXiv:1310.2918 [hep-ph]].
%%CITATION = doi:10.1007/JHEP02(2014)053;%%
%56 citations counted in INSPIRE as of 18 Jul 2018

%\cite{Yang:2016hrh}
\bibitem{Yang:2016hrh}
B.~Yang, J.~Han and N.~Liu,
%``Lepton flavor violating Higgs boson decay $h\rightarrow \mu\tau$ in the littlest Higgs model with T parity,''
Phys.\ Rev.\ D {\bf 95} (2017) no.3,  035010
doi:10.1103/PhysRevD.95.035010
[arXiv:1605.09248 [hep-ph]].
%%CITATION = doi:10.1103/PhysRevD.95.035010;%%
%8 citations counted in INSPIRE as of 13 May 2017

%\cite{Dercks:2018hgz}
\bibitem{Dercks:2018hgz}
D.~Dercks, G.~Moortgat-Pick, J.~Reuter and S.~Y.~Shim,
%``The fate of the Littlest Higgs Model with T--parity under 13 TeV LHC Data,''
JHEP {\bf 1805} (2018) 049
doi:10.1007/JHEP05(2018)049
[arXiv:1801.06499 [hep-ph]].
%%CITATION = doi:10.1007/JHEP05(2018)049;%%
%3 citations counted in INSPIRE as of 17 Jul 2018

%%%%%% Reviews %%%%%%%%

%\cite{Schmaltz:2005ky}
\bibitem{Schmaltz:2005ky}
M.~Schmaltz and D.~Tucker-Smith,
%``Little Higgs review,''
Ann.\ Rev.\ Nucl.\ Part.\ Sci.\  {\bf 55} (2005) 229
doi:10.1146/annurev.nucl.55.090704.151502
[hep-ph/0502182].
%%CITATION = doi:10.1146/annurev.nucl.55.090704.151502;%%
%720 citations counted in INSPIRE as of 22 Jul 2019

%\cite{Perelstein:2005ka}
\bibitem{Perelstein:2005ka}
M.~Perelstein,
%``Little Higgs models and their phenomenology,''
Prog.\ Part.\ Nucl.\ Phys.\  {\bf 58} (2007) 247
doi:10.1016/j.ppnp.2006.04.001
[hep-ph/0512128].
%%CITATION = doi:10.1016/j.ppnp.2006.04.001;%%
%419 citations counted in INSPIRE as of 22 Jul 2019

%\cite{Panico:2015jxa}
\bibitem{Panico:2015jxa}
G.~Panico and A.~Wulzer,
%``The Composite Nambu-Goldstone Higgs,''
Lect.\ Notes Phys.\  {\bf 913} (2016) pp.1
doi:10.1007/978-3-319-22617-0
[arXiv:1506.01961 [hep-ph]].
%%CITATION = doi:10.1007/978-3-319-22617-0;%%
%266 citations counted in INSPIRE as of 22 Jul 2019

%%%%%% Flavor %%%%%%%%

%\cite{Glashow:1970gm}
\bibitem{Glashow:1970gm}
S.~L.~Glashow, J.~Iliopoulos and L.~Maiani,
%``Weak Interactions with Lepton-Hadron Symmetry,''
Phys.\ Rev.\ D {\bf 2} (1970) 1285.
doi:10.1103/PhysRevD.2.1285
%%CITATION = doi:10.1103/PhysRevD.2.1285;%%
%5453 citations counted in INSPIRE as of 31 Jul 2017

%\cite{Isidori:2010kg}
\bibitem{Isidori:2010kg}
G.~Isidori, Y.~Nir and G.~Perez,
%``Flavor Physics Constraints for Physics Beyond the Standard Model,''
Ann.\ Rev.\ Nucl.\ Part.\ Sci.\  {\bf 60} (2010) 355
doi:10.1146/annurev.nucl.012809.104534
[arXiv:1002.0900 [hep-ph]].
%%CITATION = doi:10.1146/annurev.nucl.012809.104534;%%
%320 citations counted in INSPIRE as of 05 Aug 2018

%\cite{Raidal:2008jk}
\bibitem{Raidal:2008jk}
M.~Raidal {\it et al.},
%``Flavour physics of leptons and dipole moments,''
Eur.\ Phys.\ J.\ C {\bf 57} (2008) 13
doi:10.1140/epjc/s10052-008-0715-2
[arXiv:0801.1826 [hep-ph]].
%%CITATION = doi:10.1140/epjc/s10052-008-0715-2;%%
%342 citations counted in INSPIRE as of 05 Aug 2018

%\cite{Pich:2018njk}
\bibitem{Pich:2018njk}
A.~Pich,
%``Flavour Dynamics and Violations of the CP Symmetry,''
arXiv:1805.08597 [hep-ph].
%%CITATION = ARXIV:1805.08597;%%

%\cite{Georgi:1986ku}
\bibitem{Georgi:1986ku}
H.~Georgi,
%``The Flavor Problem,''
Phys.\ Lett.\  {\bf 169B} (1986) 231.
doi:10.1016/0370-2693(86)90657-X
%%CITATION = doi:10.1016/0370-2693(86)90657-X;%%
%82 citations counted in INSPIRE as of 30 Jun 2017

%\cite{AlvarezGaume:1983gj}
\bibitem{AlvarezGaume:1983gj}
L.~Alvarez-Gaume, J.~Polchinski and M.~B.~Wise,
%``Minimal Low-Energy Supergravity,''
Nucl.\ Phys.\ B {\bf 221} (1983) 495.
doi:10.1016/0550-3213(83)90591-6
%%CITATION = doi:10.1016/0550-3213(83)90591-6;%%
%1166 citations counted in INSPIRE as of 30 Jun 2017

%\cite{Kaplan:1991dc}
\bibitem{Kaplan:1991dc}
D.~B.~Kaplan,
%``Flavor at SSC energies: A New mechanism for dynamically generated fermion masses,''
Nucl.\ Phys.\ B {\bf 365} (1991) 259.
doi:10.1016/S0550-3213(05)80021-5
%%CITATION = doi:10.1016/S0550-3213(05)80021-5;%%
%410 citations counted in INSPIRE as of 18 Dec 2018

%\cite{Csaki:2008qq}
\bibitem{Csaki:2008qq}
C.~Csaki, C.~Delaunay, C.~Grojean and Y.~Grossman,
%``A Model of Lepton Masses from a Warped Extra Dimension,''
JHEP {\bf 0810} (2008) 055
doi:10.1088/1126-6708/2008/10/055
[arXiv:0806.0356 [hep-ph]].
%%CITATION = doi:10.1088/1126-6708/2008/10/055;%%
%169 citations counted in INSPIRE as of 18 Jan 2019

%\cite{Santiago:2008vq}
\bibitem{Santiago:2008vq}
J.~Santiago,
%``Minimal Flavor Protection: A New Flavor Paradigm in Warped Models,''
JHEP {\bf 0812} (2008) 046
doi:10.1088/1126-6708/2008/12/046
[arXiv:0806.1230 [hep-ph]].
%%CITATION = doi:10.1088/1126-6708/2008/12/046;%%
%93 citations counted in INSPIRE as of 18 Jan 2019

%\cite{Csaki:2008eh}
\bibitem{Csaki:2008eh}
C.~Csaki, A.~Falkowski and A.~Weiler,
%``A Simple Flavor Protection for RS,''
Phys.\ Rev.\ D {\bf 80} (2009) 016001
doi:10.1103/PhysRevD.80.016001
[arXiv:0806.3757 [hep-ph]].
%%CITATION = doi:10.1103/PhysRevD.80.016001;%%
%119 citations counted in INSPIRE as of 18 Jan 2019

%\cite{delAguila:2010vg}
\bibitem{delAguila:2010vg}
F.~del Aguila, A.~Carmona and J.~Santiago,
%``Neutrino Masses from an A4 Symmetry in Holographic Composite Higgs Models,''
JHEP {\bf 1008} (2010) 127
doi:10.1007/JHEP08(2010)127
[arXiv:1001.5151 [hep-ph]].
%%CITATION = doi:10.1007/JHEP08(2010)127;%%
%82 citations counted in INSPIRE as of 18 Jan 2019

%\cite{delAguila:2017ugt}
\bibitem{delAguila:2017ugt} 
F.~del Aguila, L.~Ametller, J.~I.~Illana, J.~Santiago, P.~Talavera and R.~Vega-Morales,
%``Lepton Flavor Changing Higgs decays in the Littlest Higgs Model with T--parity,''
JHEP {\bf 1708}, 028 (2017)
doi:10.1007/JHEP08(2017)028
[arXiv:1705.08827 [hep-ph]].
%%CITATION = doi:10.1007/JHEP08(2017)028;%%
%1 citations counted in INSPIRE as of 31 Aug 2017

%\cite{Borsanyi:2017zdw}
\bibitem{Borsanyi:2017zdw}
S.~Borsanyi {\it et al.} [Budapest-Marseille-Wuppertal Collaboration],
%``Hadronic vacuum polarization contribution to the anomalous magnetic moments of leptons from first principles,''
Phys.\ Rev.\ Lett.\  {\bf 121} (2018) no.2,  022002
doi:10.1103/PhysRevLett.121.022002
[arXiv:1711.04980 [hep-lat]].
%%CITATION = doi:10.1103/PhysRevLett.121.022002;%%
%10 citations counted in INSPIRE as of 17 Jul 2018

%\cite{Blum:2018mom}
\bibitem{Blum:2018mom}
T.~Blum {\it et al.} [RBC and UKQCD Collaborations],
%``Calculation of the hadronic vacuum polarization contribution to the muon anomalous magnetic moment,''
Phys.\ Rev.\ Lett.\  {\bf 121} (2018) no.2,  022003
doi:10.1103/PhysRevLett.121.022003
[arXiv:1801.07224 [hep-lat]].
%%CITATION = doi:10.1103/PhysRevLett.121.022003;%%
%9 citations counted in INSPIRE as of 17 Jul 2018  

%%%%%%%%%% EDMe %%%%%%%%%

 %\cite{Andreev:2018ayy}
\bibitem{Andreev:2018ayy}
V.~Andreev {\it et al.} [ACME Collaboration],
%``Improved limit on the electric dipole moment of the electron,''
Nature {\bf 562} (2018) no.7727,  355.
doi:10.1038/s41586-018-0599-8
%%CITATION = doi:10.1038/s41586-018-0599-8;%%
%73 citations counted in INSPIRE as of 21 May 2019

%\cite{Panico:2018hal}
\bibitem{Panico:2018hal}
G.~Panico, A.~Pomarol and M.~Riembau,
%``EFT approach to the electron Electric Dipole Moment at the two-loop level,''
JHEP {\bf 1904} (2019) 090
doi:10.1007/JHEP04(2019)090
[arXiv:1810.09413 [hep-ph]].
%%CITATION = doi:10.1007/JHEP04(2019)090;%%
%8 citations counted in INSPIRE as of 21 May 2019

%\cite{Chupp:2017rkp}
\bibitem{Chupp:2017rkp}
T.~Chupp, P.~Fierlinger, M.~Ramsey-Musolf and J.~Singh,
%``Electric dipole moments of atoms, molecules, nuclei, and particles,''
Rev.\ Mod.\ Phys.\  {\bf 91} (2019) no.1,  015001
doi:10.1103/RevModPhys.91.015001
[arXiv:1710.02504 [physics.atom-ph]].
%%CITATION = doi:10.1103/RevModPhys.91.015001;%%
%60 citations counted in INSPIRE as of 26 May 2019

%\cite{Panico:2017vlk}
\bibitem{Panico:2017vlk}
G.~Panico, M.~Riembau and T.~Vantalon,
%``Probing light top partners with CP violation,''
JHEP {\bf 1806} (2018) 056
doi:10.1007/JHEP06(2018)056
[arXiv:1712.06337 [hep-ph]].
%%CITATION = doi:10.1007/JHEP06(2018)056;%%
%9 citations counted in INSPIRE as of 23 May 2019

%%%%%%%%%%%%%%%%%%%%%%%%

%\cite{Lindner:2016bgg}
\bibitem{Lindner:2016bgg}
M.~Lindner, M.~Platscher and F.~S.~Queiroz,
%``A Call for New Physics : The Muon Anomalous Magnetic Moment and Lepton Flavor Violation,''
Phys.\ Rept.\  {\bf 731} (2018) 1
doi:10.1016/j.physrep.2017.12.001
[arXiv:1610.06587 [hep-ph]].
%%CITATION = doi:10.1016/j.physrep.2017.12.001;%%
%115 citations counted in INSPIRE as of 10 Feb 2019

%%%%%%%%%%%%%%%%%%%%%%%%

%\cite{inpreparation}
\bibitem{inpreparation} 
F.~del Aguila, L.~Ametller, J.~I.~Illana, J.M. Perez-Poyatos, 
J.~Santiago, P.~Talavera and R.~Vega-Morales,
in preparation.

%\cite{Pappadopulo:2010jx}
\bibitem{Pappadopulo:2010jx}
D.~Pappadopulo and A.~Vichi,
%``T-parity, its problems and their solution,''
JHEP {\bf 1103} (2011) 072
doi:10.1007/JHEP03(2011)072
[arXiv:1007.4807 [hep-ph]].
%%CITATION = doi:10.1007/JHEP03(2011)072;%%
%17 citations counted in INSPIRE as of 23 Dec 2018

%\cite{Tonini:2014dza}
\bibitem{Tonini:2014dza}
M.~Tonini,
%``Beyond the Standard Higgs at the LHC: present constraints on Little Higgs models and future prospects,''
DESY-THESIS-2014-038.
%%CITATION = DESY-THESIS-2014-038;%%
%2 citations counted in INSPIRE as of 30 Jul 2018

%\cite{Shim:2018ksu}
\bibitem{Shim:2018ksu}
S.~Y.~Shim,
%``Beyond the SM searches at the LHC: Little Higgs models and Effective Field Theory,''
doi:10.3204/PUBDB-2018-01793
%%CITATION = doi:10.3204/PUBDB-2018-01793;%%  

%\cite{Reuter:2018xfr}
\bibitem{Reuter:2018xfr}
J.~Reuter, D.~Dercks, G.~Moortgat-Pick and S.~Y.~Shim,
%``The fate of the Littlest Higgs with T-parity under 13 TeV LHC data,''
arXiv:1811.02268 [hep-ph].
%%CITATION = ARXIV:1811.02268;%%

%\cite{Vecchi:2013bja}
\bibitem{Vecchi:2013bja}
L.~Vecchi,
%``The Natural Composite Higgs,''
arXiv:1304.4579 [hep-ph].
%%CITATION = ARXIV:1304.4579;%%
%34 citations counted in INSPIRE as of 23 Dec 2018

%%%%%%%%% Future %%%%%%%%%%%

%\cite{Baldini:2018uhj}
\bibitem{Baldini:2018uhj}
A.~Baldini {\it et al.},
%``A submission to the 2020 update of the European Strategy for Particle Physics on behalf of the COMET, MEG, Mu2e and Mu3e collaborations,''
arXiv:1812.06540 [hep-ex].
%%CITATION = ARXIV:1812.06540;%% 

%\cite{Liventsev:2018gin}
\bibitem{Liventsev:2018gin}
D.~Liventsev [Belle II Collaboration],
%``Prospects of LFV studies at Belle II,''
PoS NuFact {\bf 2017} (2018) 118.
doi:10.22323/1.295.0118
%%CITATION = doi:10.22323/1.295.0118;%%

%\cite{Bediaga:2018lhg}
\bibitem{Bediaga:2018lhg}
R.~Aaij {\it et al.} [LHCb Collaboration],
%``Physics case for an LHCb Upgrade II - Opportunities in flavour physics, and beyond, in the HL-LHC era,''
arXiv:1808.08865.
%%CITATION = ARXIV:1808.08865;%%
%27 citations counted in INSPIRE as of 17 Jan 2019

%\cite{Benedikt:2018qee}
\bibitem{Benedikt:2018qee}
M.~Benedikt {\it et al.},
%``Future Circular Collider : Vol. 2 The Lepton Collider (FCC-ee),''
CERN-ACC-2018-0057.
%%CITATION = CERN-ACC-2018-0057;%%
%1 citations counted in INSPIRE as of 08 Feb 2019 


%\cite{Baer:2013cma}
\bibitem{Baer:2013cma}
H.~Baer {\it et al.},
%``The International Linear Collider Technical Design Report - Volume 2: Physics,''
arXiv:1306.6352 [hep-ph].
%%CITATION = ARXIV:1306.6352;%%
%680 citations counted in INSPIRE as of 09 Feb 2019


%\cite{Blanke:2007db}
\bibitem{Blanke:2007db}
M.~Blanke, A.~J.~Buras, B.~Duling, A.~Poschenrieder and C.~Tarantino,
%``Charged Lepton Flavour Violation and (g-2)(mu) in the Littlest Higgs Model with T-Parity: A Clear Distinction from Supersymmetry,''
JHEP {\bf 0705} (2007) 013
doi:10.1088/1126-6708/2007/05/013
[hep-ph/0702136].
%%CITATION = doi:10.1088/1126-6708/2007/05/013;%%
%169 citations counted in INSPIRE as of 16 Aug 2017


%%%%%%TABLE 1 REFS%%%%%%%%%

%\cite{Adam:2013mnn}
\bibitem{Adam:2013mnn}
J.~Adam {\it et al.} [MEG Collaboration],
%``New constraint on the existence of the $\mu^+ \to e^+\gamma$ decay,''
Phys.\ Rev.\ Lett.\  {\bf 110} (2013) 201801
doi:10.1103/PhysRevLett.110.201801
[arXiv:1303.0754 [hep-ex]].
%%CITATION = doi:10.1103/PhysRevLett.110.201801;%%
%510 citations counted in INSPIRE as of 03 Aug 2017

%\cite{Bellgardt:1987du}
\bibitem{Bellgardt:1987du}
U.~Bellgardt {\it et al.} [SINDRUM Collaboration],
%``Search for the Decay mu+ ---> e+ e+ e-,''
Nucl.\ Phys.\ B {\bf 299} (1988) 1.
doi:10.1016/0550-3213(88)90462-2
%%CITATION = doi:10.1016/0550-3213(88)90462-2;%%
%626 citations counted in INSPIRE as of 27 Dec 2018

%\cite{Bertl:2006up}
\bibitem{Bertl:2006up}
W.~H.~Bertl {\it et al.} [SINDRUM II Collaboration],
%``A Search for muon to electron conversion in muonic gold,''
Eur.\ Phys.\ J.\ C {\bf 47} (2006) 337.
doi:10.1140/epjc/s2006-02582-x
%%CITATION = doi:10.1140/epjc/s2006-02582-x;%%
%316 citations counted in INSPIRE as of 03 Aug 2017

%\cite{Aubert:2009ag}
\bibitem{Aubert:2009ag}
B.~Aubert {\it et al.} [BaBar Collaboration],
%``Searches for Lepton Flavor Violation in the Decays tau+- ---> e+- gamma and tau+- ---> mu+- gamma,''
Phys.\ Rev.\ Lett.\  {\bf 104} (2010) 021802
doi:10.1103/PhysRevLett.104.021802
[arXiv:0908.2381 [hep-ex]].
%%CITATION = doi:10.1103/PhysRevLett.104.021802;%%
%329 citations counted in INSPIRE as of 03 Aug 2017

%\cite{Tanabashi:2018oca}
\bibitem{Tanabashi:2018oca}
M.~Tanabashi {\it et al.} [Particle Data Group],
%``Review of Particle Physics,''
Phys.\ Rev.\ D {\bf 98} (2018) no.3,  030001.
doi:10.1103/PhysRevD.98.030001
%%CITATION = doi:10.1103/PhysRevD.98.030001;%%
%570 citations counted in INSPIRE as of 03 Dec 2018

%\cite{Nehrkorn:2017fyt}
\bibitem{Nehrkorn:2017fyt}
A.~Nehrkorn [CMS Collaboration],
%``Search for Lepton Flavor Violation in Z and Higgs decays with the CMS Experiment,''
Nucl.\ Part.\ Phys.\ Proc.\  {\bf 287-288} (2017) 160.
doi:10.1016/j.nuclphysbps.2017.03.067
%%CITATION = doi:10.1016/j.nuclphysbps.2017.03.067;%%
%1 citations counted in INSPIRE as of 27 Dec 2018

%\cite{Khachatryan:2016rke}
\bibitem{Khachatryan:2016rke}
V.~Khachatryan {\it et al.} [CMS Collaboration],
%``Search for lepton flavour violating decays of the Higgs boson to $e \tau$ and $e \mu$ in protonï¿œproton collisions at $\sqrt s=$ 8 TeV,''
Phys.\ Lett.\ B {\bf 763} (2016) 472
doi:10.1016/j.physletb.2016.09.062
[arXiv:1607.03561 [hep-ex]].
%%CITATION = doi:10.1016/j.physletb.2016.09.062;%%
%16 citations counted in INSPIRE as of 03 Aug 2017

%\cite{Akers:1995gz}
\bibitem{Akers:1995gz}
R.~Akers {\it et al.} [OPAL Collaboration],
%``A Search for lepton flavor violating Z0 decays,''
Z.\ Phys.\ C {\bf 67} (1995) 555.
doi:10.1007/BF01553981
%%CITATION = doi:10.1007/BF01553981;%%
%70 citations counted in INSPIRE as of 03 Aug 2017    

%\cite{Sirunyan:2017xzt}
\bibitem{Sirunyan:2017xzt}
A.~M.~Sirunyan {\it et al.} [CMS Collaboration],
%``Search for lepton flavour violating decays of the Higgs boson to $\mu\tau$ and e$\tau$ in proton-proton collisions at $\sqrt{s}=$ 13 TeV,''
JHEP {\bf 1806} (2018) 001
doi:10.1007/JHEP06(2018)001
[arXiv:1712.07173 [hep-ex]].
%%CITATION = doi:10.1007/JHEP06(2018)001;%%
%16 citations counted in INSPIRE as of 27 Dec 2018

%\cite{Abreu:1996mj}
\bibitem{Abreu:1996mj}
P.~Abreu {\it et al.} [DELPHI Collaboration],
%``Search for lepton flavor number violating Z0 decays,''
Z.\ Phys.\ C {\bf 73} (1997) 243.
doi:10.1007/s002880050313
%%CITATION = doi:10.1007/s002880050313;%%
%69 citations counted in INSPIRE as of 03 Aug 2017



%%%%%%TABLE 1 REFS END%%%%%%%%%


%%%%%MU to E GAMMA%%%%%%%%%%%%%%%

%\cite{delAguila:1982yu}
\bibitem{delAguila:1982yu}
F.~del Aguila and M.~J.~Bowick,
%``Suppression of Lepton Number Violation Mediated by $\Delta$ I = 0 Mass Fermions,''
Phys.\ Lett.\  {\bf 119B} (1982) 144.
doi:10.1016/0370-2693(82)90264-7
%%CITATION = doi:10.1016/0370-2693(82)90264-7;%%
%30 citations counted in INSPIRE as of 16 Aug 2017

%\cite{Hisano:1995cp}
\bibitem{Hisano:1995cp}
J.~Hisano, T.~Moroi, K.~Tobe and M.~Yamaguchi,
%``Lepton flavor violation via right-handed neutrino Yukawa couplings in supersymmetric standard model,''
Phys.\ Rev.\ D {\bf 53} (1996) 2442
doi:10.1103/PhysRevD.53.2442
[hep-ph/9510309].
%%CITATION = doi:10.1103/PhysRevD.53.2442;%%
%663 citations counted in INSPIRE as of 16 Aug 2017

%\cite{Illana:2000ic}
\bibitem{Illana:2000ic}
J.~I.~Illana and T.~Riemann,
%``Charged lepton flavor violation from massive neutrinos in Z decays,''
Phys.\ Rev.\ D {\bf 63} (2001) 053004
doi:10.1103/PhysRevD.63.053004
[hep-ph/0010193].
%%CITATION = doi:10.1103/PhysRevD.63.053004;%%
%90 citations counted in INSPIRE as of 16 Aug 2017  

%\cite{Illana:2002tg}
\bibitem{Illana:2002tg}
J.~I.~Illana and M.~Masip,
%``Lepton flavor violation in Z and lepton decays in supersymmetric models,''
Phys.\ Rev.\ D {\bf 67} (2003) 035004
doi:10.1103/PhysRevD.67.035004
[hep-ph/0207328].
%%CITATION = doi:10.1103/PhysRevD.67.035004;%%
%35 citations counted in INSPIRE as of 16 Aug 2017 

%\cite{Arganda:2005ji}
\bibitem{Arganda:2005ji}
E.~Arganda and M.~J.~Herrero,
%``Testing supersymmetry with lepton flavor violating tau and mu decays,''
Phys.\ Rev.\ D {\bf 73} (2006) 055003
doi:10.1103/PhysRevD.73.055003
[hep-ph/0510405].
%%CITATION = doi:10.1103/PhysRevD.73.055003;%%
%177 citations counted in INSPIRE as of 16 Aug 2017

%%%%%%%%%%%%%%%%%%%%%%%%%%



%%%%%TOOLS%%%%%%%%%%%%%%%

%\cite{Mertig:1990an}
\bibitem{Mertig:1990an}
R.~Mertig, M.~Bohm and A.~Denner,
%``FEYN CALC: Computer algebraic calculation of Feynman amplitudes,''
Comput.\ Phys.\ Commun.\  {\bf 64} (1991) 345.
doi:10.1016/0010-4655(91)90130-D
%%CITATION = doi:10.1016/0010-4655(91)90130-D;%%
%626 citations counted in INSPIRE as of 16 Aug 2017

%\cite{Shtabovenko:2016sxi}
\bibitem{Shtabovenko:2016sxi}
V.~Shtabovenko, R.~Mertig and F.~Orellana,
%``New Developments in FeynCalc 9.0,''
Comput.\ Phys.\ Commun.\  {\bf 207} (2016) 432
doi:10.1016/j.cpc.2016.06.008
[arXiv:1601.01167 [hep-ph]].
%%CITATION = doi:10.1016/j.cpc.2016.06.008;%%
%66 citations counted in INSPIRE as of 16 Aug 2017

%\cite{Passarino:1978jh}
\bibitem{Passarino:1978jh}
G.~Passarino and M.~J.~G.~Veltman,
%``One Loop Corrections for e+ e- Annihilation Into mu+ mu- in the Weinberg Model,''
Nucl.\ Phys.\ B {\bf 160} (1979) 151.
doi:10.1016/0550-3213(79)90234-7
%%CITATION = doi:10.1016/0550-3213(79)90234-7;%%
%2107 citations counted in INSPIRE as of 13 May 2017  

%%%%%%%%%%%%%%%%%%%%%%%%%%% 

%%%%%% Phase space %%%%%%%%%%% 

%\cite{Ilakovac:1994kj}
\bibitem{Ilakovac:1994kj}
A.~Ilakovac and A.~Pilaftsis,
%``Flavor violating charged lepton decays in seesaw-type models,''
Nucl.\ Phys.\ B {\bf 437} (1995) 491
doi:10.1016/0550-3213(94)00567-X
[hep-ph/9403398].
%%CITATION = doi:10.1016/0550-3213(94)00567-X;%%
%298 citations counted in INSPIRE as of 05 Aug 2018

%%%%%%%%%%%%%%%%%%%%%%%%%%% 

%%%%%MU to E Conversion: LHT%%%%%%%%%%%%%%%

%\cite{Kitano:2002mt}
\bibitem{Kitano:2002mt}
R.~Kitano, M.~Koike and Y.~Okada,
%``Detailed calculation of lepton flavor violating muon electron conversion rate for various nuclei,''
Phys.\ Rev.\ D {\bf 66} (2002) 096002
Erratum: [Phys.\ Rev.\ D {\bf 76} (2007) 059902]
doi:10.1103/PhysRevD.76.059902, 10.1103/PhysRevD.66.096002
[hep-ph/0203110].
%%CITATION = doi:10.1103/PhysRevD.76.059902, 10.1103/PhysRevD.66.096002;%%
%213 citations counted in INSPIRE as of 16 Sep 2017

%\cite{Suzuki:1987jf}
\bibitem{Suzuki:1987jf}
T.~Suzuki, D.~F.~Measday and J.~P.~Roalsvig,
%``Total Nuclear Capture Rates for Negative Muons,''
Phys.\ Rev.\ C {\bf 35} (1987) 2212.
doi:10.1103/PhysRevC.35.2212
%%CITATION = doi:10.1103/PhysRevC.35.2212;%%
%303 citations counted in INSPIRE as of 25 Dec 2018  

%\cite{Abusalma:2018xem}
\bibitem{Abusalma:2018xem}
F.~Abusalma {\it et al.} [Mu2e Collaboration],
%``Expression of Interest for Evolution of the Mu2e Experiment,''
arXiv:1802.02599 [physics.ins-det].
%%CITATION = ARXIV:1802.02599;%%
%7 citations counted in INSPIRE as of 25 Dec 2018

%\cite{Angelique:2018svf}
\bibitem{Angelique:2018svf}
J.-C.~Ang\'elique {\it et al.},
%``COMET - A submission to the 2020 update of the European Strategy for Particle Physics on behalf of the COMET collaboration,''
arXiv:1812.07824 [hep-ex].
%%CITATION = ARXIV:1812.07824;%%

%%%%%%%%%%%%%%%%%%%%%%%%%%%

%%%%%%%% Higgs potential %%%%%%%%

%\cite{Coleman:1973jx}
\bibitem{Coleman:1973jx}
S.~R.~Coleman and E.~J.~Weinberg,
%``Radiative Corrections as the Origin of Spontaneous Symmetry Breaking,''
Phys.\ Rev.\ D {\bf 7} (1973) 1888.
doi:10.1103/PhysRevD.7.1888
%%CITATION = doi:10.1103/PhysRevD.7.1888;%%
%3863 citations counted in INSPIRE as of 13 May 2017

%\cite{Han:2003wu}
\bibitem{Han:2003wu} 
T.~Han, H.~E.~Logan, B.~McElrath and L.~T.~Wang,
%``Phenomenology of the little Higgs model,''
Phys.\ Rev.\ D {\bf 67}, 095004 (2003)
doi:10.1103/PhysRevD.67.095004
[hep-ph/0301040].
%%CITATION = doi:10.1103/PhysRevD.67.095004;%%
%552 citations counted in INSPIRE as of 03 Sep 2017

%%%%%%%%%%%%%%%%%%%%%%%%%%%

%\cite{Aaboud:2018pii}
\bibitem{Aaboud:2018pii}
M.~Aaboud {\it et al.} [ATLAS Collaboration],
%``Combination of the searches for pair-produced vector-like partners of the third-generation quarks at $\sqrt{s} =$ 13 TeV with the ATLAS detector,''
Phys.\ Rev.\ Lett.\  {\bf 121} (2018) no.21,  211801
doi:10.1103/PhysRevLett.121.211801
[arXiv:1808.02343 [hep-ex]].
%%CITATION = doi:10.1103/PhysRevLett.121.211801;%%
%14 citations counted in INSPIRE as of 25 Dec 2018

%%%%%%%%%%%%%%%%%%%%%%%%%%%

%\cite{Kou:2018nap}
\bibitem{Kou:2018nap}
E.~Kou {\it et al.} [Belle II Collaboration],
%``The Belle II Physics Book,''
arXiv:1808.10567 [hep-ex].
%%CITATION = ARXIV:1808.10567;%%
%64 citations counted in INSPIRE as of 16 Dec 2018 

%%%%%%%%%%%%%%%%%%%%%%%%%%%

%\cite{Barbieri:1987fn}
\bibitem{Barbieri:1987fn}
R.~Barbieri and G.~F.~Giudice,
%``Upper Bounds on Supersymmetric Particle Masses,''
Nucl.\ Phys.\ B {\bf 306} (1988) 63.
doi:10.1016/0550-3213(88)90171-X
%%CITATION = doi:10.1016/0550-3213(88)90171-X;%%
%1155 citations counted in INSPIRE as of 03 Dec 2018

%%%%%%%%%%%% FUTURE LIMITS %%%%%%%%%%%%%%%

%\cite{Baldini:2018nnn}
\bibitem{Baldini:2018nnn}
A.~M.~Baldini {\it et al.} [MEG II Collaboration],
%``The design of the MEG II experiment,''
Eur.\ Phys.\ J.\ C {\bf 78} (2018) no.5,  380
doi:10.1140/epjc/s10052-018-5845-6
[arXiv:1801.04688 [physics.ins-det]].
%%CITATION = doi:10.1140/epjc/s10052-018-5845-6;%%
%23 citations counted in INSPIRE as of 16 Dec 2018

%\cite{Nakao:2018hip}
\bibitem{Nakao:2018hip}
M.~Nakao {\it et al.},
%``Results from Pilot Run for MEG II Positron Timing Counter,''
Springer Proc.\ Phys.\  {\bf 213} (2018) 237
doi:10.1007/978-981-13-1316-5-44
[arXiv:1808.07279 [physics.ins-det]].
%%CITATION = doi:10.1007/978-981-13-1316-5_44;%% 

%\cite{Perrevoort:2018ttp}
\bibitem{Perrevoort:2018ttp}
A.~K.~Perrevoort [Mu3e Collaboration],
%``The Rare and Forbidden: Testing Physics Beyond the Standard Model with Mu3e,''
arXiv:1812.00741 [hep-ex].
%%CITATION = ARXIV:1812.00741;%%

%\cite{Blondel:2013ia}
\bibitem{Blondel:2013ia}
A.~Blondel {\it et al.},
%``Research Proposal for an Experiment to Search for the Decay $\mu \to eee$,''
arXiv:1301.6113 [physics.ins-det].
%%CITATION = ARXIV:1301.6113;%%
%198 citations counted in INSPIRE as of 16 Dec 2018

%\cite{Teshima:2018msm}
\bibitem{Teshima:2018msm}
N.~Teshima,
%``$\mu$-$e$ conversion experiments at J-PARC,''
arXiv:1811.05671 [physics.ins-det].
%%CITATION = ARXIV:1811.05671;%%


%\cite{Aaij:2014azz}
\bibitem{Aaij:2014azz}
R.~Aaij {\it et al.} [LHCb Collaboration],
%``Search for the lepton flavour violating decay ?$^{?}$ ? ?$^{?}$ ?$^{+}$ ?$^{?}$,''
JHEP {\bf 1502} (2015) 121
doi:10.1007/JHEP02(2015)121
[arXiv:1409.8548 [hep-ex]].
%%CITATION = doi:10.1007/JHEP02(2015)121;%%
%55 citations counted in INSPIRE as of 09 Feb 2019


%\cite{DeBruyn:2017aqu}
\bibitem{DeBruyn:2017aqu}
K.~De Bruyn [ATLAS and CMS and LHCb Collaborations],
%``Lepton Flavour Violation in Tau Decays: Results and Prospects at the LHC,''
Nucl.\ Part.\ Phys.\ Proc.\  {\bf 287-288} (2017) 164.
doi:10.1016/j.nuclphysbps.2017.03.068
%%CITATION = doi:10.1016/j.nuclphysbps.2017.03.068;%%
%1 citations counted in INSPIRE as of 08 Feb 2019

%%%%%%%%%%%%%%%%%%%%%%%%%

%\cite{Ellis:2002fe}
\bibitem{Ellis:2002fe}
J.~R.~Ellis, J.~Hisano, M.~Raidal and Y.~Shimizu,
%``A New parametrization of the seesaw mechanism and applications in supersymmetric models,''
Phys.\ Rev.\ D {\bf 66} (2002) 115013
doi:10.1103/PhysRevD.66.115013
[hep-ph/0206110].
%%CITATION = doi:10.1103/PhysRevD.66.115013;%%
%236 citations counted in INSPIRE as of 08 Feb 2019

%\cite{Babu:2002et}
\bibitem{Babu:2002et}
K.~S.~Babu and C.~Kolda,
%``Higgs mediated tau ---> 3 mu in the supersymmetric seesaw model,''
Phys.\ Rev.\ Lett.\  {\bf 89} (2002) 241802
doi:10.1103/PhysRevLett.89.241802
[hep-ph/0206310].
%%CITATION = doi:10.1103/PhysRevLett.89.241802;%%
%182 citations counted in INSPIRE as of 22 Dec 2018

%\cite{Brignole:2004ah}
\bibitem{Brignole:2004ah}
A.~Brignole and A.~Rossi,
%``Anatomy and phenomenology of mu-tau lepton flavor violation in the MSSM,''
Nucl.\ Phys.\ B {\bf 701} (2004) 3
doi:10.1016/j.nuclphysb.2004.08.037
[hep-ph/0404211].
%%CITATION = doi:10.1016/j.nuclphysb.2004.08.037;%%
%221 citations counted in INSPIRE as of 08 Feb 2019

%\cite{Dassinger:2007ru}
\bibitem{Dassinger:2007ru}
B.~M.~Dassinger, T.~Feldmann, T.~Mannel and S.~Turczyk,
%``Model-independent analysis of lepton flavour violating tau decays,''
JHEP {\bf 0710} (2007) 039
doi:10.1088/1126-6708/2007/10/039
[arXiv:0707.0988 [hep-ph]].
%%CITATION = doi:10.1088/1126-6708/2007/10/039;%%
%48 citations counted in INSPIRE as of 08 Feb 2019

%\cite{Dam:2018rfz}
\bibitem{Dam:2018rfz}
M.~Dam,
%``Tau-lepton Physics at the FCC-ee circular e$^+$e$^-$ Collider,''
arXiv:1811.09408 [hep-ex].
%%CITATION = ARXIV:1811.09408;%%


%%%%%%%%%          %%%%%%%%%%%%

%\cite{Branco:1999fs}
\bibitem{Branco:1999fs}
G.~C.~Branco, L.~Lavoura and J.~P.~Silva,
%``CP Violation,''
Int.\ Ser.\ Monogr.\ Phys.\  {\bf 103} (1999) 1.
%%CITATION = IMPHA,103,1;%%
%146 citations counted in INSPIRE as of 29 Aug 2017 

%\cite{Barr:1990vd}
\bibitem{Barr:1990vd}
S.~M.~Barr and A.~Zee,
%``Electric Dipole Moment of the Electron and of the Neutron,''
Phys.\ Rev.\ Lett.\  {\bf 65} (1990) 21
Erratum: [Phys.\ Rev.\ Lett.\  {\bf 65} (1990) 2920].
doi:10.1103/PhysRevLett.65.2920, 10.1103/PhysRevLett.65.21
%%CITATION = doi:10.1103/PhysRevLett.65.2920, 10.1103/PhysRevLett.65.21;%%
%455 citations counted in INSPIRE as of 26 May 2019

\end{thebibliography}
\end{document}